\documentclass[review]{elsarticle}
\graphicspath{ {./figures/} }
\usepackage[hidelinks]{hyperref}
\usepackage{amssymb,amsmath,array}
\usepackage[table,svgnames]{xcolor}
\usepackage{algorithm}
\usepackage[noend]{algpseudocode}
\usepackage{multirow}
\usepackage[section]{placeins}
\usepackage{float}
\usepackage{verbatim} 
\usepackage{subcaption}
\usepackage{graphicx}
\usepackage{adjustbox}
\usepackage[normalem]{ulem}
\usepackage{cancel}
\usepackage{placeins}
\usepackage{apalike}
\usepackage{caption}
\restylefloat{figure}
\restylefloat{table}

\journal{Expert Systems with Applications}

\biboptions{authoryear}
\let\today\relax
\begin{document}
\begin{frontmatter}










\title{Generalizing Logic-based Explanations for Machine Learning Classifiers via Optimization}

\author{Francisco Mateus Rocha Filho}
\ead{francisco.mateus.rocha06@aluno.ifce.edu.br}

\author{Ajalmar Rêgo da Rocha Neto}
\ead{ajalmar@ifce.edu.br}

\author{Thiago Alves Rocha\corref{cor1}}
\ead{thiago.alves@ifce.edu.br}

\cortext[cor1]{Corresponding author.}
\address{Instituto Federal de Educação, Ciência e Tecnologia do Ceará (IFCE), Fortaleza, Brazil}

\begin{abstract}

Machine learning models support decision-making, yet the reasons behind their predictions are opaque. Clear and reliable explanations help users make informed decisions and avoid blindly trusting model outputs. However, many existing explanation methods fail to guarantee correctness. Logic-based approaches ensure correctness but often offer overly constrained explanations, limiting coverage. Recent work addresses this by incrementally expanding explanations while maintaining correctness. This process is performed separately for each feature, adjusting both its upper and lower bounds. However, this approach faces a trade-off: smaller increments incur high computational costs, whereas larger ones may lead to explanations covering fewer instances. To overcome this, we propose two novel methods. Onestep builds upon this prior work, generating explanations in a single step for each feature and each bound, eliminating the overhead of an iterative process. \textit{Twostep} takes a gradual approach, improving coverage. Experimental results show that Twostep significantly increases explanation coverage—by up to 72.60\% on average across datasets—compared to Onestep and, consequently, to prior work.

%

\end{abstract}


\end{frontmatter}

\section{Introduction}
\label{introduction}

The increased integration of artificial intelligence (AI) into daily life has led to the more widespread use of machine learning (ML) models~\citep{lecun2015deep,biswas2023role}. Some of these models, such as Support Vector Machines (SVM) and Multilayer Perceptrons (MLP), are used for classification tasks, making decisions based on input data, and providing important information to human specialists in complex decision-making scenarios~\citep{ramkumar2021ecg,yang2018automatic}. As a result, it is crucial that the outputs generated by these models can be trusted and supported by guarantees of correctness. Concepts of eXplainable AI (XAI) can be used to better understand the answers given by such AI systems, with explainability being fundamental for delivering trustworthy models \citep{ignatiev19onValidating,ignatiev2020towards,marques2022delivering, samek2019towards}. This can be used for generating explanations that lead to a better understanding of the decision-making process of the models.

In this work, we consider instance-based explanations \citep{ignatiev2020towards}. These can provide interpretable insights regarding the decision logic used by the model to determine an answer, highlighting the relevant features of the given instance that influenced the output. The linking of the output to specific instances and their features enables a better understanding of how the model operates and may lead to a more comprehensive analysis of the decision-making process.

Most efforts in XAI lack formal guarantees of correctness, making it difficult to determine whether the models can be fully trusted. Model-agnostic methods for generating explanations, such as LIME \citep{Ribeiro2016Lime}, SHAP \citep{Lundberg2017Shap}, and Anchors \citep{Ribeiro2018Anchors}, often rely on heuristic techniques, which can lead to incorrect model interpretations. These usually generate explanations by exploring the local instance space of the given data, which may not be thorough enough and fail to represent the true behavior of the model \citep{ignatiev2020onFormal}. Moreover, they can be often proven wrong through counterexamples, exposing their contradictions. Further evidence has been found to support the assertion that these methods lack guarantees of correctness \citep{dimanov2020you,ignatiev19onValidating,ignatiev2020towards}. This makes such explanations difficult to trust in high-stakes applications (e.g., healthcare and finance), where incorrect conclusions lead to severe repercussions. Therefore, there is a need for alternative methods that ensure correctness.

In recent years, there has been a surge of research dedicated to investigating logic-based XAI for ML models, such as neural networks, naive bayes, random forests, decision trees, and boosted trees~\citep{audemard2022preferred,ignatiev2022using,ignatiev2019abduction,izza2020explaining,marques2020explaining,rocha2024logic}. Unlike heuristic approaches, logic-based methods ensure that the resulting explanations are provably correct, thereby enhancing trustworthiness~\citep{ignatiev2020towards}. Moreover, logic-based approaches provide non-redundant explanations. This is particularly important because concise explanations seem easier for humans to interpret \citep{miller1956magical}, helping them focus on the most important features. By ensuring that explanations are both correct and non-redundant, logic-based explanations are crucial for applications where explainability is not only useful but necessary, such as high-stake scenarios including legal decision-making, medical diagnostics, and loan approval~\citep{ignatiev2020towards,ignatiev2020onFormal}.

These logic-based explanations are commonly referred to as \textit{abductive explanations} \citep{ignatiev2019abduction} or \textit{sufficient reasons} \citep{shih2018symbolic}. An abductive explanation can be seen as a subset of features that together form a rule. Whenever this rule is applied, the predictions of the model remain the same \citep{ignatiev2019abduction,shih2018symbolic}. However, most rules obtained by recent work are a set of restricted values of the form $feature = value$ and therefore are only satisfied for a narrow set of points in the instance space. This type of explanation usually returns very limited information and may not relay much overall insight about the model. For example, stating that a medical analysis can be applied to patients with a height of 180 cm and an age of 56 is less informative than affirming it is also valid for heights between 175 cm and 185 cm and ages between 50 and 60. Therefore, an explanation that includes features associated with ranges of values is a lot more insightful, since a single explanation can be applied to a variety of cases. Such more general explanations can be highly beneficial in real-world applications. For instance, in finance, these explanations could be utilized in predictive models for risk assessment or loan approval. By providing more general explanations for financial features such as income, credit history, or spending behavior, financial institutions could gain deeper insights into the rationale behind AI-driven decisions. Moreover, by guaranteeing that the explanations are both correct and non-redundant, institutions could provide insights that are easier to interpret, which could ultimately lead to a more trustworthy financial system.

Despite the recent development of logic-based XAI methods, there is a lack of works dealing with guarantees of correctness and minimality that also focus on more general explanations. While Anchors \citep{Ribeiro2018Anchors} returns explanations with some degree of generalization, it is a heuristic method for generating explanations and therefore has some issues regarding giving correct answers. A recent work \citep{izza2023delivering} defined an \textit{inflated abductive explanation} as a set of features, each associated with a set of values that always includes the value of the instance being explained. Moreover, for any of the allowed values associated with features, the prediction must remain consistent with the instance being explained. The method presented by \cite{izza2023delivering} focuses on inflating feature ranges within an abductive explanation. It achieves this by incrementally adjusting the lower and upper bounds for each feature, ensuring correctness throughout the process. For every feature, the algorithm sets the initial lower and upper bound values the feature can assume equal to the original value in a given abductive explanation. Then, the algorithm seeks to inflate the upper bound by adding a small value iteratively, checking in each iteration if the predictions hold with the updated value. If it does not, the last value is discarded, with the previous one being set as the upper bound of the explanation. This process is repeated similarly for the lower bound. However, it may potentially face challenges regarding time cost for computing explanations. Small increments can lead to a large quantity of iterations, therefore increasing the cost for computing explanations. Furthermore, large increments may lead the algorithm to precociously find that the predictions may not hold. In other words, it could result in explanations with narrow feature ranges. Consequently, this may lead to a loss of coverage, restricting the instance space in which the explanation holds.

Given the importance of logic-based explainability and generalization for trustworthy ML models, this work proposes two novel approaches for inflating logic-based explanations, building on earlier work \citep{izza2023delivering}. Our first method, Onestep, uses optimization to determine explanation ranges \textit{in a single step} for each bound (upper and lower) and for each feature, avoiding the inefficiencies of incremental expansion from earlier work. Our second method, Twostep, addresses a key limitation of Onestep, where the expansion of some features can excessively narrow the ranges of others, leading to overly restrictive and less general explanations. By controlling the expansion of certain features first, Twostep allows for a more balanced and flexible range expansion, improving the generalization of explanations.

To evaluate these approaches, we focus on SVM and MLP as ML models in a classification problem scenario. We compute abductive explanations by solving a set of logical constraints, specifically, boolean combinations of linear constraints. This is achieved through first-order logic (FOL) since the resultant first-order formulas can be represented as a Mixed-Integer Linear Programming (MILP) model, which can be solved with MILP solvers. Finally, we expand explanations by attributing a range of values to each of the features. While our experiments focus on SVM and MLP, the proposed approaches can be adapted to other classifiers.

We compare Twostep and Onestep across 12 different datasets, evaluating them based on four key metrics: computation time, coverage on the original dataset, coverage on synthetic data, and range width. These metrics are crucial for understanding both the efficiency and the generalization capability of each method. In particular, coverage on synthetic data is important, as real datasets may not contain enough instances to properly evaluate the generalization capability of each method. In our comparison, Onestep serves as a reference point, improving upon the method proposed in earlier work \citep{izza2023delivering} by being more efficient and avoiding range overshooting. By comparing Twostep with Onestep, we assess not only the improvements brought by Twostep but also how it compares to prior work in terms of explanation generalization. Experimental results show that Twostep produces more general explanations, covering up to 72.60\% more instances while increasing computation time by 55.65\%.


The paper is organized as follows. Section \ref{Background} introduces the notations and definitions used in this work. Section \ref{Generating_Explanations} defines an algorithm for computing abductive explanations. Section~\ref{contributions} presents the core contributions of this paper, focusing on our two novel approaches for inflating abductive explanations: Onestep and Twostep. Furthermore, Section \ref{Experiments} depicts experiments, results, and discussions regarding the proposed algorithms, with SVM and MLP as target classifiers. Finally, conclusions and future work are described in Section~\ref{Conclusions}.

\section{Background}\label{Background}

\noindent\textbf{Classification problems}. In machine learning, classification problems are defined over a set of $n$ features $\mathcal{F} = \{f_1, ..., f_n\}$ and a set of $\mathcal{N}$ classes $\mathcal{K} = \{c_1, c_2,...,c_\mathcal{N}\}$. In this work, we consider that each feature $f_i \in \mathcal{F}$ takes its values $x_i$ from the domain of real numbers. Moreover, each feature $f_i$ has an upper bound $u_i$ and a lower bound $l_i$ such that $l_i \leq x_i \leq u_i$, and its domain is the closed interval $[l_i, u_i]$. This is represented as a set of domain constraints $D = \{l_1 \leq f_1 \leq u_1,\text{ }l_2 \leq f_2 \leq u_2, ..., l_n \leq f_n \leq u_n \}$. Furthermore, the notation $\mathbf{x} = \{f_1 = x_1, f_2 = x_2, ..., f_n = x_n\}$ represents a specific point or instance such that each $x_i$ is in the domain of $f_i$. 

A classifier $\mathcal{C}$ is a function that maps elements in the feature space into the set of classes $\mathcal{K}$. For example, $\mathcal{C}$ can map instance $\{f_1 = x_1, f_2 = x_2, ..., f_n = x_n\}$ to class $c_1$. A classifier is usually obtained by a training process, given as input a training set $\{ \mathbf{x}_i,y_i\}^{\mathcal{L}}_{i=1}$ where $\mathbf{x}_i \in \mathbb{R}^n$ is an input vector or instance, $y_i \in \mathcal{K}$ is the respective class label, and $\mathcal{L}$ is the number of training patterns. Then, for each input vector $\mathbf{x}_i$, its input values $x_{i, 1}$, $x_{i, 2}$, ..., $x_{i, n}$ are in the domain of corresponding features $f_1$, ..., $f_n$, respectively. Throughout the work, we consider the following well-known classifiers: support vector machines and multilayer perceptrons with \textit{rectified linear unit} ($ReLU$) \citep{nair2010rectified} as activation functions. In what follows, we give some details on these classifiers. \\

\noindent\textbf{Support Vector Machine}. The Support Vector Machine (SVM)~\citep{cortes1995support} is a supervised machine learning model, often used for classification problems as Support Vector Classifier (SVC). It separates the data by using the concept of an optimal separating hyperplane defined in~\citep{cortes1995support}. On a ${\rm I\!R}^n$ space, the hyperplane is defined by a set of points $\mathbf{x}$ that satisfies $\mathbf{w} \cdot \mathbf{x} + b = 0$, where $\mathbf{w} \in {\rm I\!R}^n$ is the optimal weight vector, $\mathbf{x} \in {\rm I\!R}^n $ is a feature vector with $n$ features and an intercept (bias) $b \in {\rm I\!R}$. Thus, the decision function used for classifying input instances $\mathbf{x}$ is defined as $f(\mathbf{x}) = \mathbf{w} \cdot \mathbf{x} + b$.

In this work, we consider linear SVCs in a binary classification scenario. Therefore, assuming the set of classes $\mathcal{K} = \{c_1, c_2\}$, such that $c_1 = +1$,\text{ }$c_2 = -1$, the prediction result $\hat{y} \in \mathcal{K}$ for an input instance $\mathbf{x}$ is computed as follows:\\
\begin{equation}\label{svm_prediction}
    \hat{y}= 
    \begin{cases}
        \begin{aligned}
         +1, \quad \text{if } f(\mathbf{x}) \geq 0,\\
         -1, \quad \text{if } f(\mathbf{x}) < 0. 
        \end{aligned}    
    \end{cases}
\end{equation}\\

\noindent\textbf{Multilayer Perceptron}. The Multilayer Perceptron (MLP) \citep{bishop1995neural} is a feedforward Neural Network (NN) that can be used for dealing with classification problems. An NN is composed of \textcolor{black}{$K+1$} layers of neurons. Each layer $k \in \{0, 1, ..., K\}$ is composed of $n_k$ neurons, numbered from 1 to $n_k$. Let $\mathbf{x}^k \in \mathbb{R}^{n_k}$ be the output vector of the $k$-th layer and $x^{k}_{j}$ be the output of the $j$-th neuron of that layer, with $j \in \{1,...,n_k\}$. The output values $\mathbf{x}^{k} = [x^{k}_{1},x^{k}_{2},...,x^{k}_{n_k}]$ of the neurons in a given layer $k$ are computed through the output values $\mathbf{x}^{k-1}$ of the previous layer. Assuming layer 0 as the input layer and $\mathbf{x}^{0}$ the input vector, such a process is repeated for each following layer, up to $K$ layers.

\textcolor{black}{Between each pair of consecutive layers, there is a weight matrix $W^{k-1} = [\mathbf{w}^{k-1}_{1}, \mathbf{w}^{k-1}_{2},...,\mathbf{w}^{k-1}_{n_k}]$ that connects layers $k-1$ and $k$}. Moreover, an intermediate value vector $\mathbf{y}^{k} = [y^{k}_{1},\text{ } y^{k}_{2},\text{ }...,\text{ }y^{k}_{j}]$ is computed by the neurons in layer $k$ through output values $\mathbf{x}^{k-1}$, the weights \textcolor{black}{matrix $W^{k-1}$} , and the respective bias vector $\mathbf{b}^{k-1}$. The computation of the output value of each neuron is computed by applying a non-linear activation function, such as the \textit{rectified linear unit} ($ReLU$) \citep{nair2010rectified}, on $\mathbf{y}^{k}$, pointwise. Moreover, the last layer $K$ is composed of \textcolor{black}{$\mathcal{N}$ neurons, one for each class.} Therefore, the following is obtained:

\begin{equation}
\label{eq:neuron_vector_output}
    \begin{aligned}
        &\mathbf{y}^{k} = W^{k-1}\mathbf{x}^{k-1} + \mathbf{b}^{k-1}\\
        &\mathbf{x}^{k} = ReLU(\mathbf{y^{k}}).
    \end{aligned}
\end{equation}
where \textcolor{black}{$x^{k}_{j} = max(0,\text{ } y^{k}_{j})$, for $j \in \{1,...,n_k\}$.} As such, each scalar element $x^{k}_j$ can be computed through
\begin{equation}
\label{eq:neuron_scalar_output}
    \begin{aligned}
        &{y}^{k}_j = {\mathbf{w}^{k-1}_j}^{\intercal}\mathbf{x}^{k-1} + {b}^{k-1}_j\\
        &{x}^{k}_j = ReLU(y^{k}_j),
    \end{aligned}
\end{equation}
\textcolor{black}{where $^\intercal$ is the transpose operation.}

The predicted output $\hat{y}$ is the $argmax$ result over the neuron values in the last layer $K$, computed as follows:
\begin{equation}
\label{eq:mlp_prediction}
    \begin{aligned}
        &\mathbf{\hat{y}} = argmax(\sigma(W^{K-1}\mathbf{x}^{K-1} + \mathbf{b}^{K-1})),
    \end{aligned}
\end{equation}
where $\sigma$ is the softmax function.\\

\noindent\textbf{Heuristic methods for XAI}. LIME, SHAP, and Anchors are some of the most well-known heuristic methods for generating explanations regarding ML models. These are model-agnostic approaches that generate local explanations, without taking into account the instance space as a whole \citep{Lundberg2017Shap,Ribeiro2016Lime,Ribeiro2018Anchors}. This gives a margin for explanations to fail, where the explanation does not hold when applied and leads to predictions with different classes. Moreover, they are passive to contain irrelevant elements that could otherwise be removed while still maintaining the correctness of the answer~\citep{ignatiev2020onFormal}, increasing redundancy.

Anchors have been shown as a superior version to LIME, having a better accuracy with the resultant explanations~\citep{Ribeiro2018Anchors}. The obtained explanations are decision rules designed to highlight which features of a given instance are sufficient for a classifier to make a certain prediction while being intuitive and easy to understand. For example, consider an instance $\mathbf{x} = \{$\textit{sepal\_length} $ = 4.0$, \textit{sepal\_width} $ = 1.8$, \textit{petal\_length} $ = 3.7$, \textit{petal\_width} $ = 1.2\}$ of the well known Iris dataset predicted as class \textit{versicolor} by a classifier $\mathcal{C}$. An example of an explanation obtained by Anchors is the decision rule:
$$\textbf{IF } \textit{sepal\_width} \leq 3.0 \textbf{ AND } \textit{petal\_length} > 1.5 \textbf{ THEN } \textit{versicolor}.$$

Such an explanation has some degree of generalization since it can be applied over a range of values instead of a specific point in the instance space. In this case, it is applied to all instances that follow the described rule. However, the lack of guarantees regarding correctness and minimality in such a heuristic method reduces the reliability of the generated explanations.\\

\noindent\textbf{First-Order Logic}. We use first-order logic (FOL) \citep{kroening2016decision} for generating explanations with guarantees of correctness, through quantifier-free first-order formulas over the theory of linear real arithmetic. First-order variables are allowed to assume values from the real numbers. This enables the formulation of the following formulas:

\begin{equation}
        \begin{aligned}
             \varphi, \psi &:= s \mid (\varphi \wedge \psi) \mid (\varphi \vee \psi) \mid (\neg \varphi) \mid (\varphi \to \psi),\\
             s &:= \sum^n_{i=1} a_i z_i \leq e \mid \sum^n_{i=1} a_i z_i < e,
        \end{aligned}    
\end{equation}
where $\varphi$ and $\psi$ are quantifier-free first-order formulas and $s$ represents the atomic formulas such that $n \geq 1$, each $z_i$ is a first-order variable and each $e$ and $a_i$ are concrete real numbers. By this definition, $(2.5z_1 + 3.1z_2 \geq 6) \wedge (z_1=1 \vee z_1=2) \wedge (z_1=2 \to z_2 \leq 1.1)$ is considered a formula. Observe that we allow standard abbreviations as $\neg (2.5z_1 + 3.1z_2 < 6)$ for $2.5z_1 + 3.1z_2 \geq 6$. Also note that we allow other letters for variables instead of $z_i$, such as $f_i$, $s_i$, $x_i^k$, $o_i$, and $r_i$.

By assuming the semantics of formulas over the domain of real numbers, an \textit{assignment} $\mathcal{A}$ for a formula $\varphi$ is a mapping from the first-order variables of $\varphi$ to elements in the domain of real numbers. $\mathcal{A}$ \textit{satisfies} a formula $\varphi$ if $\varphi$ is true under $\mathcal{A}$. For example, $\{z_1 \mapsto 2,\text{ } z_2 \mapsto 1.05\}$ satisfies the formula $(2.5z_1 + 3.1z_2 \geq 6) \wedge (z_1=1 \vee z_1=2) \wedge (z_1=2 \to z_2 \leq 1.1)$, whereas $\{z_1 \mapsto 2.3,\text{ } z_2 \mapsto 1\}$ does not.

If there exists an $\mathcal{A}$ that satisfies a formula $\varphi$, then $\varphi$ is \textit{satisfiable}. Then, the formula in the example above is satisfiable since $\{z_1 \mapsto 2, z_2 \mapsto 1.05\}$ satisfies the formula. The notion of satisfiability can be augmented to sets of formulas $\Gamma$. A set of first-order formulas is satisfiable if there is an $\mathcal{A}$ that makes all formulas in $\Gamma$ true simultaneously. For example, $\{(2.5z_1 + 3.1z_2 \geq 6), (z_1=1 \vee z_1=2), (z_1=2 \to z_2 \leq 1.1)\}$ is satisfiable given that $\{z_1 \mapsto 2, z_2 \mapsto 1.05\}$ jointly satisfies each one of the formulas in the set. It is well known that, for all sets of formulas $\Gamma$ and all formulas $\varphi$ and $\psi$,
\begin{equation}\label{or_unsat}
        \begin{aligned}
             \Gamma \cup \{\varphi \vee \psi\} \text{ is unsatisfiable iff } & \Gamma \cup \{\varphi\} \text{ is unsatisfiable and }\\
             & \Gamma \cup \{\psi\} \text{ is unsatisfiable.}
        \end{aligned}  
\end{equation}

For a given $\varphi$ and a set of formulas $\Gamma$, a \textit{logical consequence} can be denoted through the used of the notation 
$\Gamma \models \varphi$. This means that every assignment $\mathcal{A}$ that satisfies all formulas in $\Gamma$ also satisfies $\varphi$. As an illustrative example, let $\Gamma$ be $\{z_1 = 2, z_2 \geq 1\}$ and $\varphi$ be $(2.5z_1 + z_2 \geq 5) \wedge (z_1=1 \vee z_1=2)$. Then, $\Gamma \models \varphi$ since each satisfying assignment for all formulas in $\Gamma$ also satisfies $\varphi$. Moreover, it is widely known that, for all sets of formulas $\Gamma$ and all formulas $\varphi$,
\begin{equation}\label{entailment_and_unsat}
        \begin{aligned}
             \Gamma \models \varphi \text{ iff } \Gamma \cup \{\neg \varphi\} \text{ is unsatisfiable.} 
        \end{aligned}  
\end{equation}

For instance, there is no assignment $\mathcal{A}$ that satisfies $\{ z_1 = 2, z_2 \geq 1, \neg((2.5z_1 + z_2 \geq 5) \wedge (z_1=1 \vee z_1=2)) \}$ since an $\mathcal{A}$ that satisfies $(z_1 = 2 \wedge z_2 \geq 1)$ also satisfies $(2.5z_1 + z_2 \geq 5) \wedge (z_1=1 \vee z_1=2)$ and, therefore, does not satisfy $\neg((2.5z_1 + z_2 \geq 5) \wedge (z_1=1 \vee z_1=2))$.

We say that two first-order formulas $\varphi$ and $\psi$ are equivalent if, for each assignment $\mathcal{A}$, both $\varphi$ and $\psi$ are true under $\mathcal{A}$ or both are false under $\mathcal{A}$. We use the notation $\varphi \equiv \psi$ to represent that $\varphi$ and $\psi$ are equivalent. For example, $\neg ((z_1 + z_2 \leq 2) \wedge z_1 \geq 1)$ $\equiv$ $(\neg(z_1 + z_2 \leq 2) \vee \neg (z_1 \geq 1))$. Besides, these two formulas are also equivalent to $((z_1 + z_2 > 2) \vee z_1 < 1)$.

\section{Computing Logic-based Explanations}\label{Generating_Explanations}
As mentioned earlier, most methods for explaining ML models are heuristic, which means that their explanations cannot be fully trusted due to a lack of correctness guarantees. In this section, we present a method that addresses this issue by encoding classifiers as first-order formulas, as demonstrated in \citep{ignatiev2019abduction}. In that work, given an instance to be explained and and its corresponding prediction, an abductive explanation is a subset-minimal set of features and their associated values. Moreover, the abductive explanation ensures that if the features assume the specified values, they will always produce the same prediction. For illustration of the method presented in \citep{ignatiev2019abduction}, we use a linear SVC and MLP as examples of classifiers.

\noindent\textbf{Linear SVC}. Given a trained linear SVC in a binary classification setting, defined by a weight vector $\mathbf{w}$ and a bias $b$, we represent the classifier with the following first-order formula:
\begin{equation}\label{P_positive}
        \begin{aligned}
             P = (\sum^n_{i=1} w_i f_i  + b \geq 0).
        \end{aligned}    
\end{equation}

This formula holds for all points $\mathbf{x}$ classified as $c = +1$ by the SVC. For an instance $\mathbf{x} = \{f_1 = x_1, f_2 = x_2, ..., f_n = x_n\}$ where the domain of the features is defined by $D = \{l_1 \leq f_1 \leq u_1, l_2 \leq f_2 \leq u_2, ..., l_n \leq f_n \leq u_n \}$, an explanation is a subset $E \subseteq \mathbf{x}$ such that $E \cup D \models P$. Every assignment that satisfies all the formulas in $E \cup D$ must also satisfy $P$, ensuring that the prediction remains unchanged. Since we aim to obtain minimal explanations, $E$ must be such that, for all strict subsets $E' \subset E$, it holds that $E' \cup D \not\models P$. \textcolor{black}{It is important to note that the values in $E$ correspond to the original values of the instance $\mathbf{x}$, while the values of features in $\mathbf{x} - E$ can take any value within the domain constraints D without affecting the prediction}.

For instances predicted in other classes, the approach remains similar. For example, a first-order formula $P$ for instances classified as $c = -1$ is defined as
\begin{equation}\label{P_negative}
        \begin{aligned}
             P = (\sum^n_{i=1} w_i f_i  + b < 0).
        \end{aligned}    
\end{equation}


Following the ideas outlined by \cite{ignatiev2019abduction}, we use Algorithm~\ref{algorithm1} to compute an abductive explanation for a given $\mathbf{x}$, $D$, and $P$. For a predicted class $c$ by the classifier, $P$ is defined accordingly. The algorithm starts with $E$ initialized as $\mathbf{x}$ and iteratively removes features. If for a feature $f_i = x_i \in E$ it holds that $(E \setminus \{f_i = x_i\}) \cup D \models P$, then the value $x_i$ of feature $f_i$ is unnecessary for ensuring the prediction and $f_i = x_i$ is discarded. Otherwise, it remains in $E$. This process is performed for all features. At the end of the algorithm, it is ensured that the explanation is correct and independent of the values of the removed features.

By the result of (\ref{entailment_and_unsat}), entailments of the form $(E \setminus \{f_i = x_i\}) \cup D \models P$ in Algorithm \ref{algorithm1} can be verified through the testing of whether the set of first-order formulas $(E \setminus \{f_i = x_i\}) \cup D \cup \{\neg P\}$ is unsatisfiable. Moreover, since $(E \setminus \{f_i = x_i\}) \cup D$ is a set of linear constraints and $P$ is a linear constraint, the unsatisfiability can be efficiently checked using a linear programming (LP) solver. If the set of linear constraints $(E \setminus \{f_i = x_i\}) \cup D \cup \{\neg P\}$ has a solution, then $(E \setminus \{f_i = x_i\}) \cup D \not\models P$. From a computational complexity viewpoint, the linear programming problem is solvable in polynomial time. Then, computing abductive explanations for linear SVCs is also achievable in polynomial time, requiring a linear number of LP solver calls (proportional to the number of features).

\begin{algorithm}[H]
\caption{Computing an abductive explanation} \label{algorithm1}
\begin{algorithmic}
    \State \textbf{Input:} instance $\mathbf{x}$ predicted as $c$, domain constraints $D$, formula $P$ according to prediction $c$
    \State \textbf{Output:} abductive explanation $E$
    \State $E \gets \mathbf{x}$
    \For{$f_i=x_i \in E$}
        \If{$(E \backslash \{f_i=x_i\}) \cup D \models P$}
            \State $E \gets E \backslash \{f_i=x_i\}$
        \EndIf
    \EndFor
    \State \Return $E$
\end{algorithmic}
\end{algorithm}

\noindent\textbf{Multilayer Perceptron}. We follow the concepts of representing a neural network model as a Mixed Integer Linear Program (MILP) model depicted in \citep{fischetti2018deep}. For each layer $1 \leq k \leq K-1$, the $ReLU$ output $x^{k}_j$ of $j$th neuron is obtained through the following linear constraints:
\begin{equation}
    \begin{aligned}
        \sum^{n_{k-1}}_{i=1}w^{k-1}_{ij}x^{k-1}_{i} + b^{k-1}_{j} = x^{k}_j - s^{k}_j, \quad x^{k}_j\geq 0, \quad s^{k}_j\geq 0.
    \end{aligned}
\end{equation}

We also must impose that at least one the two terms $x^{k}_j$ and $s^{k}_j$ must be zero. Then, we introduce a binary variable $z^{k}_j\in \{0, 1\}$, imposing the logical implications
\begin{equation*}
    \begin{aligned}
        z^{k}_j=1 \to x^{k}_j \leq 0&\\
        z^{k}_j=0 \to s^{k}_j \leq 0&.
    \end{aligned}
\end{equation*}

If $z^{k}_{j}$ is equal to $1$, the $\mathrm{ReLU}$ output $x^{k}_{j}$ is $0$ and $- s^{k}_{j}$ is equal to the $\mathrm{ReLU}$ input. Otherwise, the output $x^{k}_{j}$ is equal to the $\mathrm{ReLU}$ input and $s^{k}_{j}$ is equal to $0$. The resultant formulation of the MLP is presented in constraints (\ref{eq:MILP_Conditions_1})-(\ref{eq:indicator2}):
\begin{equation}\label{eq:MILP_Conditions_1}
    \left.\begin{aligned}
        \sum^{n_{k-1}}_{i=1}w^{k-1}_{ij}x^{k-1}_{i} + b^{k-1}_{j} = x^{k}_{j} - s^{k}_{j}&\\
        x^{k}_{j} \geq 0&\\
        s^{k}_{j} \geq 0&\\
        z^{k}_{j} \in \{0,1\}&\\
        z^{k}_{j} = 1 \to x^{k}_{j} \leq 0&\\
        z^{k}_{j} = 0 \to s^{k}_{j} \leq 0&
\end{aligned}\right\} k = 1,...,K, \quad j = 1,...,n_k
\end{equation}
\begin{align}
     &l_i \leq x^0_{i} \leq u_i,\quad i = 1, ..., n \label{eq:input}\\
     &o_i = \sum_{j=1}^{n_{K-1}} w^{K}_{i,j} x^{K-1}_i + b^{K}_{i},\quad i = 1, ..., \mathcal{N} \label{eq:indicator2}
\end{align}

In the formulation, weights $w^{k-1}_{ij}$ and biases $b^{k}_{j}$ are given fixed paremeters. Moreover, each variable $x^{0}_{i}$ represents an input feature $f_i$, which has lower and upper bounds $l_i$, $u_i$, respectively, defined by the domain of the features. We assume constraints (\ref{eq:MILP_Conditions_1})-(\ref{eq:indicator2}) represent a formula $D$. Note that, in this case for MLP classifiers, $D$ consists of the domain constraints and also the formulation of the neural network. Given an instance $\mathbf{x}$ classified as class $c_{j^{'}}$ in a multiclass classification context, we define a first-order formula $P$ as:
\begin{equation}
    \begin{aligned}
       P = \bigwedge_{j\neq j^{'}} o_{j^{'}}>o_j.
    \end{aligned}
\end{equation}
This formula $P$ represents that the value of the output neuron $o_{j^{'}}$ associated with class $c_{j^{'}}$ is higher than the values of the other output neurons. Note that we do not need to consider the softmax function as it does not change the maximum value of the last layer. Similarly to what is proposed in \citep{ignatiev2019abduction}, the generated explanation for $\mathbf{x}$ is a subset $E \subseteq \mathbf{x}$ such that $E \cup D \models P$, with every assignment that satisfies all the formulas in $E \cup D$ also having to satisfy $P$. Again, $E$ must be minimal, i.e., for all subsets $E' \subset E$, it must be the case that $E' \cup D \not\models P$. Therefore, we must further elaborate $P$ while taking into account these requirements.

Through (\ref{entailment_and_unsat}), $E \cup D \models P$ iff $E \cup D \cup \{\neg P\}$ is unsatisfiable. The constraints in $E \cup D$ are accepted by modern MILP solvers. We represent $\neg P$ as
\begin{equation}
    \begin{aligned}
        \neg &(o_{j'} > o_1 \wedge ... \wedge o_{j'} > o_\mathcal{N}),\quad\text{ which is equivalent to}\\
        &(o_{j'} \leq o_1 \vee ... \vee o_{j'} \leq o_\mathcal{N})\text{.}
    \end{aligned}
\end{equation}
Therefore, to represent this disjunction as constraints accepted by modern MILP solvers, we add a binary variable $r_j$ for each output neuron $o_j$ such that $j \neq j'$, defined as follows:
\begin{equation}
    \sum_{j \neq j'} r_j \geq 1
\end{equation}
\begin{equation}\label{Loser_Neurons}
    \left.\begin{aligned}
        r_{j} = 1 \to o_{j'} \leq o_{j}&\\
        r_j \in \{0, 1\}&
    \end{aligned}\right\}j \in \{1,...,\mathcal{N}\},\quad j\neq j'.
\end{equation}

Therefore, we can check whether $E \cup D \cup \{\neg P\}$ is unsatisfiable by modern MILP solvers. Furthermore, we also can use Algorithm \ref{algorithm1} for computing abductive explanations for MLP classifiers.

\section{Inflating Logic-based Explanations via Optimization}\label{contributions}
An abductive explanation $E \subseteq \mathbf{x}$, obtained as in Section \ref{Generating_Explanations}, contains features and their corresponding values that will lead to the same class, independently of the remaining features in $\mathbf{x} - E$.  However, it is potentially the case that variations of values in some features in $E$ still result in the same decision. Therefore, the features present in $E$ are the targets for expansion. An \emph{inflated abductive explanation} consists of a set of features, each associated with a range of values, such that the prediction remains unchanged for any of the allowed values.

\begin{figure}[!htpb]
\caption{\label{fig:Inflated_ranges_example} Example of how the range of a feature $f_i$ is computed with an incremental approach.}
\centering
\resizebox{1\textwidth}{!}{
\includegraphics[]{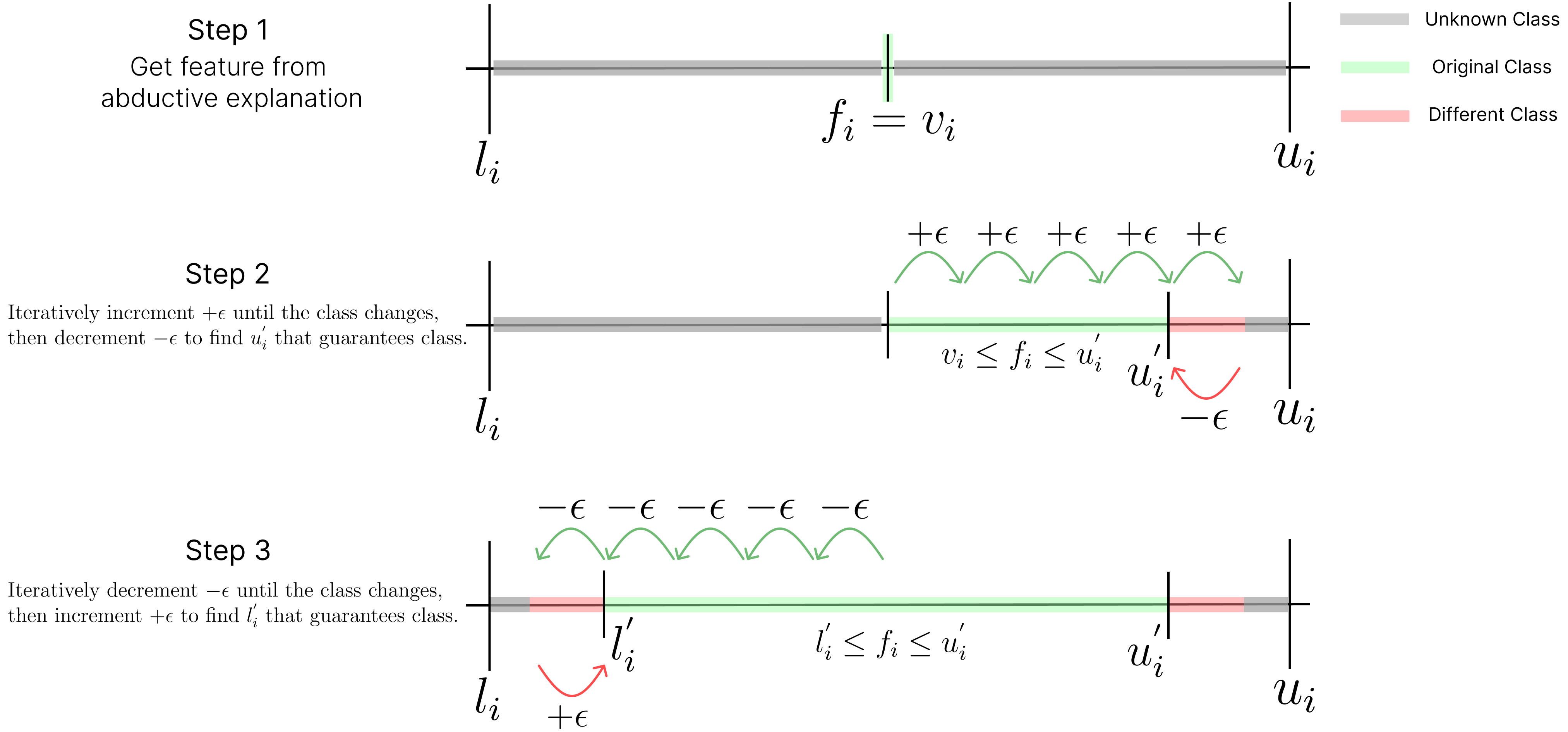}
}
\end{figure}

Earlier work \citep{izza2023delivering} proposed a method that expands feature ranges, starting from an abductive explanation, by iteratively adjusting the upper and lower bounds of each feature, while ensuring the same prediction. This strategy is illustrated in Figure~\ref{fig:Inflated_ranges_example}. However, small increments can lead to a high number of iterations, thus increasing computational cost. Large increments, on the other hand, may lead to overshooting, causing a premature conclusion that the prediction does not hold, which results in narrower feature ranges.

Building on this, we propose a more efficient approach to compute inflated abductive explanations. Our first novel approach, named \textbf{Onestep}, has the core difference in the use of optimization to find the maximum and minimum values of the ranges \textit{in one step} instead of an incremental approach. Our tactic enables minimizing the issues derived from the size of the increments used in earlier work \citep{izza2023delivering}.

We define the explanation expansion task as a sequence of optimization problems. In the initial step, given $f_i=x_i$ in an abductive explanation $E$, it involves finding $u'_i$ such that $x_i \leq u'_i \leq u_i$ and $(E\setminus\{f_i=x_i\})\cup \{f_i \leq u'_i\} \cup D \models P$, i.e., replacing $f_i=x_i$ with $f_i \leq u'_i$ does not alter the predictions, ensuring that they remain in the same class. We determine $u'_i$ by finding the lowest value $t^{*}_i$ above the original value $x_i$ such that the prediction changes. The following is the description of the optimization model for finding $t^{*}_i$:
\begin{equation}\label{eq:first_upperbound}
    \begin{aligned}
        min& \quad f_i\\
        s.t.& \quad x_i \leq f_i \leq u_i\\
        & \quad f_j = x_j, \quad \text{for all } f_j = x_j \in E \text{ with } j \neq i \\
        & \quad l_g \leq f_g \leq u_g, \quad \text{for all } f_g = x_g \in \mathbf{x} \setminus E\\
        & \quad \mathcal{R}
    \end{aligned}
\end{equation}
whereas $x_i$ and $x_j$ are the values of features $f_i,f_j$, respectively, $f_j = x_j \in E$ are the features in the abductive explanation whose ranges have not been found yet, and $f_g = x_g \in \mathbf{x} \setminus E$ are the features outside of the abductive explanation $E$, and $\mathcal{R}$ is a constraint imposing that the prediction does not hold. This constraint depends on the classifier. For SVC, if the prediction result is class $c = +1$, then $\mathcal{R}$ is $P$ as in (\ref{P_negative}), i.e., $P = (\sum^n_{i=1} w_i f_i  + b < 0)$. Otherwise, if $c = -1$ then $\mathcal{R}$ is $P$ as in (\ref{P_positive}). A similar process is done for MLP.

\begin{figure}[!htpb]
\caption{\label{fig:Onestep_ranges_example} Example of how the range of a feature $f_i$ is computed with Onestep.}
\centering
\resizebox{1\textwidth}{!}{
\includegraphics[]{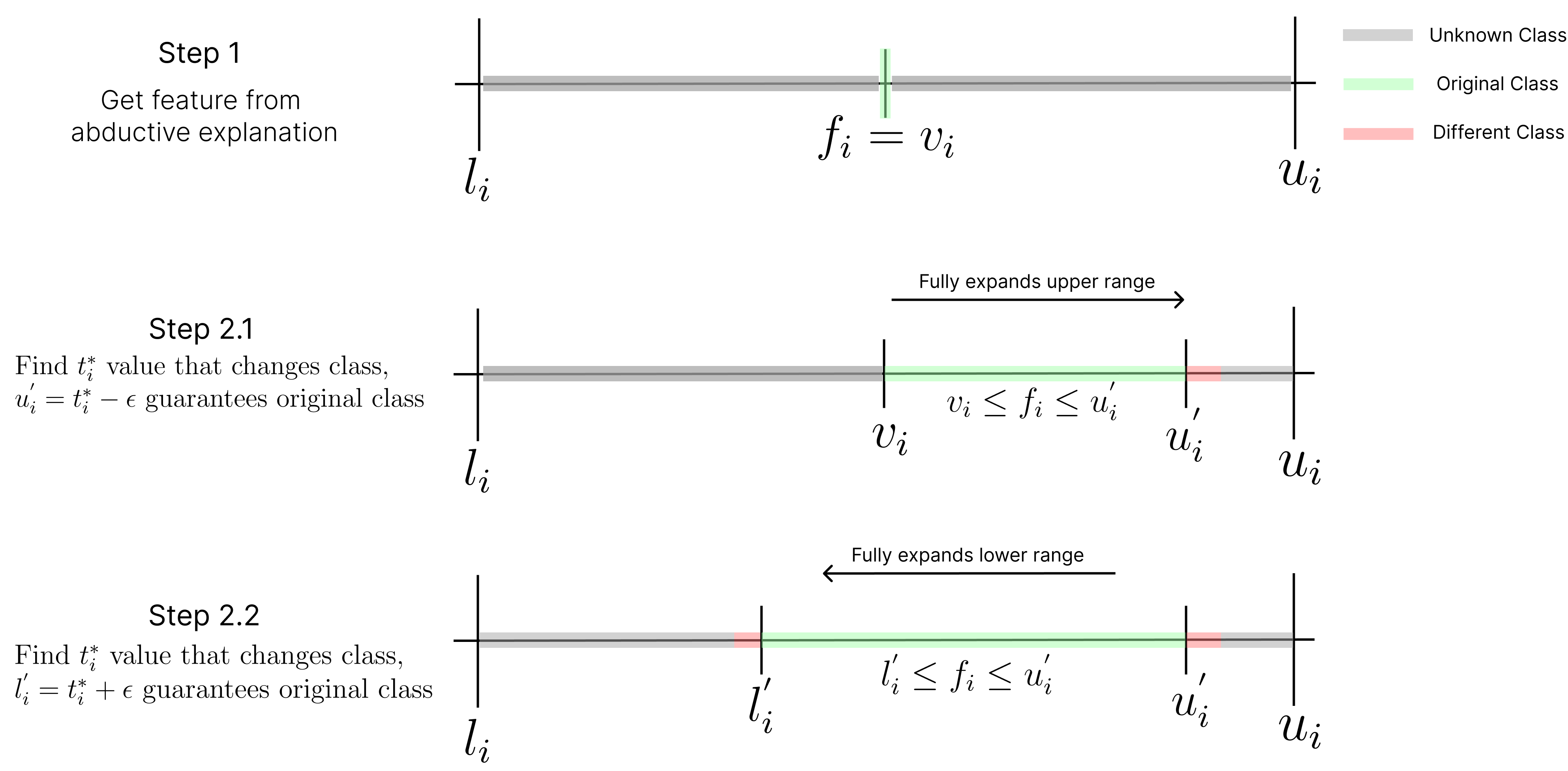}
}
\end{figure}

If the set of constraints in the problem is satisfiable, the optimal solution value $t_{i}^{*}$ for feature $f_i$ is found. Thus, it is the lowest value above the original value, i.e., $x_i < t_{i}^{*} \leq u_i$ that changes the prediction result. Therefore, we set $u'_i = t_{i}^{*} - \varepsilon$, with $\varepsilon$ assuming a small positive real value. Moreover, if $t_{i}^{*} - \varepsilon < x_i$ then we set $u'_i = x_i$, guaranteeing that $x_i \leq u'_i$. The value obtained from $t_{i}^{*} - \varepsilon$ is the highest value (up to an imprecision of $\varepsilon$) that $f_i$ can assume ensuring that the prediction hold for all values in $[x_i, u'_i]$. Otherwise, if the set of constraints is unsatisfiable, then there is no value for $f_i$ such that $x_i < f_i \leq u_i$ that changes the prediction. Therefore, we set $u'_i = u_i$. A similar process is done for the lower bound $l'_{i}$:

\begin{equation}\label{eq:first_lowerbound}
    \begin{aligned}
        max& \quad f_i \\
        s.t.& \quad l_i \leq f_i \leq x_i\\
        & \quad f_j = x_j, \quad \text{for all } f_j = x_j \in E \text{ with } j \neq i\\
        & \quad l_g \leq f_g \leq u_g, \quad \text{for all } f_g = x_g \in \mathbf{x} - E \\
        &\quad \mathcal{R}
    \end{aligned}
\end{equation}
If the problem is satisfiable, the optimal value $t_{i}^{*}$ for $f_i$ is found. Therefore, $t_{i}^{*}$, which is the highest value below the original value, i.e., $l_i \leq t_{i}^{*} < x_i$ that makes the prediction change. In this case, we set $l'_{i} = t_{i}^{*} + \varepsilon$. Then, $l_{i}^{'}$ is the lowest value that makes the prediction hold. Moreover, if $t_{i}^{*} + \varepsilon > x_i$ then we set $l'_i = x_i$, guaranteeing that $l'_{i} \leq x_i$. Otherwise, if the problem is unsatisfiable, we set $l'_i = l_i$. Therefore, the prediction result holds if the feature assumes any value within the range $[l'_i, u'_i]$, with every other feature $f_j$ such that $f_j = x_j \in E$ still set to their values in the abductive explanation. Figure \ref{fig:Onestep_ranges_example} shows a simplified illustration of how the range of a feature $f_i$ is computed with Onestep.

Next, for every other feature, we find the new maximum and lowest values of the range through 
\begin{equation}\label{eq:subsequent_upperbound}
    \begin{aligned}
        min& \quad f_i\\
        s.t.& \quad l'_h \leq f_h \leq u'_h, \quad \text{for all } f_h \text{ such that } f_h = x_j \in E \text{ and } h < i\\
        & \quad x_i \leq f_i \leq u_i,\\
        & \quad f_j = x_j, \quad \text{for all } f_j \text{ such that } f_j = x_j \in E \text{ and } i < j \leq n\\
        & \quad l_g \leq f_g \leq u_g, \quad \text{for all } f_g \text{ such that } f_g = x_g \in \mathbf{x} - E\\
        & \quad \mathcal{R}
    \end{aligned}
\end{equation}
and
\begin{equation}\label{eq:subsequent_lowerbound}
    \begin{aligned}
        max& \quad f_i\\
        s.t.& \quad l'_h \leq f_h \leq u'_h, \quad \text{for all } f_h \text{ such that } f_h = x_j \in E \text{ and } h < i\\
        & \quad l_i \leq f_i \leq x_i,\\
        & \quad f_j = x_j, \quad \text{for all } f_j \text{ such that } f_j = x_j \in E \text{ and }  i < j \leq n\\
        & \quad l_g \leq f_g \leq u_g, \quad \text{for all } f_g \text{ such that } f_g = x_g \in \mathbf{x} - E\\
        & \quad \mathcal{R}
    \end{aligned}
\end{equation}
respectively. Note that $l'_h$ and $ u'_h$, for all $f_h \in E$ such that $h < i$, are the previously found ranges that make the predictions hold. Our method Onestep is shown in Algorithm \ref{Onestep_algorithm}. The method \emph{solve} in the algorithm directly computes the value of $u'_i$, as described earlier. 



\begin{algorithm}[H]
\scriptsize
\caption{Onestep} \label{Onestep_algorithm}
\begin{algorithmic}
    \State \textbf{Input:} instance $\mathbf{x}$ predicted as $c$, domain constraints $D$, constraint $\mathcal{R}$ according to $c$, abductive explanation $E$
    \State \textbf{Output:} inflated abductive explanation $E_{xp}$
    \State $h \gets \{\}$
    \State $E_{xp} \gets \{\}$
    
    \State let $f_i = x_i \in E$
    
    \State min\_opt\_model\ $\gets$ create\_first\_min\_model($\mathbf{x}$, $D$, $\mathcal{R}$, $E$, $f_i = x_i$)\Comment{\ref{eq:first_upperbound}}
    
    \State max\_opt\_model\ $\gets$ create\_first\_max\_model($\mathbf{x}$, $D$, $\mathcal{R}$, $E$, $f_i = x_i$) \Comment{\ref{eq:first_lowerbound}}
    \State $u'_i \gets$ solve(min\_opt\_model)
    \State $l'_i \gets$ solve(max\_opt\_model)  
    
    \State $h \gets h \cup \{f_i = x_i\} $
    \State $E_{xp} \gets E_{xp} \cup \{l'_i \leq f_i \leq u'_i\}$
    
    \For{$f_i=x_i \in E$ such that $f_i=x_i \not\in h$}
        
            \State min\_opt\_model\ $\gets$ create\_min\_model($\mathbf{x}$, $D$, $\mathcal{R}$, $E_{xp}$, $f_i=x_i$, $E$)
            \Comment{\ref{eq:subsequent_upperbound}}
            
            \State max\_opt\_model\ $\gets$ create\_max\_model($\mathbf{x}$, $D$, $\mathcal{R}$, $E_{xp}$, $f_i=x_i$, $E$)
            \Comment{\ref{eq:subsequent_lowerbound}}
            \State $u'_i \gets$ solve(min\_opt\_model)
            \State $l'_i \gets$ solve(max\_opt\_model)
            
            \State $h \gets h \cup \{f_i = x_i\} $
            \State $E_{xp} \gets E_{xp} \cup \{l'_i \leq f_i \leq u'_i\}$
        
    \EndFor
    \State \Return $E_{xp}$
    
\end{algorithmic}
\end{algorithm}

Our second method, named \textbf{Twostep}, follows a similar approach to Onestep, with the main difference being how the range of values is set for each feature $f_i$. While Onestep can find ranges that guarantee the prediction, some features may have been expanded too much, which limits the possible ranges for other features. As a result, these other features are restricted to a very small range. Moreover, it is possible that $l^{'}_i = u^{'}_i = x_i$, that is, the range could not be expanded, due to any other value making the prediction change.

Our method Twostep is a strategy for reducing the rate of such occurrences. This involves reducing the range of some features to allow the expansion of other ranges.  We define $l''_i$ and $u''_i$ as lower and upper limits of a subrange within $[l'_i,$ $u'_i]$, i.e., $l'_i \leq l''_i$, $u''_i \leq u'_i$, and $l''_i \leq u''_i$. Moreover, we set $l''_i \leq f_i \leq u''_i$, referring to all values the feature $f_i$ can assume within the subrange. 

\begin{figure}[!htpb]
\caption{\label{fig:Twostep_ranges_example} Illustration of how a partial range for $f_i$ is computed with Twostep.}
\centering
\resizebox{1\textwidth}{!}{
\includegraphics[]{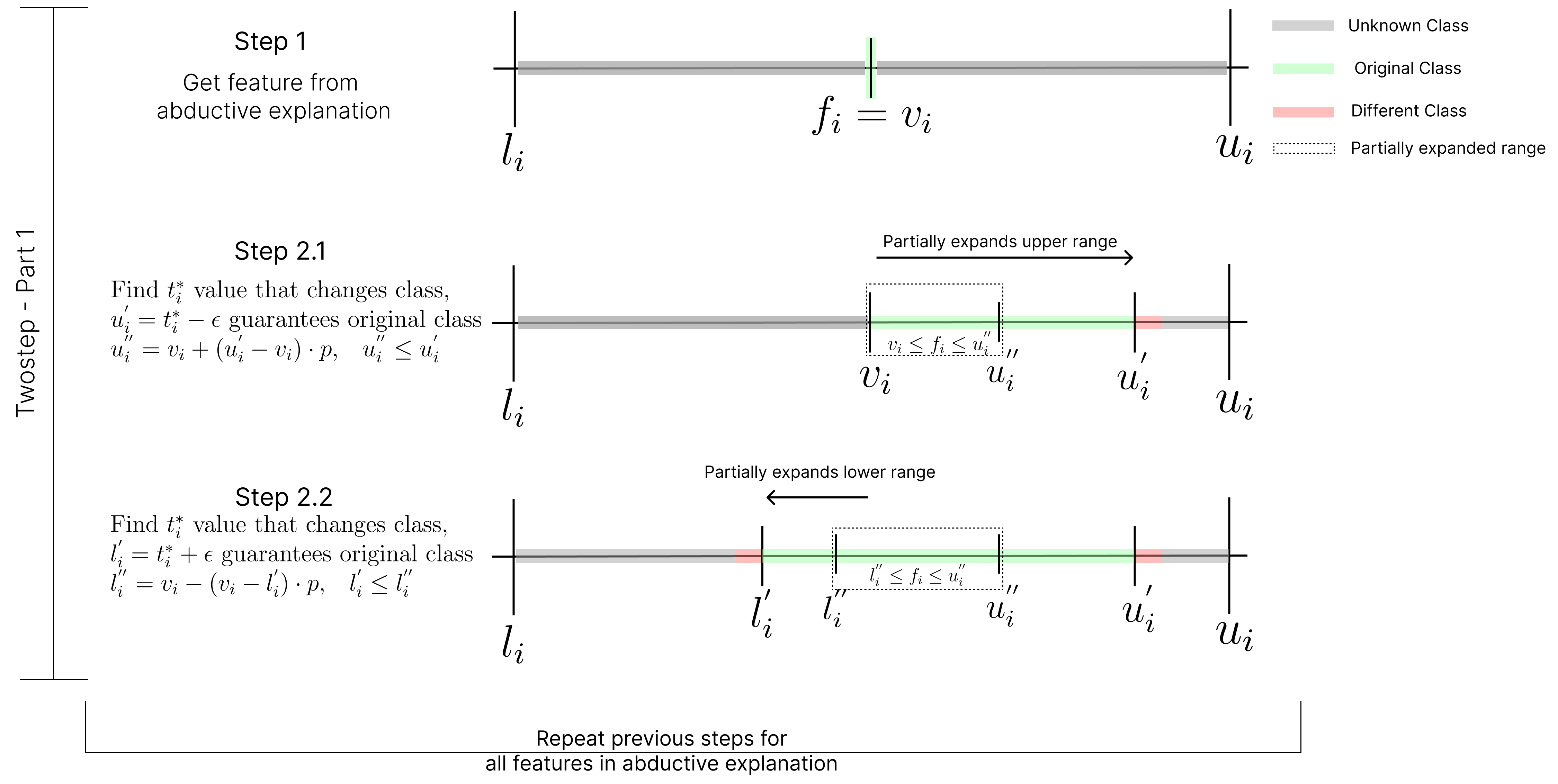}
}
\end{figure}

In the initial phase of Twostep, we determine $l'_i$ and $u'_i$ as depicted for Onestep. Then, we set $l''_i = x_i - (x_i - l'_i) \cdot p$ and $u''_i = x_i + (u'_i - x_i) \cdot p$, where $p$ is a given parameter with value $0 < p \leq 1$ that controls the size of the subrange. Note that $p = 1$ is equivalent to Onestep. For example, if $p = 0.5$, then $l''_i = x_i - (x_i - l'_i) \cdot 0.5$ and $u''_i = x_i + (u'_i - x_i) \cdot 0.5$, that is, the lower and upper limits of the subrange are equivalent to 50\% of the respective original range limits. Figure \ref{fig:Twostep_ranges_example} illustrates the initial phase of Twostep for a feature $f_i$.

The second step of our algorithm comprises expanding the found ranges while ensuring the predictions hold. Therefore, we can employ the following optimization problems for inflating the ranges while considering the found values for $u''_i$ and $l''_i$ in the first step of our method:
\begin{equation}\label{eq:twostep_upperbound}
    \begin{aligned}
        min& \quad f_i\\
        s.t.& \quad l'_h \leq f_h \leq u'_h, \quad \text{for all } f_h \text{ such that } f_h = x_j \in E \text{ and } h < i\\
        & \quad u''_i \leq f_i \leq u_i,\\
        & \quad l''_j \leq f_j \leq u''_j, \quad \text{for all } f_j \text{ such that } f_j = x_j \in E \text{ and } i < j \leq n\\
        & \quad l_g \leq f_g \leq u_g, \quad \text{for all } f_g \text{ such that } f_g = x_g \in \mathbf{x} - E\\
        & \quad \mathcal{R}
    \end{aligned}
\end{equation}
\begin{equation}\label{eq:twostep_lowerbound}
    \begin{aligned}
        max& \quad f_i\\
        s.t.& \quad l'_h \leq f_h \leq u'_h, \quad \text{for all } f_h \text{ such that } f_h = x_j \in E \text{ and } h < i\\
        & \quad l_i \leq f_i \leq l''_i,\\
        & \quad l''_j \leq f_j \leq u''_j, \quad \text{for all } f_j \text{ such that } f_j = x_j \in E \text{ and }  i < j \leq n\\
        & \quad l_g \leq f_g \leq u_g, \quad \text{for all } f_g \text{ such that } f_g = x_g \in \mathbf{x} - E\\
        & \quad \mathcal{R}.
    \end{aligned}
\end{equation}

Algorithm \ref{Twostep_algorithm} describes the method Twostep. The first step of the algorithm involves finding $u'_i$ and $l'_i$ for the first feature through (\ref{eq:first_upperbound}) and (\ref{eq:first_lowerbound}). Then, we obtain the intermediate bounds $u''_i$ and $l''_i$ by reducing $u'_i$ and $l'_i$ according to $p$. We employ (\ref{eq:subsequent_upperbound}) and (\ref{eq:subsequent_lowerbound}) to find $u'_i$ and $l'_i$, and therefore $u''_i$ and $l''_i$, for each of the subsequent features. The second step of the algorithm comprises inflating the found ranges as far as possible while still ensuring the predictions hold, using (\ref{eq:twostep_upperbound}-\ref{eq:twostep_lowerbound}) for each feature. The algorithm outputs an inflated abductive explanation $E_{xp}$, containing the updated range of values each feature can assume without changing the prediction.

\begin{algorithm}[H]
\scriptsize
\caption{Twostep} \label{Twostep_algorithm}
\begin{algorithmic}
    \State \textbf{Input:} instance $\mathbf{x}$ predicted as $c$, domain constraints $D$, restriction $\mathcal{R}$ according to $c$, abductive explanation $E$, parameter $p$
    \State \textbf{Output:} inflated abductive explanation $E_{xp}$
    \State $h \gets \{\}$\Comment{Step 1}
    \State $E_{in} \gets \{\}$
    \State let $f_i = x_i \in E$
    \State min\_opt\_model\ $\gets$ create\_first\_min\_model($\mathbf{x}$, $D$, $\mathcal{R}$, $E$, $f_i = x_i$)\Comment{\ref{eq:first_upperbound}}
    \State max\_opt\_model\ $\gets$ create\_first\_max\_model($\mathbf{x}$, $D$, $\mathcal{R}$, $E$, $f_i = x_i$) \Comment{\ref{eq:first_lowerbound}}
    \State $u'_i$, $l'_i\gets$ solve(min\_opt\_model),\quad solve(max\_opt\_model)
    \State $u''_i,$ $l''_i \gets x_i + (u'_i - x_i) \cdot p$,\quad $x_i - (x_i - l'_i) \cdot p$
    \State $h \gets h \cup \{f_i = x_i\} $
    \State $E_{in} \gets E_{in} \cup \{l''_i \leq f_i \leq u''_i\}$
    
    \For{$f_i=x_i \in E$ such that $f_i=x_i \not\in h$}
            \State min\_opt\_model\ $\gets$ create\_min\_model($\mathbf{x}$, $D$, $\mathcal{R}$, $E_{in}$, $f_i = x_i$)
            \Comment{\ref{eq:subsequent_upperbound}}
            \State max\_opt\_model\ $\gets$ create\_max\_model($\mathbf{x}$, $D$, $\mathcal{R}$, $E_{in}$, $f_i = x_i$)
            \Comment{\ref{eq:subsequent_lowerbound}}
            \State $u'_i$, $l'_i\gets$ solve(min\_opt\_model),\quad solve(max\_opt\_model)
            \State $u''_i,$ $l''_i \gets x_i + (u'_i - x_i) \cdot p$,\quad $x_i - (x_i - l'_i) \cdot p$
            \State $h \gets h \cup \{f_i = x_i\} $
            \State $E_{in} \gets E_{in} \cup \{l''_i \leq f_i \leq u''_i\}$
    \EndFor
    \State $h \gets \{\}$\Comment{Step 2}
    \State $E_{xp} \gets \{\}$
    \For{$l''_i \leq f_i \leq u''_i\in E_{in}$ such that $l''_i \leq f_i \leq u''_i \not\in h$}
            \State min\_opt\_model\ $\gets$ create\_min\_model($\mathbf{x}$, $D$, $\mathcal{R}$, $E_{xp}$, $l''_i \leq f_i \leq u''_i$, $E_{in}$)
            \Comment{\ref{eq:twostep_upperbound}}
            \State max\_opt\_model\ $\gets$ create\_max\_model($\mathbf{x}$, $D$, $\mathcal{R}$, $E_{xp}$, $l''_i \leq f_i \leq u''_i$, $E_{in}$)
            \Comment{\ref{eq:twostep_lowerbound}}
            \State $u'_i$, $l'_i\gets$ solve(min\_opt\_model),\quad solve(max\_opt\_model)
            \State $h \gets h \cup \{l''_i \leq f_i \leq u''_i\}$
            \State $E_{xp} \gets E_{xp} \cup \{l'_i \leq f_i \leq u'_i\}$
    \EndFor
    
    \State \Return $E_{xp}$
    
\end{algorithmic}
\end{algorithm}

\section{Experiments}\label{Experiments}

In this work, we used a total of 12 datasets in our experiments\footnote{Code available at \href{https://github.com/franciscomateus0119/GExp}{https://github.com/franciscomateus0119/GExp}}: Blood Transfusion, Vertebral Column, Parkinsons, Ionosphere, Banknote Authentication, Climate Model Simulation Crashes, Glass, and User Knowledge Modeling, which are available at the UCI machine learning repository\footnote{\href{https://archive.ics.uci.edu/ml/datasets}{https://archive.ics.uci.edu/ml/datasets}}; Pima Indians Diabetes, available on Kaggle\footnote{\href{https://www.kaggle.com/datasets/uciml/pima-indians-diabetes-database}{https://www.kaggle.com/datasets/uciml/pima-indians-diabetes-database}}; Iris, Breast Cancer Wisconsin, and Wine, available through the scikit-learn package\footnote{\href{https://github.com/scikit-learn/scikit-learn/tree/main/sklearn/datasets/data}{https://github.com/scikit-learn/scikit-learn/tree/main/sklearn/datasets/data}}. All features were scaled to the range $[0, 1]$, as a standard procedure.


To ensure binary classification in the context of the linear SVC, classes in the Iris, Wine, User Knowledge Modeling, and Glass datasets were binarized. For the MLP, the original version of these datasets was used, i.e., with three classes for both Iris and Wine, four classes for User Knowledge Modeling, and six classes for Glass. The remaining datasets already contain two classes. Any categorical data has been removed since we are assuming features have continuous values. A summary of the datasets, with the said modifications, is presented in Table \ref{tab:Datasets Details}.


\begin{table}[htbp]
\caption{\textcolor{black}{Datasets details.}}
\label{tab:Datasets Details}
\rowcolors{2}{white}{lightgray} 
\centering 
\resizebox{1.0\textwidth}{!}{
    \begin{tabular}{|c|c|c|c|}\hline
    Dataset                         & Acronym & Number of Features & Instances \\\hline
    Iris                            & IRIS & 4 & 150 \\
    Blood Transfusion               & BLDT & 4 & 748 \\
    Banknote Authentication         & BANK& 4& 1372\\
    User Knowledge Modeling         & UKMO& 5& 403\\
    Vertebral Column                & VRTC & 6 & 310 \\
    Pima                            & PIMA & 8 & 768 \\
    Glass                           & GLAS& 9& 214\\
    Wine                            & WINE & 13 & 178 \\
    Climate Model Simulation Crashes& CLIM& 18& 540\\
    Parkinsons                      & PARK & 22 & 195 \\
    Breast Cancer Wisconsin         & BRCW & 30 & 569 \\
    Ionosphere                      & IONS & 34 & 351\\
    \hline
    \end{tabular}
}
\end{table}

\textbf{The SVM classifier}. For each dataset, a linear SVC was trained based on 25\% of the original data. A regularization parameter $C=1$ was used, along with stratified sampling. Other parameters were set to their default value, as stated within the scikit-learn package. The average accuracy of the linear SVM classifiers trained across the datasets was 87\%.

\textbf{The MLP classifier}. For each dataset, an MLP was trained based on 25\% of the original data. A single hidden layer with the number of neurons equal to the input layer size was used, along with ReLU as activation function. On the output layer, a softmax activation was used. Models were compiled with Adam Optimizer. Each model was trained with 400 epochs, a batch size of 32, and an early stopping patience of 40, saving only the best model. The training was done with stratified sampling. The average accuracy of the MLP classifiers trained across the datasets was 83\%.

It is important to note that, in the context of our experiment, the proportion of instances used for training is not the primary focus, as our explanation methods aim to explain the trained models rather than the data itself. In our evaluation, we generate explanations for the test instances, meaning that a larger number of test instances allows for a more comprehensive assessment of the explanation methods.



\textbf{Our approaches.} Both Onestep and Twostep prototype implementations were written in Python, following Algorithms \ref{Onestep_algorithm} and \ref{Twostep_algorithm}. In our comparison, Onestep serves as a reference point, as it improves upon the method proposed in earlier work \citep{izza2023delivering} by being more efficient and avoiding range overshooting. Thus, comparing Twostep with Onestep allows us to assess not only the improvements brought by Twostep but also how it compares to earlier work \citep{izza2023delivering} in terms of explanation generalization. The CPLEX MILP solver was chosen to check the unsatisfiability of sets of first-order sentences and to solve the optimization problems in (\ref{eq:first_upperbound})-(\ref{eq:twostep_lowerbound}). The modeling of the optimization problems in (\ref{eq:first_upperbound})-(\ref{eq:twostep_lowerbound}) was done with the DOcplex Python Modeling API. We evaluated Twostep using the parameter values $p \in \{0.25,0.50,0.75\}$ to compute the results. 


\textbf{Evaluation.} Since we used 25\% of each dataset for training, the remaining 75\% served as the target for generating explanations. To compare each approach, we evaluated the following metrics:
\begin{itemize}
    \item \textbf{Computation time}: The mean elapsed time for computing explanations.
    
    \item \textbf{Explanation range width}: The mean range width of the explanations, where the range width of an explanation is computed as:
    
    \begin{equation}
        \frac{\sum\limits_{l^{'}_i \leq f_i \leq u^{'}_i \in E_{xp}} (u^{'}_i - l^{'}_i)}{|E_{xp}|}
    \end{equation}
    
    \item \textbf{Explanation coverage}: To assess how well the generated explanations generalize, we analyze two types of coverage:
    \begin{enumerate}
        \item \textbf{Dataset coverage} – The proportion of original dataset instances covered by each explanation, i.e., instances whose feature values fall within the computed explanation ranges. We compute the average coverage across all generated explanations to obtain a measure of how well each method generalizes over the dataset. We also conduct a per-instance comparison, calculating the percentage improvement in coverage that Twostep provides over Onestep for each test instance.

        \item \textbf{Synthetic data coverage} – To further assess the performance of the Onestep and Twostep methods, we also focus on the proportion of artificially generated instances covered by the explanations. This second metric is particularly relevant, as real datasets may not contain enough instances to properly evaluate the generalization capability of each method. Similar to dataset coverage analysis, we compare the percentage improvement in coverage of Twostep over Onestep for a per-instance basis. For each original instance $\mathbf{x}=\{f_1=v_1,f_2=v_2,…,f_n=v_n\}$, we generate one explanation using Onestep and another using Twostep. Then, we generate 100 artificial instances by perturbing each feature value within a controlled range. Specifically, each artificial instance takes the form $\mathbf{x'}=\{f_1 = v_1',f_2 = v_2', ... ,f_n = v'_n\}$, where $v_i' \in [v_i - d, v_i + d]$, and $d$ defines the magnitude of the perturbation. The choice of dd is crucial. If $d$ is too large, particularly for high-dimensional datasets, both methods may fail to cover any of the artificial instances. Conversely, if $d$ is too small, and the dataset does not have too many dimensions, it is possible that both methods will cover all artificial instances in many cases, which may not provide a clear distinction in their generalization performance. A value of $d=0.1$ is used for most datasets. However, for BRCW and IONS, which have a higher dimensionality, a smaller value of $d=0.001$ is applied.
    \end{enumerate}
\end{itemize}

It is important to note that the correctness and minimality of the generated explanations are guaranteed by the Onestep and Twostep methods. Then, there is no need to assess the fidelity or accuracy of the explanations with respect to the trained models.

        
        
        


\subsection{Results of the Experiments}


In this subsection, we present the results of the experiments conducted to compare the Onestep and Twostep methods. Computation time results are shown in Table \ref{tab:Time_SVM} and Table \ref{tab:Time_MLP}, for the SVM and MLP models, respectively. Dataset coverage results are presented in Table \ref{tab:Coverage_SVM} and Table \ref{tab:Coverage_MLP}, for SVM and MLP, respectively. A per-instance comparison for dataset coverage is illustrated in Figures \ref{fig:SVM_relative_coverage} and \ref{fig:MLP_relative_coverage} for the SVM and MLP classifiers, respectively. These histograms show the number of cases where Twostep achieves higher coverage than Onestep. Additionally, we analyze the distribution of coverage improvement across different percentage intervals to better understand the magnitude of improvement. Synthetic data coverage results are displayed in Figures \ref{fig:SVM_artificial_coverage} and \ref{fig:MLP_artificial_coverage}, also in histogram form. Explanation range width results can be found in Table \ref{tab:Sum_of_Ranges_SVM} and Table \ref{tab:Sum_of_Ranges_MLP}.

\subsubsection{Results for Computation Time}
As expected, Twostep tends to be slower than Onestep, with an increase of up to approximately 155.72\% (1.941 seconds vs. 4.949 seconds) in the IONS dataset when using MLP as the classifier, as shown in Table \ref{tab:Time_MLP}. This slowdown occurs because Twostep requires additional computational steps to generate explanations. Therefore, in situations where computational resources are limited, the performance of Twostep might be constrained, especially in larger datasets or when faster response times are necessary. However, the benefits of enhanced coverage could outweigh the additional computational cost in many practical applications, particularly in domains where explanation quality is prioritized over computational efficiency.

For the SVM results presented in Table \ref{tab:Time_SVM}, the execution time increase of Twostep over Onestep is less pronounced across all datasets. The difference ranges from 13.29\% (0.013 seconds vs. 0.015 seconds) on the BLDT dataset to 56.40\% (0.086 seconds vs. 0.135 seconds) on the BRCW dataset. This is likely due to the underlying linear programming problem being simpler, allowing for faster computations of explanations in both methods.



\begin{table}[htbp]
\caption{Average execution time results for SVM. $\#M$ demotes the mean execution time along with the standard deviation. $\%M$ represents the percentage increase in average execution time of Twostep compared to Onestep.}
\label{tab:Time_SVM}
\centering
\resizebox{1.0\textwidth}{!}{
    \begin{tabular}{| c|c|c |c|c |c|c |c |}
    \hline
    \multicolumn{1}{|c|}{\multirow{3}{*}{Dataset}} & 
    \multicolumn{1}{c|}{\multirow{2}{*}{Onestep}} & 
    \multicolumn{6}{c|}{Twostep} \\  
    \cline{3-8}& & \multicolumn{2}{c|}{T-Step $0.25$} & \multicolumn{2}{c|}{T-Step $0.50$} & \multicolumn{2}{c|}{T-Step $0.75$} \\
    \cline{2-8}& $\#M$& $\#M$& $\%M$& $\#M$& $\%M$& $\#M$& $\%M$\\ \hline
                        
\rowcolor{lightgray} IRIS & 0.017 $\pm$ 0.003& 0.022 $\pm$ 0.004& 29.41& 0.022 $\pm$ 0.004& 29.41& 0.022 $\pm$ 0.004& 29.41\\\hline
\rowcolor{white} BLDT & 0.014 $\pm$ 0.004&  0.015 $\pm$ 0.004& 7.14& 0.015 $\pm$ 0.003& 7.14& 0.015 $\pm$ 0.004& 7.14\\\hline
\rowcolor{lightgray}BANK& 0.017 $\pm$ 0.004&  0.023 $\pm$ 0.004& 35.29& 0.023 $\pm$ 0.004& 35.29& 0.023 $\pm$ 0.004& 35.29\\\hline
\rowcolor{white} UKMO& 0.018 $\pm$ 0.005&  0.023 $\pm$ 0.006& 27.78& 0.023 $\pm$ 0.006& 27.78& 0.023 $\pm$ 0.006& 27.78\\\hline
\rowcolor{lightgray}VRTC & 0.021 $\pm$ 0.005&  0.028 $\pm$ 0.007& 33.33& 0.028 $\pm$ 0.007& 33.33& 0.028 $\pm$ 0.007& 33.33\\\hline
\rowcolor{white} PIMA & 0.026 $\pm$ 0.005&  0.036 $\pm$ 0.006& 38.46& 0.036 $\pm$ 0.007& 33.33& 0.037 $\pm$ 0.007& 42.31\\\hline
\rowcolor{lightgray}GLAS& 0.025 $\pm$ 0.004&  0.034 $\pm$ 0.007& 36.00& 0.033 $\pm$ 0.006& 32.00& 0.033 $\pm$ 0.005& 32.00\\\hline
\rowcolor{white} WINE & 0.037 $\pm$ 0.005&  0.053 $\pm$ 0.009& 43.24& 0.054 $\pm$ 0.010& 42.11& 0.052 $\pm$ 0.009& 40.54\\\hline
\rowcolor{lightgray}CLIM& 0.050 $\pm$ 0.007&  0.072 $\pm$ 0.011& 44.00& 0.074 $\pm$ 0.011& 45.10& 0.072 $\pm$ 0.011& 44.00\\\hline
\rowcolor{white} PARK & 0.052 $\pm$ 0.011&  0.072 $\pm$ 0.019& 38.46& 0.074 $\pm$ 0.020& 42.31& 0.073 $\pm$ 0.020& 40.38\\\hline
\rowcolor{lightgray}BRCW & 0.087 $\pm$ 0.008&  0.136 $\pm$ 0.016& 56.32& 0.133 $\pm$ 0.014& 54.65& 0.134 $\pm$ 0.014& 54.02\\\hline
\rowcolor{white} IONS & 0.099 $\pm$ 0.012&  0.149 $\pm$ 0.019& 50.51& 0.148 $\pm$ 0.019& 54.17& 0.148 $\pm$ 0.019& 54.17\\\hline
    \end{tabular} 
}
\\
      \makebox[\width]{}
\end{table}


\begin{table}[H]
\caption{Average execution time results for MLP. $\#M$ demotes the mean execution time along with the standard deviation. $\%M$ represents the percentage increase in average execution time of Twostep compared to Onestep.}
\label{tab:Time_MLP}
\centering
\resizebox{1.0\textwidth}{!}{
    \begin{tabular}{ |c|c|c |c|c |c|c |c| }
    \hline
    \multicolumn{1}{|c|}{\multirow{3}{*}{Dataset}} & 
    \multicolumn{1}{c|}{\multirow{2}{*}{Onestep}} & 
    \multicolumn{6}{c|}{Twostep} \\  
    \cline{3-8}& & \multicolumn{2}{c|}{T-Step $0.25$} & \multicolumn{2}{c|}{T-Step $0.50$} & \multicolumn{2}{c|}{T-Step $0.75$} \\
    \cline{2-8}& $\#M$& $\#M$& $\%M$& $\#M$& $\%M$& $\#M$& $\%M$\\ \hline
                        
\rowcolor{lightgray} IRIS & 0.16 $\pm$ 0.02& 0.21 $\pm$ 0.04& 28.83& 0.20 $\pm$ 0.03& 21.30& 0.21 $\pm$ 0.04& 26.51\\\hline
\rowcolor{white} BLDT & 0.14 $\pm$ 0.01&  0.16 $\pm$ 0.02& 14.71& 0.16 $\pm$ 0.02& 13.87& 0.16 $\pm$ 0.02& 14.49\\\hline
\rowcolor{lightgray}BANK& 0.16 $\pm$ 0.02&  0.20 $\pm$ 0.03& 24.84& 0.20 $\pm$ 0.03& 26.58& 0.20 $\pm$ 0.03& 25.48\\\hline
\rowcolor{white} UKMO& 0.18 $\pm$ 0.02&  0.23 $\pm$ 0.04& 28.25& 0.23 $\pm$ 0.04& 32.39& 0.24 $\pm$ 0.05& 33.33\\\hline
\rowcolor{lightgray}VRTC & 0.18 $\pm$ 0.02&  0.24 $\pm$ 0.05& 32.07& 0.24 $\pm$ 0.04& 26.98& 0.24 $\pm$ 0.04& 28.42\\\hline
\rowcolor{white} PIMA & 0.26 $\pm$ 0.07&  0.38 $\pm$ 0.17& 43.56& 0.39 $\pm$ 0.08& 48.29& 0.39 $\pm$ 0.07& 47.33\\\hline
\rowcolor{lightgray}GLAS& 0.34 $\pm$ 0.07&  0.52 $\pm$ 0.12& 55.65& 0.50 $\pm$ 0.10& 54.15& 0.51 $\pm$ 0.15& 50.29\\\hline
\rowcolor{white} WINE & 0.45 $\pm$ 0.06&  0.71 $\pm$ 0.09& 55.85& 0.71 $\pm$ 0.06& 62.90& 0.71 $\pm$ 0.07& 63.16\\\hline
\rowcolor{lightgray}CLIM& 0.53 $\pm$ 0.12&  0.81 $\pm$ 0.18& 51.69& 0.84 $\pm$ 0.27& 58.52& 0.80 $\pm$ 0.13& 58.58\\\hline
\rowcolor{white} PARK & 0.80 $\pm$ 0.14&  1.31 $\pm$ 0.28& 62.64& 1.39 $\pm$ 0.33& 71.06& 1.40 $\pm$ 0.35& 74.62\\\hline
\rowcolor{lightgray}BRCW & 1.70 $\pm$ 0.34&  3.02 $\pm$ 0.56& 78.36& 3.20 $\pm$ 0.62& 93.52& 3.42 $\pm$ 0.73& 79.77\\\hline
\rowcolor{white} IONS & 1.94 $\pm$ 0.75&  4.02 $\pm$ 1.22& 107.32& 4.55 $\pm$ 1.64& 141.01& 4.95 $\pm$ 1.71& 155.76\\\hline
    \end{tabular} 
}
\\
      \makebox[\width]{}
\end{table}


\subsubsection{Results for Dataset Coverage}

\begin{table}[htbp]
\caption{Average coverage results for SVM. $\#M$ denotes the mean dataset coverage with standard deviation. $\%M$ represents the percentage improvement in mean dataset coverage of Twostep over Onestep.}
\label{tab:Coverage_SVM}
\centering
\resizebox{1.0\textwidth}{!}{
    \begin{tabular}{ |c|c|c |c|c |c|c |c| }
    \hline
    \multicolumn{1}{|c|}{\multirow{3}{*}{Dataset}} & 
    \multicolumn{1}{c|}{\multirow{2}{*}{Onestep}} & 
    \multicolumn{6}{c|}{Twostep} \\  
    \cline{3-8}& & \multicolumn{2}{c|}{T-Step $0.25$} & \multicolumn{2}{c|}{T-Step $0.50$} & \multicolumn{2}{c|}{T-Step $0.75$} \\
    \cline{2-8}& $\#M$& $\#M$& $\%M$& $\#M$& $\%M$& $\#M$& $\%M$\\ \hline
                        
\rowcolor{lightgray} IRIS & 30.41 $\pm$ 19.51& 35.04 $\pm$ 15.35& 15.23& 35.56 $\pm$ 14.62& 16.94& 34.44 $\pm$ 15.87& 13.25\\\hline
\rowcolor{white} BLDT & 519.92 $\pm$ 94.91&  520.85 $\pm$ 90.60& 0.18& 521.02 $\pm$ 89.80& 0.21& 520.89 $\pm$ 90.41& 0.19\\\hline
\rowcolor{lightgray}BANK& 36.03 $\pm$ 36.37&  49.60 $\pm$ 36.08& 37.66& 51.00 $\pm$ 36.77& 41.55& 46.57 $\pm$ 36.92& 29.25\\\hline
\rowcolor{white} UKMO& 38.29 $\pm$ 28.95&  37.81 $\pm$ 27.69& -1.25& 37.76 $\pm$ 27.57& -1.38& 38.10 $\pm$ 28.12& -0.50\\\hline
\rowcolor{lightgray}VRTC & 27.51 $\pm$ 22.43&  27.89 $\pm$ 22.51& 1.38& 27.83 $\pm$ 22.71& 1.16& 27.75 $\pm$ 22.68& 0.87\\\hline
\rowcolor{white} PIMA & 12.07 $\pm$ 13.60&  15.52 $\pm$ 15.19& 28.58& 14.35 $\pm$ 14.39& 18.89& 13.02 $\pm$ 13.93& 7.87\\\hline
\rowcolor{lightgray}GLAS& 24.66 $\pm$ 22.73&  26.19 $\pm$ 23.25& 6.20& 26.45 $\pm$ 23.36& 7.26& 25.98 $\pm$ 23.23& 5.35\\\hline
\rowcolor{white} WINE & 1.19 $\pm$ 0.59&  1.22 $\pm$ 0.64& 2.52& 1.23 $\pm$ 0.66& 3.36& 1.22 $\pm$ 0.62& 2.52\\\hline
\rowcolor{lightgray}CLIM& 1.01 $\pm$ 0.10&  1.01 $\pm$ 0.11& 0.00& 1.01 $\pm$ 0.11& 0.00& 1.01 $\pm$ 0.10& 0.00\\\hline
\rowcolor{white} PARK & 2.20 $\pm$ 2.54&  2.21 $\pm$ 2.47& 0.45& 2.22 $\pm$ 2.47& 0.91& 2.20 $\pm$ 2.50& 0.00\\\hline
\rowcolor{lightgray}BRCW & 1.04 $\pm$ 0.28&  1.06 $\pm$ 0.40& 1.92& 1.06 $\pm$ 0.40& 1.92& 1.06 $\pm$ 0.38& 1.92\\\hline
\rowcolor{white} IONS & 1.02 $\pm$ 0.15&  1.02 $\pm$ 0.15& 0.00& 1.02 $\pm$ 0.15& 0.00& 1.02 $\pm$ 0.1& 0.00\\\hline
    \end{tabular} 
}
\\
      \makebox[\width]{}
\end{table}

\begin{table}[htbp]
\caption{Average coverage results for MLP. $\#M$ denotes the mean dataset coverage with standard deviation. $\%M$ represents the percentage improvement in mean dataset coverage of Twostep over Onestep.}
\label{tab:Coverage_MLP}
\centering
\resizebox{1.0\textwidth}{!}{
    \begin{tabular}{ |c|c|c |c|c |c|c |c| }
    \hline
    \multicolumn{1}{|c|}{\multirow{3}{*}{Dataset}} & 
    \multicolumn{1}{c|}{\multirow{2}{*}{Onestep}} & 
    \multicolumn{6}{c|}{Twostep} \\  
    \cline{3-8}& & \multicolumn{2}{c|}{T-Step $0.25$} & \multicolumn{2}{c|}{T-Step $0.50$} & \multicolumn{2}{c|}{T-Step $0.75$} \\
    \cline{2-8}& $\#M$& $\#M$& $\%M$& $\#M$& $\%M$& $\#M$& $\%M$\\ \hline
                        
\rowcolor{lightgray} IRIS & 19.57 $\pm$ 14.48& 21.81 $\pm$ 12.21& 11.45& 22.76 $\pm$ 11.81& 16.30& 22.65 $\pm$ 12.95& 15.74\\\hline
\rowcolor{white} BLDT & 149.55 $\pm$ 113.63&  157.78 $\pm$ 122.77& 5.50& 162.46 $\pm$ 127.68& 8.63& 157.19 $\pm$ 122.89& 5.11\\\hline
\rowcolor{lightgray}BANK& 25.15 $\pm$ 29.50&  33.34 $\pm$ 31.56& 32.56& 33.85 $\pm$ 32.21& 34.59& 31.05 $\pm$ 31.59& 23.46\\\hline
\rowcolor{white} UKMO& 12.85 $\pm$ 10.93&  13.91 $\pm$ 10.70& 8.25& 14.46 $\pm$ 10.99& 12.53& 14.14 $\pm$ 10.88& 10.04\\\hline
\rowcolor{lightgray}VRTC & 12.94 $\pm$ 13.24&  14.34 $\pm$ 13.61& 10.82& 14.41 $\pm$ 13.73& 11.36& 13.74 $\pm$ 13.57& 6.18\\\hline
\rowcolor{white} PIMA & 7.42 $\pm$ 8.65&  8.90 $\pm$ 9.41& 19.95& 8.77 $\pm$ 9.63& 18.19& 8.10 $\pm$ 9.20& 9.16\\\hline
\rowcolor{lightgray}GLAS& 1.46 $\pm$ 1.43&  2.52 $\pm$ 2.34& 72.60& 2.27 $\pm$ 2.09& 55.48& 1.75 $\pm$ 1.69& 19.86\\\hline
\rowcolor{white} WINE & 1.00 $\pm$ 0.00&  1.00 $\pm$ 0.00& 0.00& 1.01 $\pm$ 0.09& 1.00& 1.00 $\pm$ 0.00& 0.00\\\hline
\rowcolor{lightgray}CLIM& 1.05 $\pm$ 0.22&  1.05 $\pm$ 0.23& 0.00& 1.05 $\pm$ 0.23& 0.00& 1.05 $\pm$ 0.23& 0.00\\\hline
\rowcolor{white} PARK & 1.03 $\pm$ 0.22&  1.03 $\pm$ 0.22& 0.00& 1.03 $\pm$ 0.20& 0.00& 1.03 $\pm$ 0.20& 0.00\\\hline
\rowcolor{lightgray}BRCW & 1.06 $\pm$ 0.32&  1.07 $\pm$ 0.34& 0.94& 1.07 $\pm$ 0.33& 0.94& 1.07 $\pm$ 0.33& 0.94\\\hline
\rowcolor{white} IONS & 1.01 $\pm$ 0.09&  1.01 $\pm$ 0.09& 0.00& 1.01 $\pm$ 0.09& 0.00& 1.01 $\pm$ 0.09& 0.00\\\hline
    \end{tabular} 
}
\\
      \makebox[\width]{}
\end{table}

Regarding the dataset coverage results for SVM in Table \ref{tab:Coverage_SVM}, Twostep has demonstrated the ability to achieve equal or higher coverage compared to Onestep for most datasets. However, the UKMO dataset is an exception, where Twostep shows up to 1.38\% less coverage on average. For the BANK dataset, Twostep covers up to 41.55\% more instances than Onestep. Moreover, the results for the PIMA dataset show that Twostep can cover up to 28.58\% more instances than Onestep. However, for the BLDT, VRTC, CLIM, PARK, BRCW, and IONS datasets, Twostep achieves minimal to no improvements in coverage compared to Onestep.

Similarly, the dataset coverage results for MLP in Table~\ref{tab:Coverage_MLP} show that Twostep achieves equal or higher coverage than Onestep. Again for the PIMA and BANK datasets, Twostep shows a significant improvement in dataset coverage, with 19.95\% and 34.59\% more instances covered, respectively. Particularly, for the GLAS dataset, Twostep covers an impressive 73.71\% more instances than Onestep, demonstrating its potential to provide more general explanations. However, for the WINE, CLIM, PARK, BRCW, and IONS datasets, Twostep shows either milder improvements or no improvements at all, suggesting that providing broader coverage is more challenging for these datasets.

We now analyze the coverage results from an explanationwise perspective. Specifically, for each test instance, both methods generate an explanation, and we compare the improvement in dataset coverage of the explanation generated by Twostep relative to the one generated by Onestep. In this analysis, we fixed the Twostep parameter at $p=0.25$.


\begin{figure}[H]
    \centering
    \subcaptionbox{IRIS dataset coverage}{\includegraphics[width=0.49\textwidth]{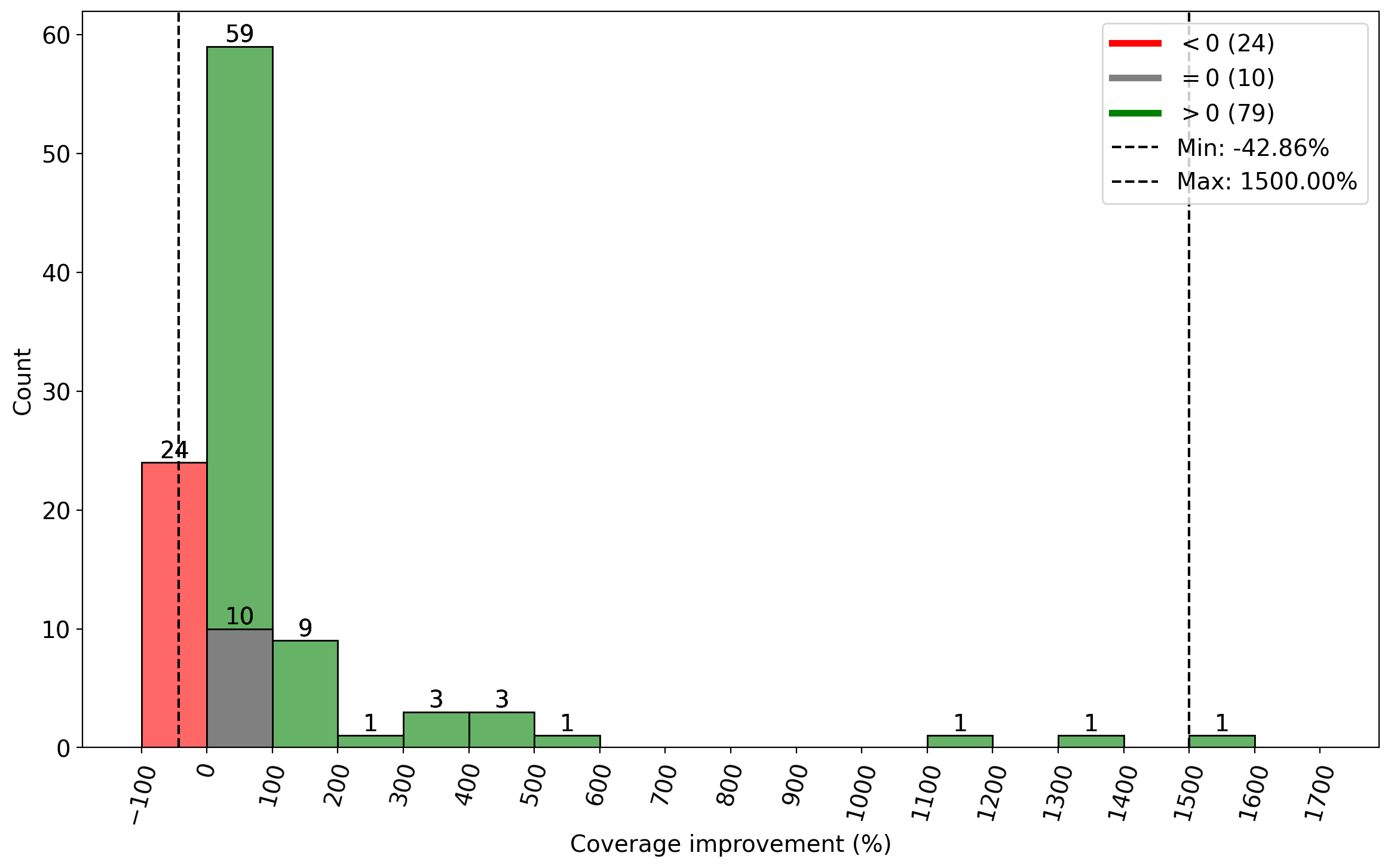}} 
    \subcaptionbox{BLDT dataset coverage}{\includegraphics[width=0.49\textwidth]{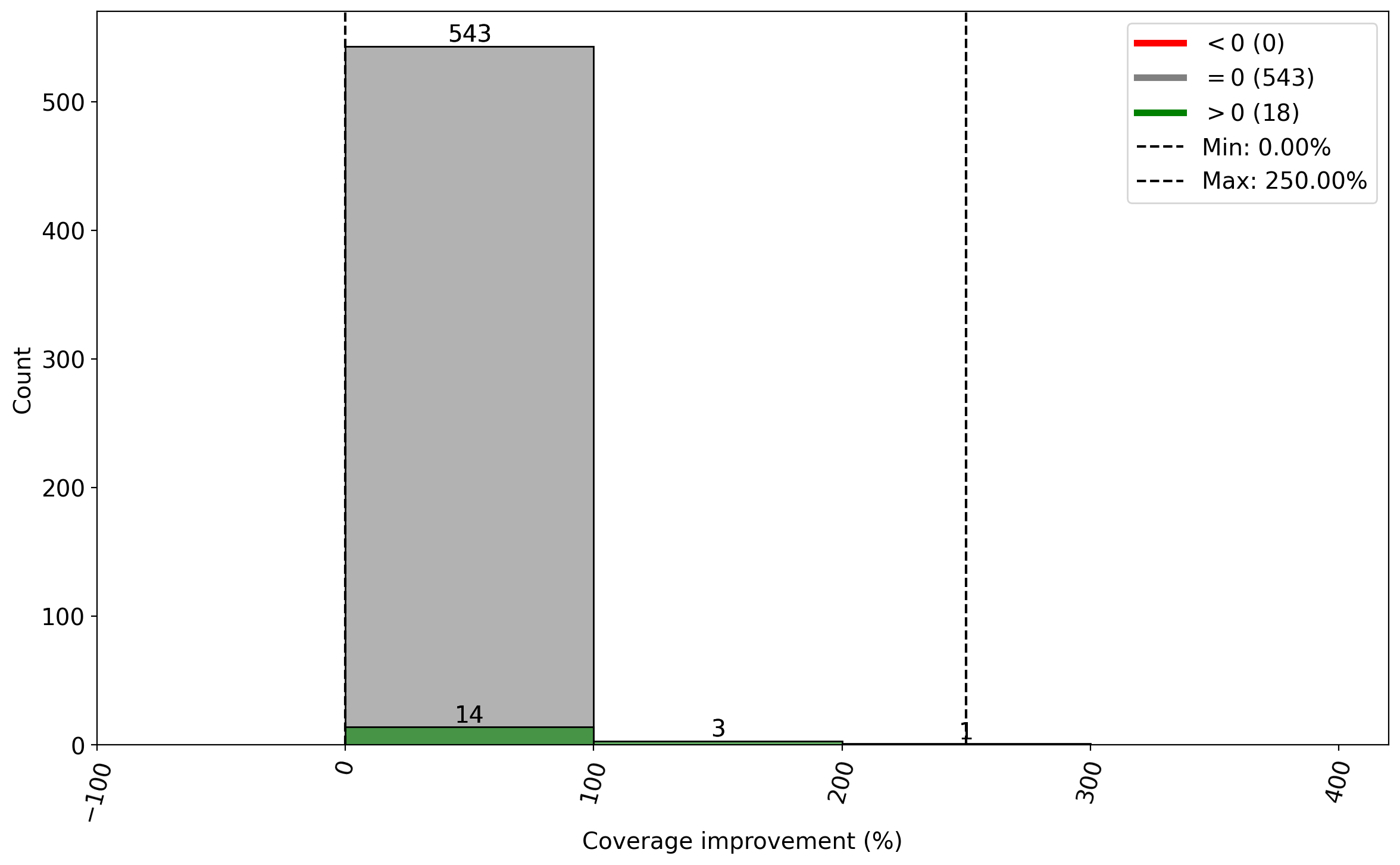}} \\
    \subcaptionbox{BANK dataset coverage}{\includegraphics[width=0.49\textwidth]{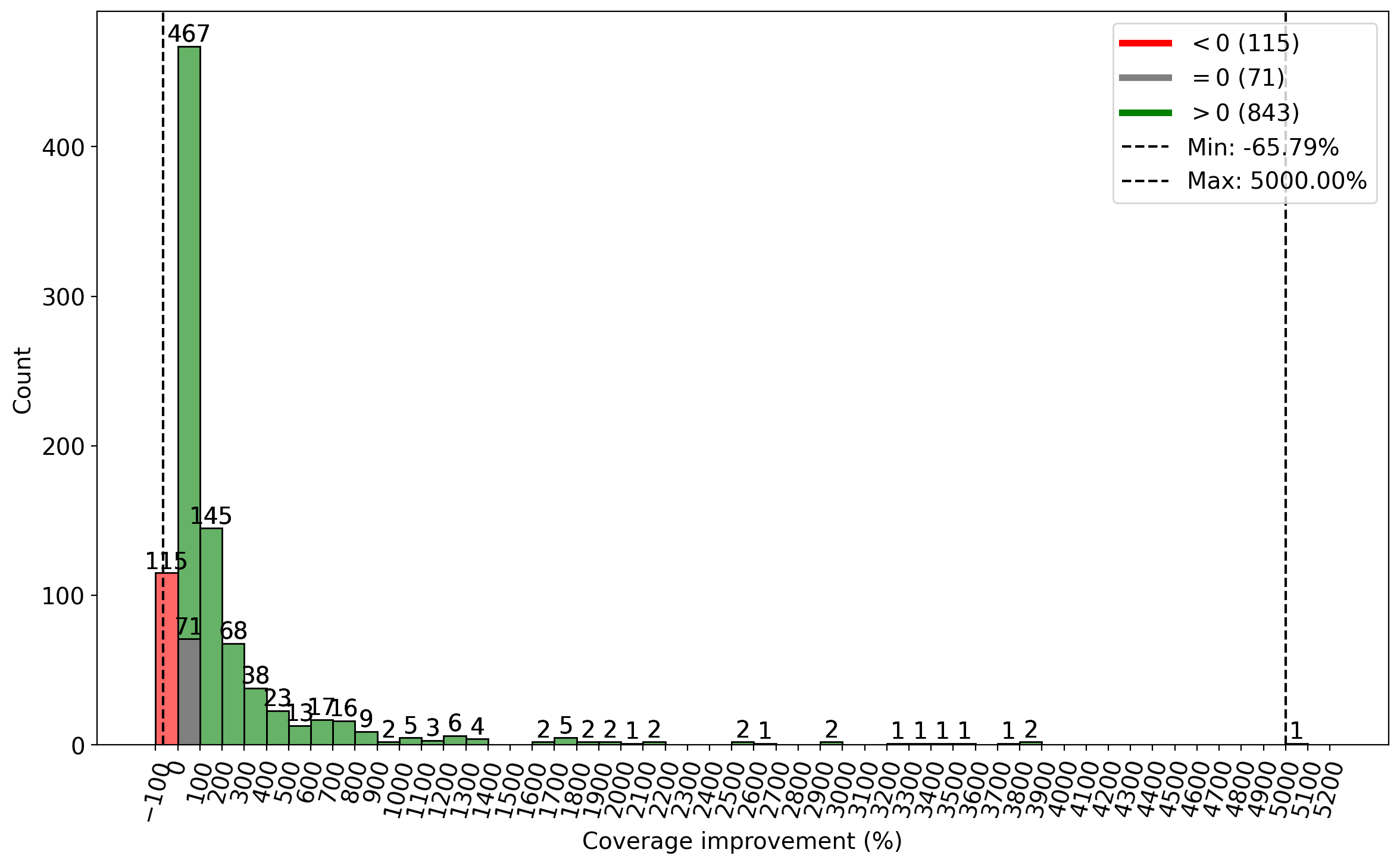}} 
    \subcaptionbox{UKMO dataset coverage}{\includegraphics[width=0.49\textwidth]{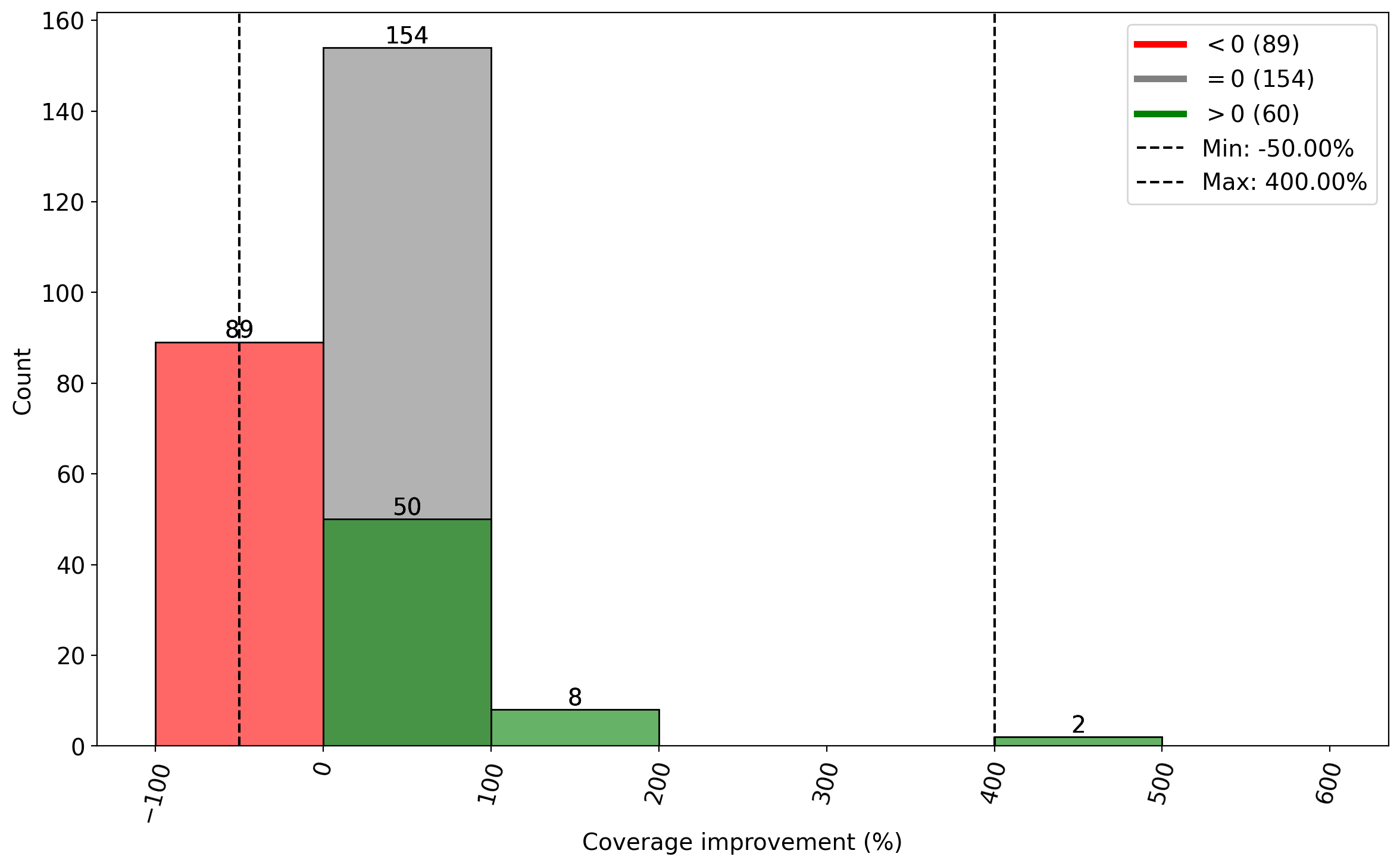}} \\
    
    \caption{Distribution of dataset coverage improvement (\%) achieved by Twostep over Onestep for SVM, with Twostep parameter fixed at $p=0.25$. Red bars represent cases with worsened coverage, gray bars represent cases with same coverage, and the green bars represent cases with improved coverage. The best improvement and the worst deterioration are highlighted.}
    \label{fig:SVM_relative_coverage}
\end{figure}

\begin{figure}[H]
    \ContinuedFloat  
    \centering
    \subcaptionbox{VRTC dataset coverage}{\includegraphics[width=0.49\textwidth]{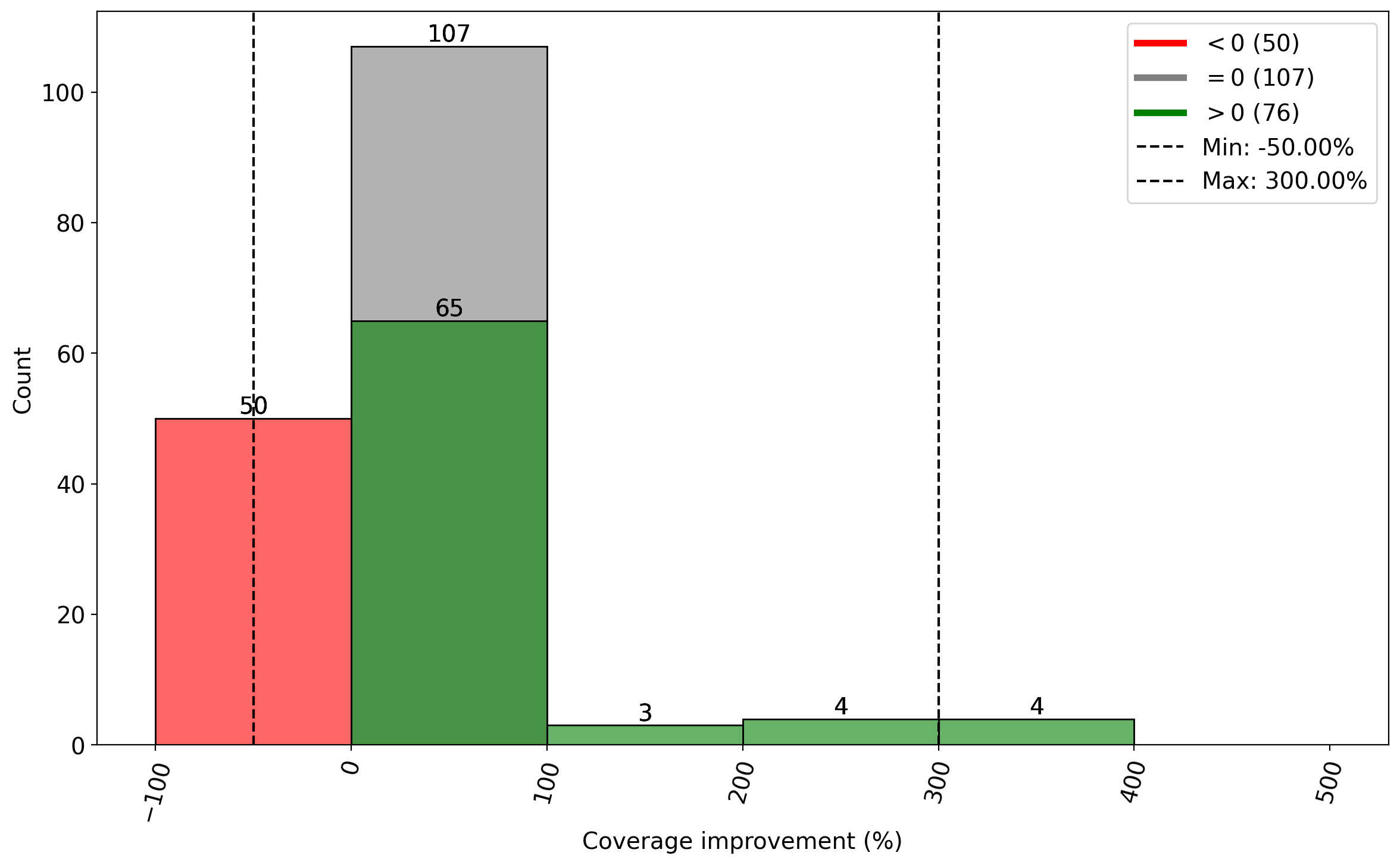}} 
    \subcaptionbox{PIMA dataset coverage}{\includegraphics[width=0.49\textwidth]{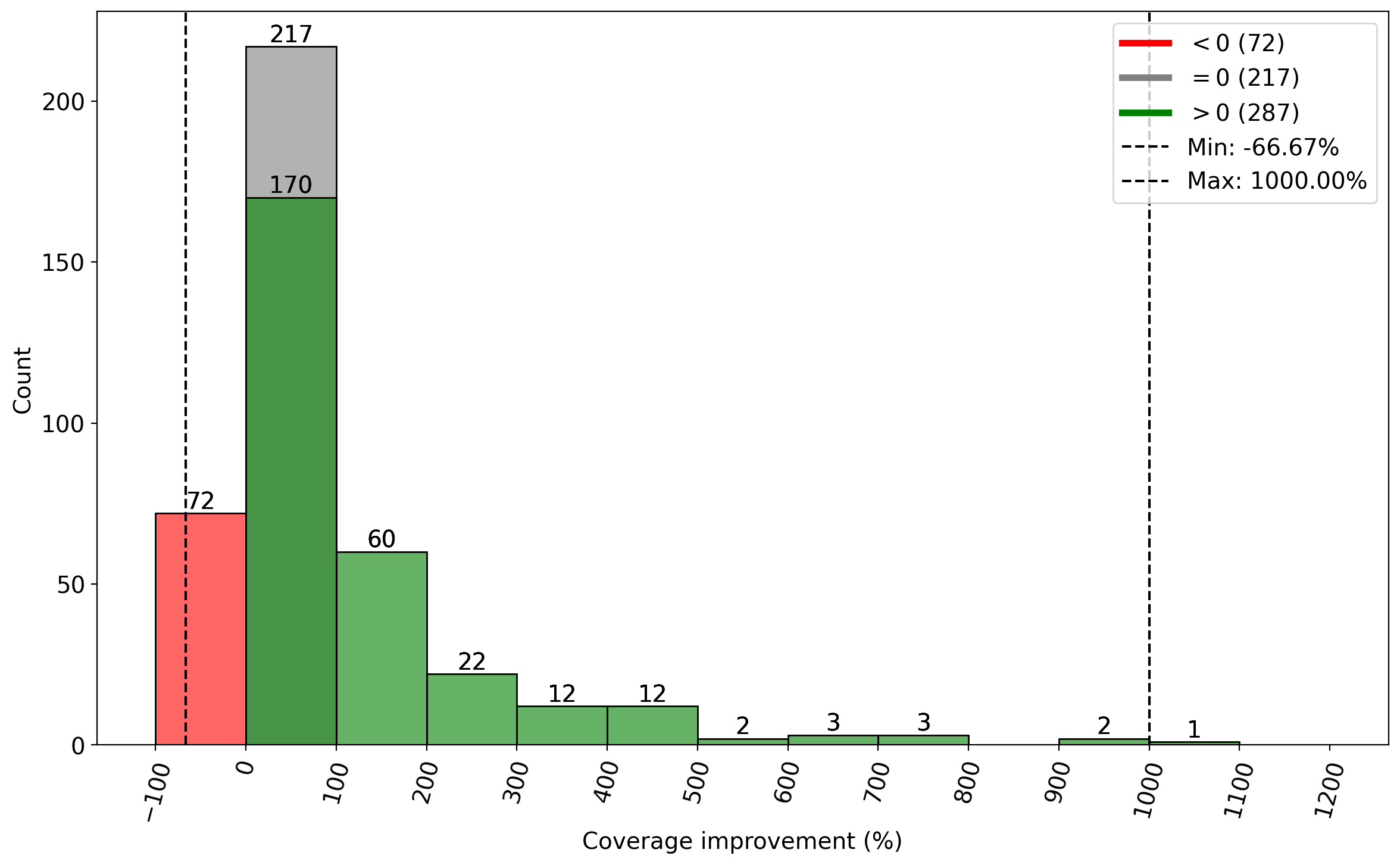}} \\
    \subcaptionbox{GLAS dataset coverage}{\includegraphics[width=0.49\textwidth]{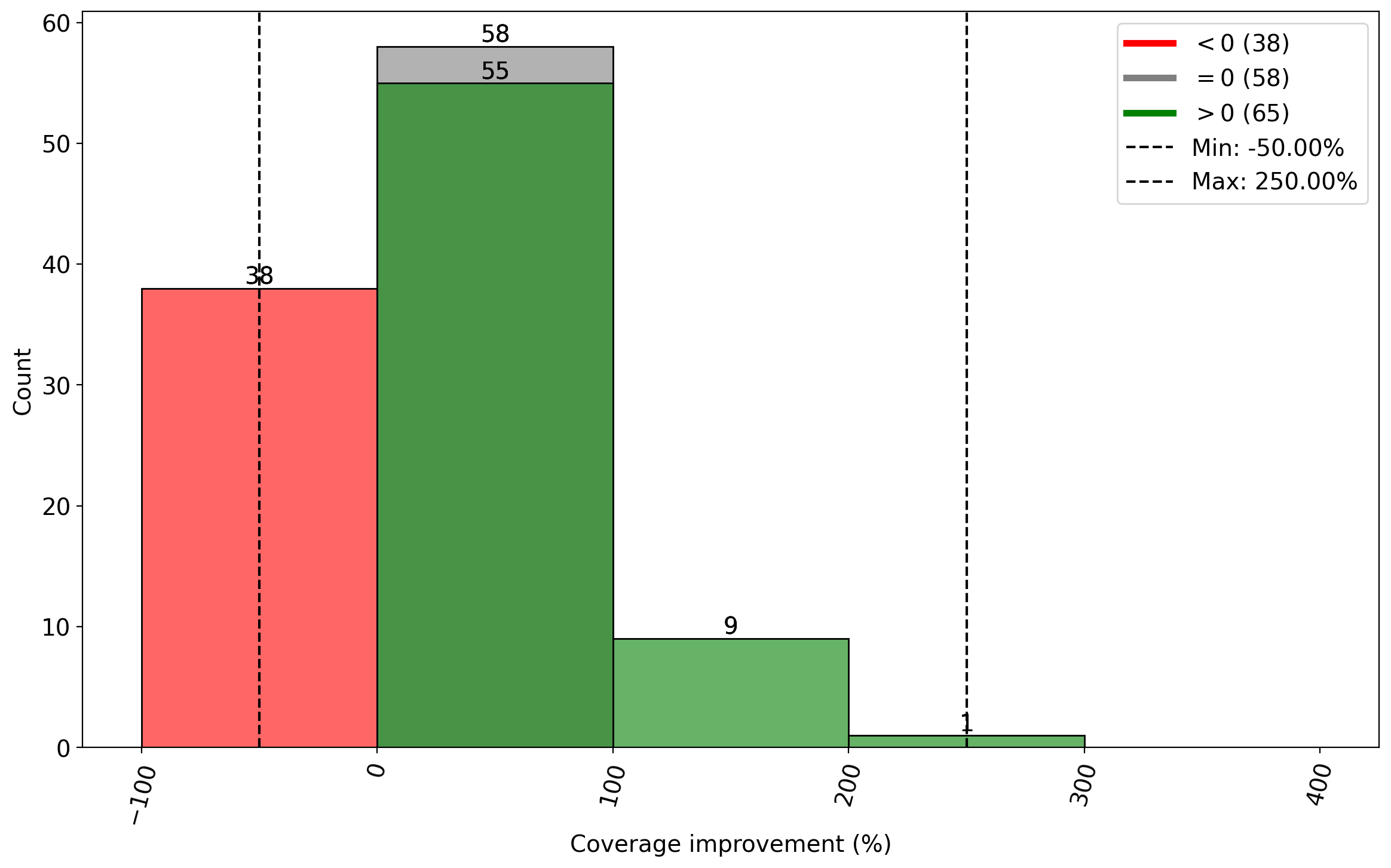}} 
    \subcaptionbox{WINE dataset coverage}{\includegraphics[width=0.49\textwidth]{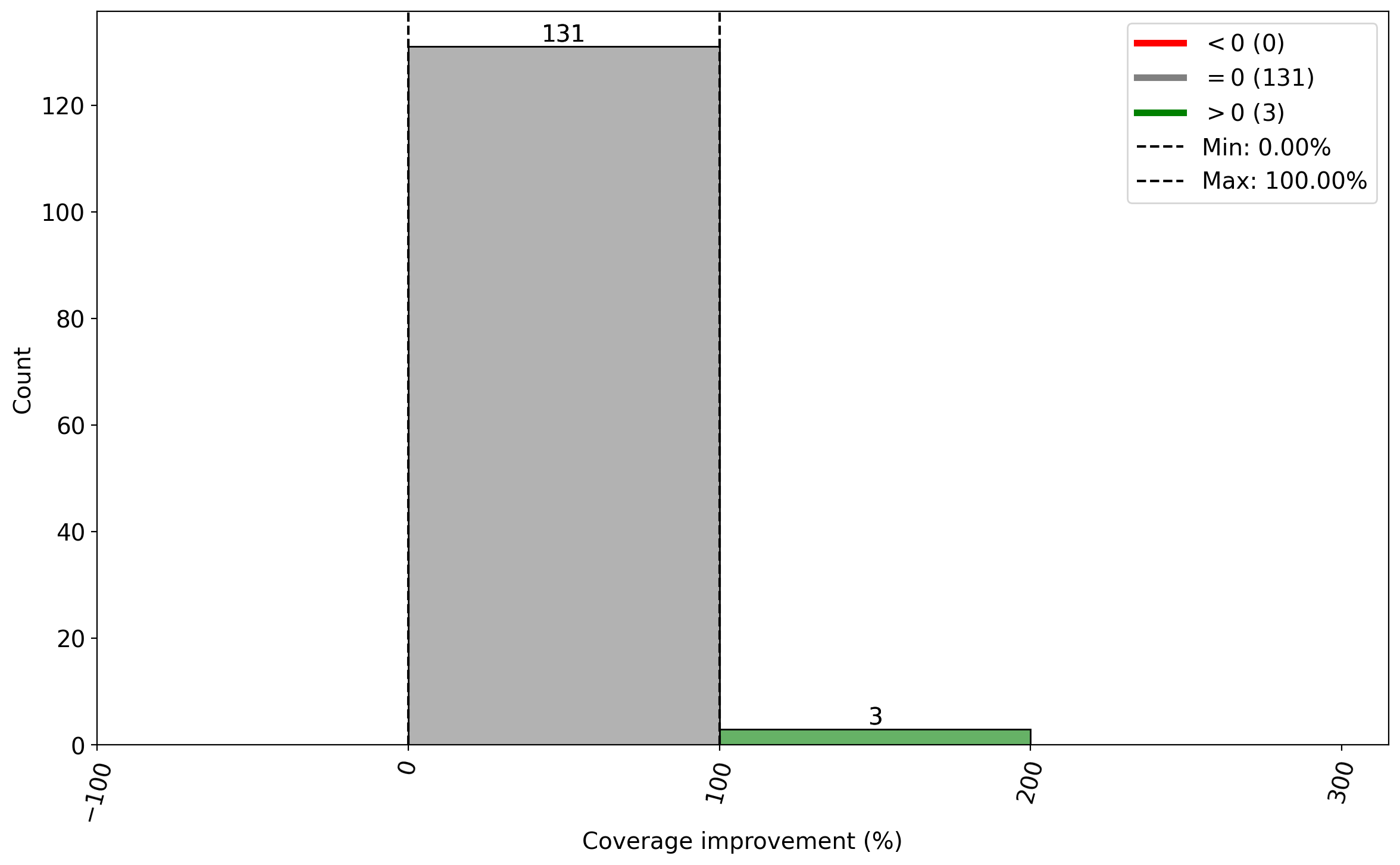}} \\
    \subcaptionbox{CLIM coverage}{\includegraphics[width=0.49\textwidth]{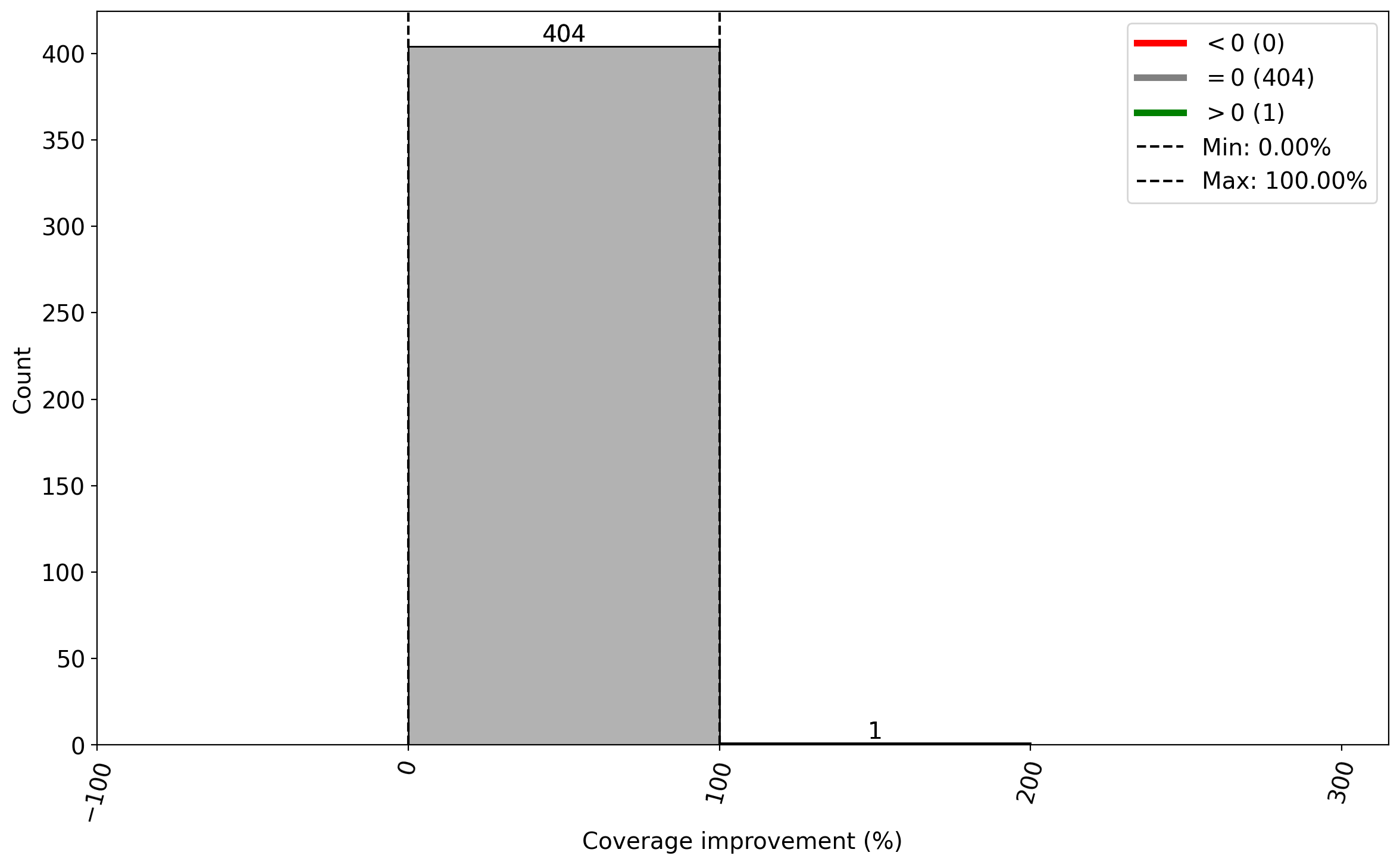}} 
    \subcaptionbox{PARK coverage}{\includegraphics[width=0.49\textwidth]{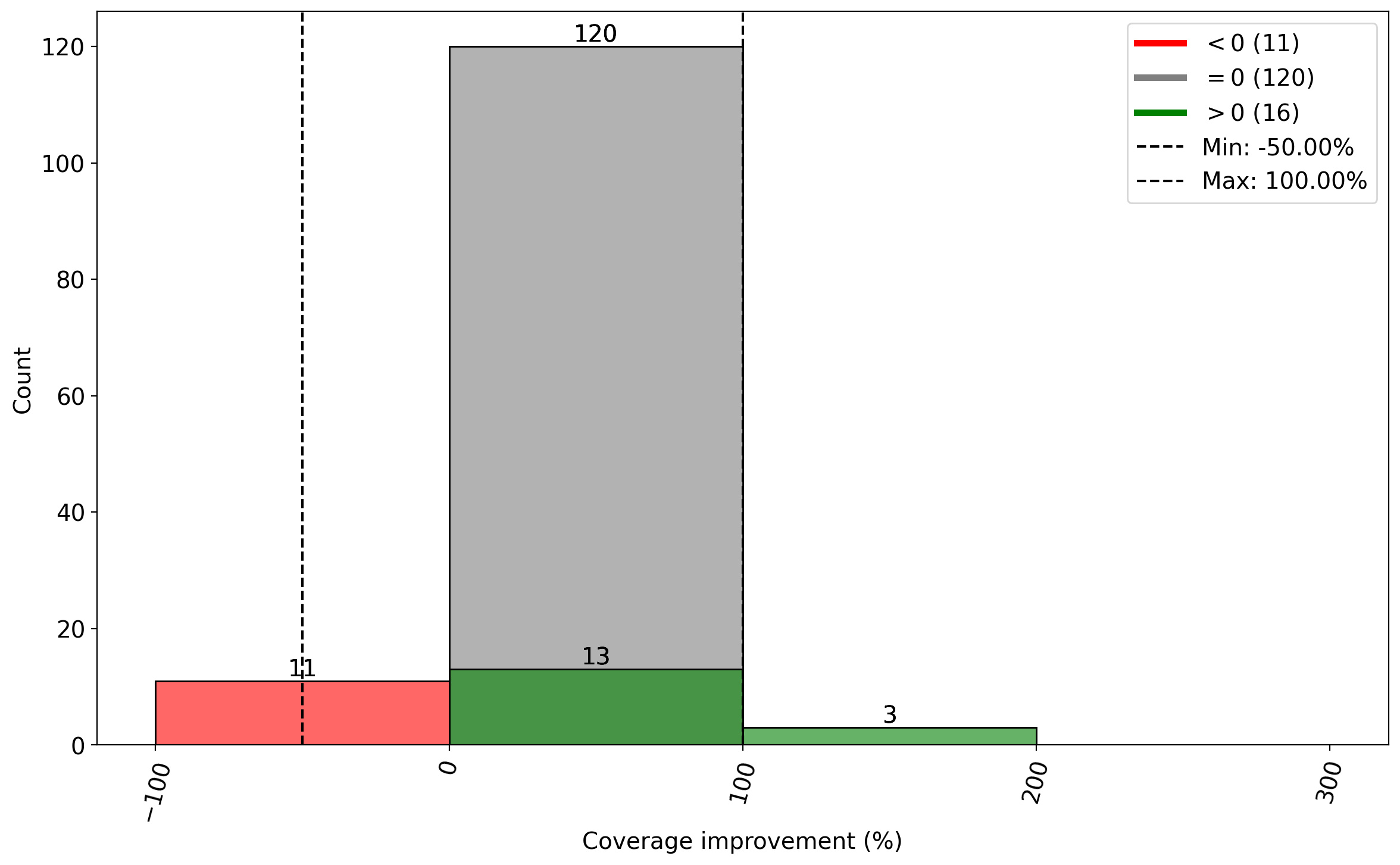}} \\
    \subcaptionbox{BRCW coverage}{\includegraphics[width=0.49\textwidth]{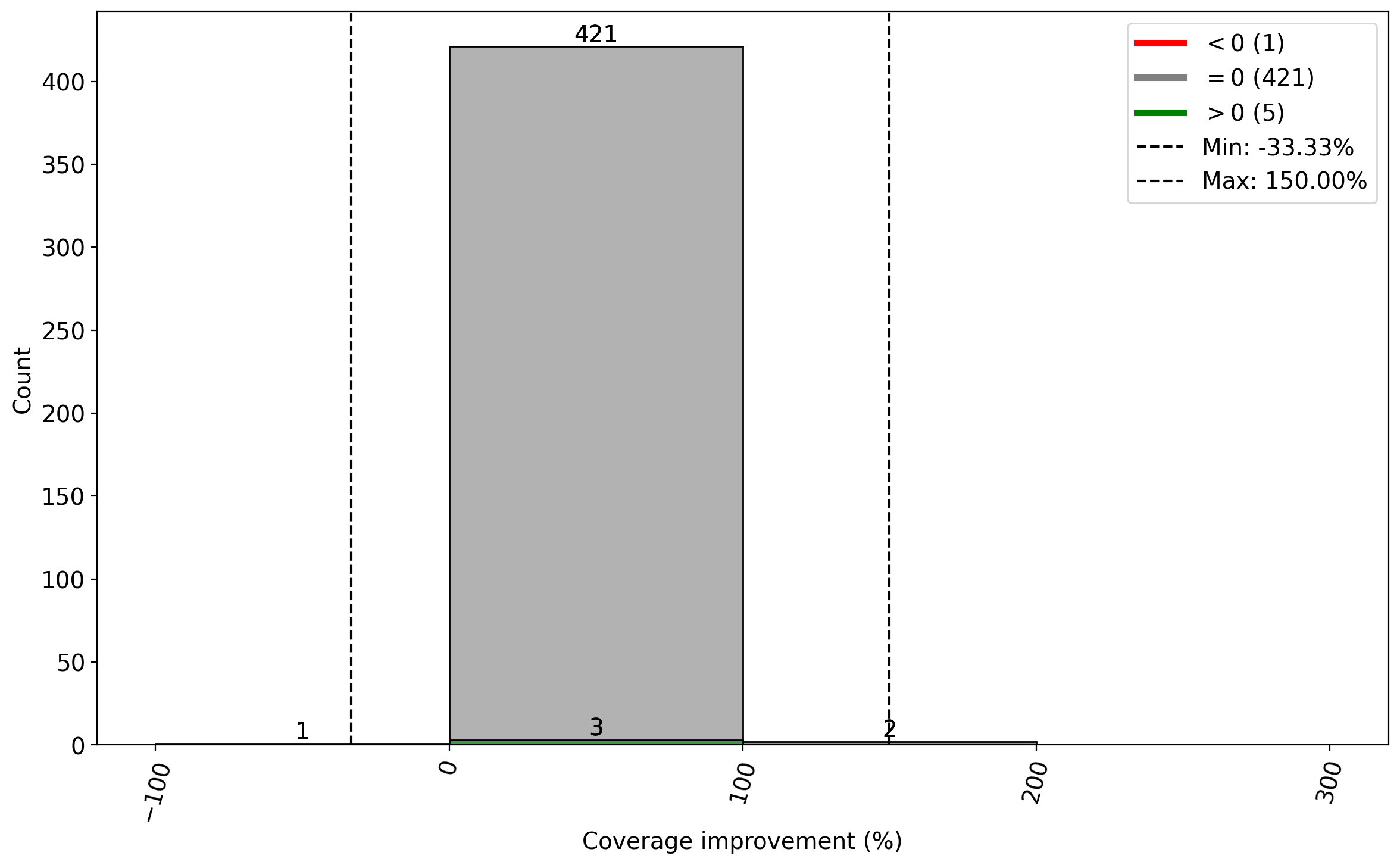}} 
    \subcaptionbox{IONS coverage}{\includegraphics[width=0.49\textwidth]{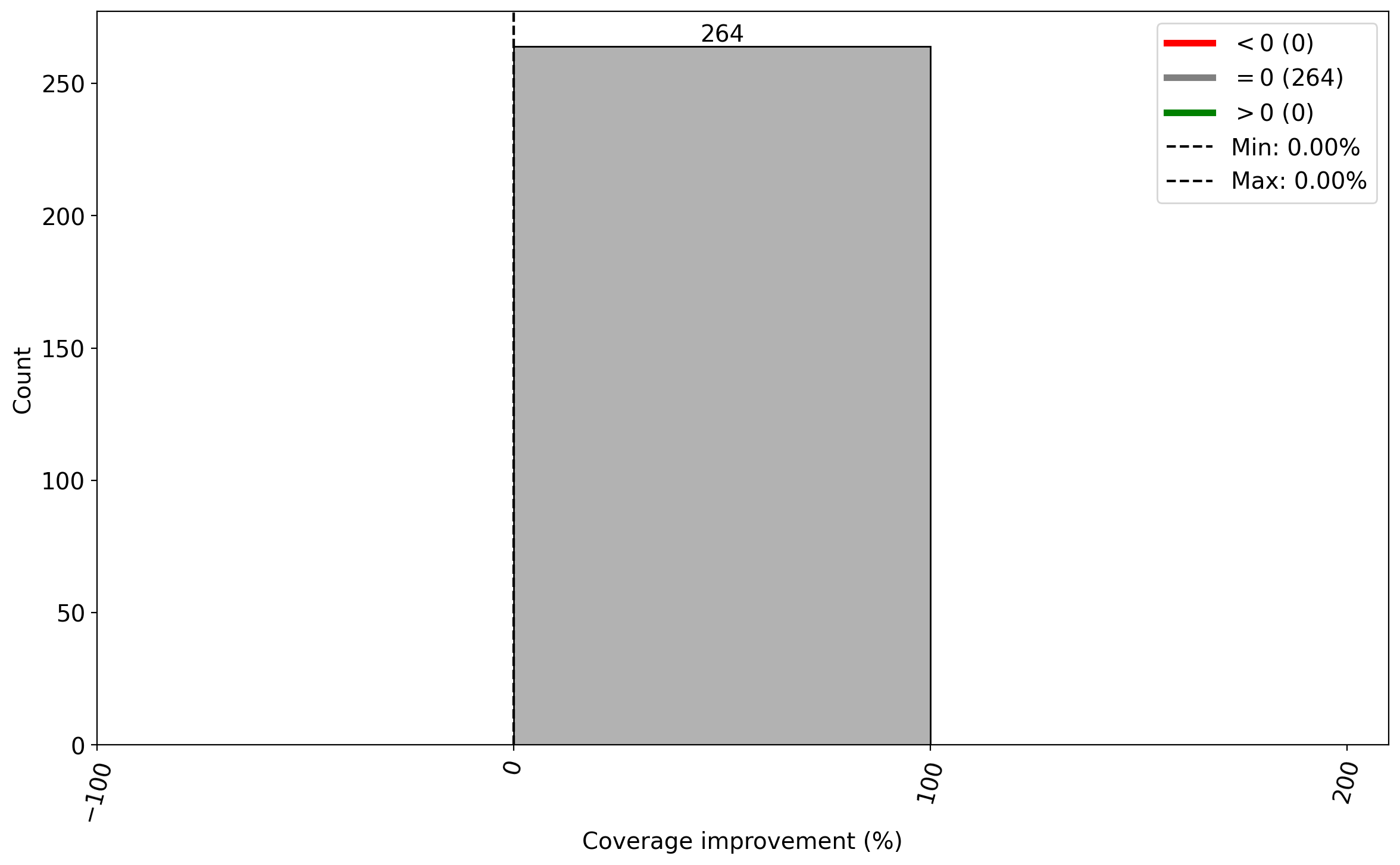}} \\
\end{figure}

The results for SVM in Figure~\ref{fig:SVM_relative_coverage} highlight the significant improvements achieved by Twostep over Onestep. For the BANK dataset, Twostep covered from 65.79\% fewer to an astounding 5000\% more instances. Similarly, for the IRIS dataset, Twostep achieved an impressive improvement, covering from 42.86\% fewer up to 1500\% more instances.
In the UKMO dataset, Twostep explanations ranged from covering 50\% fewer to 400\% more instances than Onestep. Although Twostep explanations covered fewer instances in 89 cases for this dataset, they covered more in 60 cases, and in 154 cases, both methods achieved the same coverage. Moreover, Twostep consistently achieved equal or better coverage than Onestep for the BLDT, WINE, and CLIM datasets, with no cases of coverage loss. One exception is the IONS dataset, where Twostep and Onestep covered the same number of instances in each explanation.

\begin{figure}[H]
    \centering
    \subcaptionbox{IRIS dataset coverage}{\includegraphics[width=0.49\textwidth]{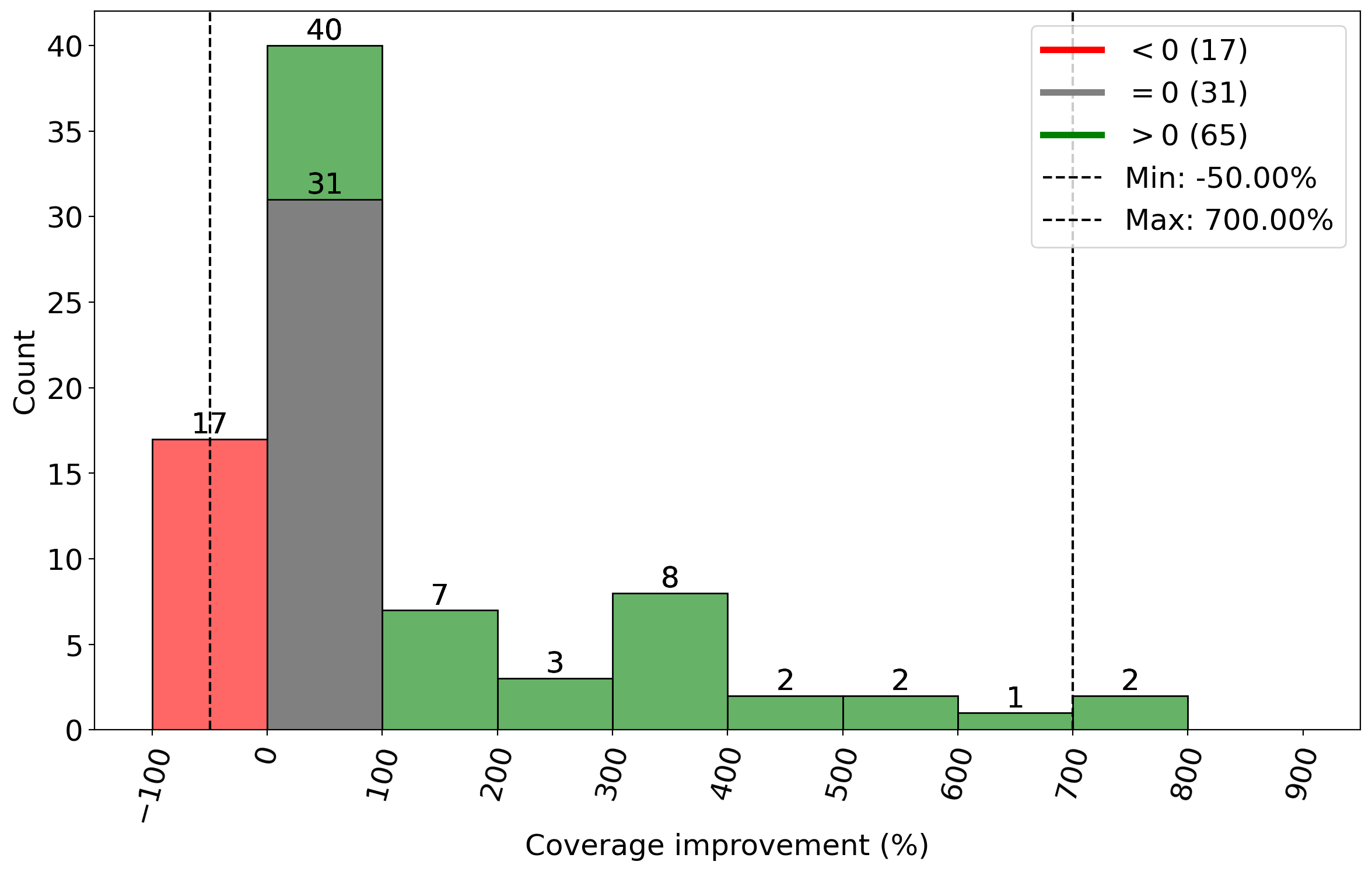}} 
    \subcaptionbox{BLDT dataset coverage}{\includegraphics[width=0.49\textwidth]{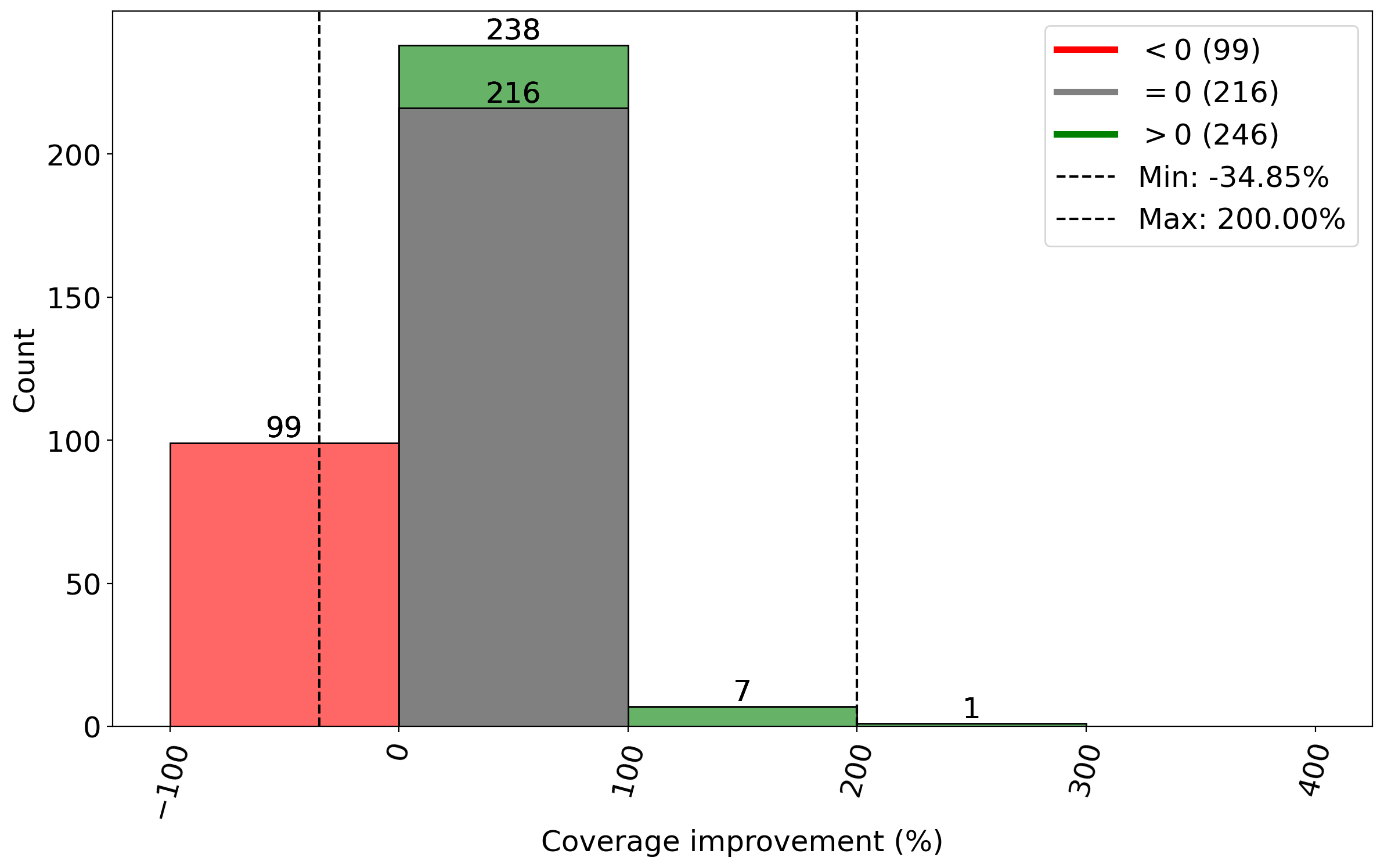}} \\
    \subcaptionbox{BANK dataset coverage}{\includegraphics[width=0.49\textwidth]{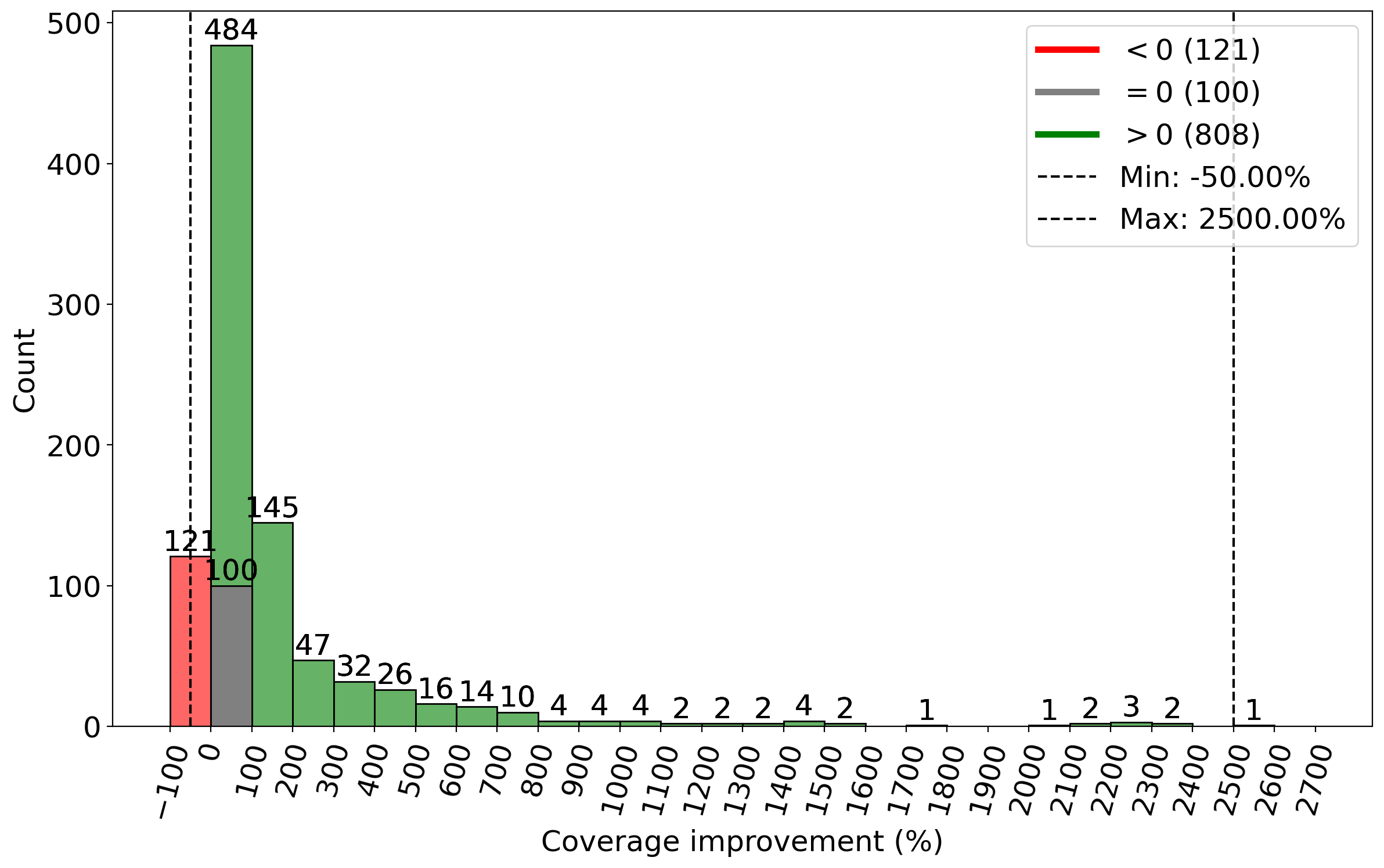}} 
    \subcaptionbox{UKMO dataset coverage}{\includegraphics[width=0.49\textwidth]{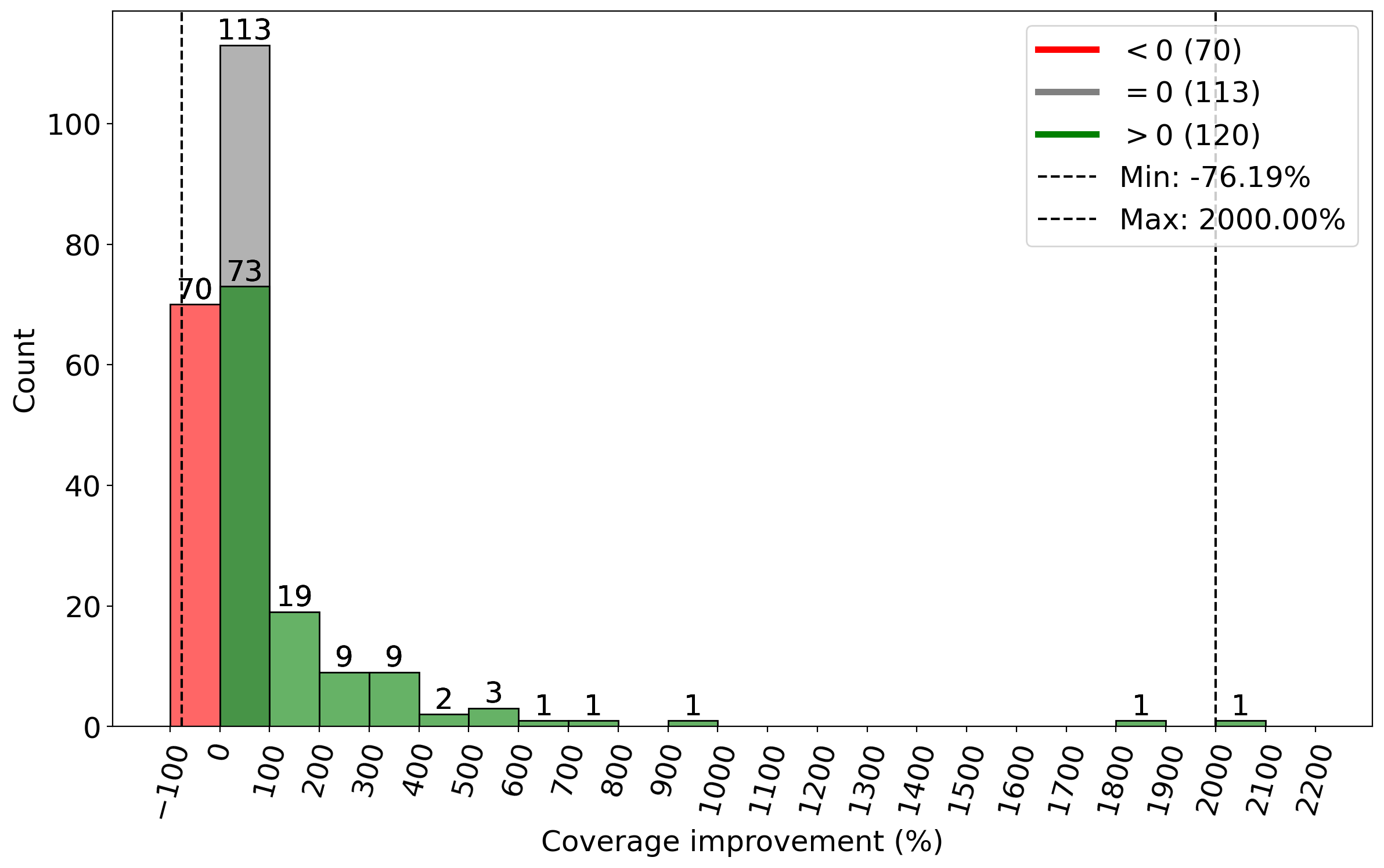}} \\
    
    \caption{Distribution of dataset coverage improvement (\%) achieved by Twostep over Onestep for MLP, with Twostep parameter fixed at $p=0.25$. Red bars represent cases with worsened coverage, gray bars represent cases with same coverage, and green bars represent cases with improved coverage. The best improvement and the worst deterioration are highlighted.}
    \label{fig:MLP_relative_coverage}
\end{figure}

\begin{figure}[H]
    \ContinuedFloat  
    \centering
    \subcaptionbox{VRTC dataset coverage}{\includegraphics[width=0.49\textwidth]{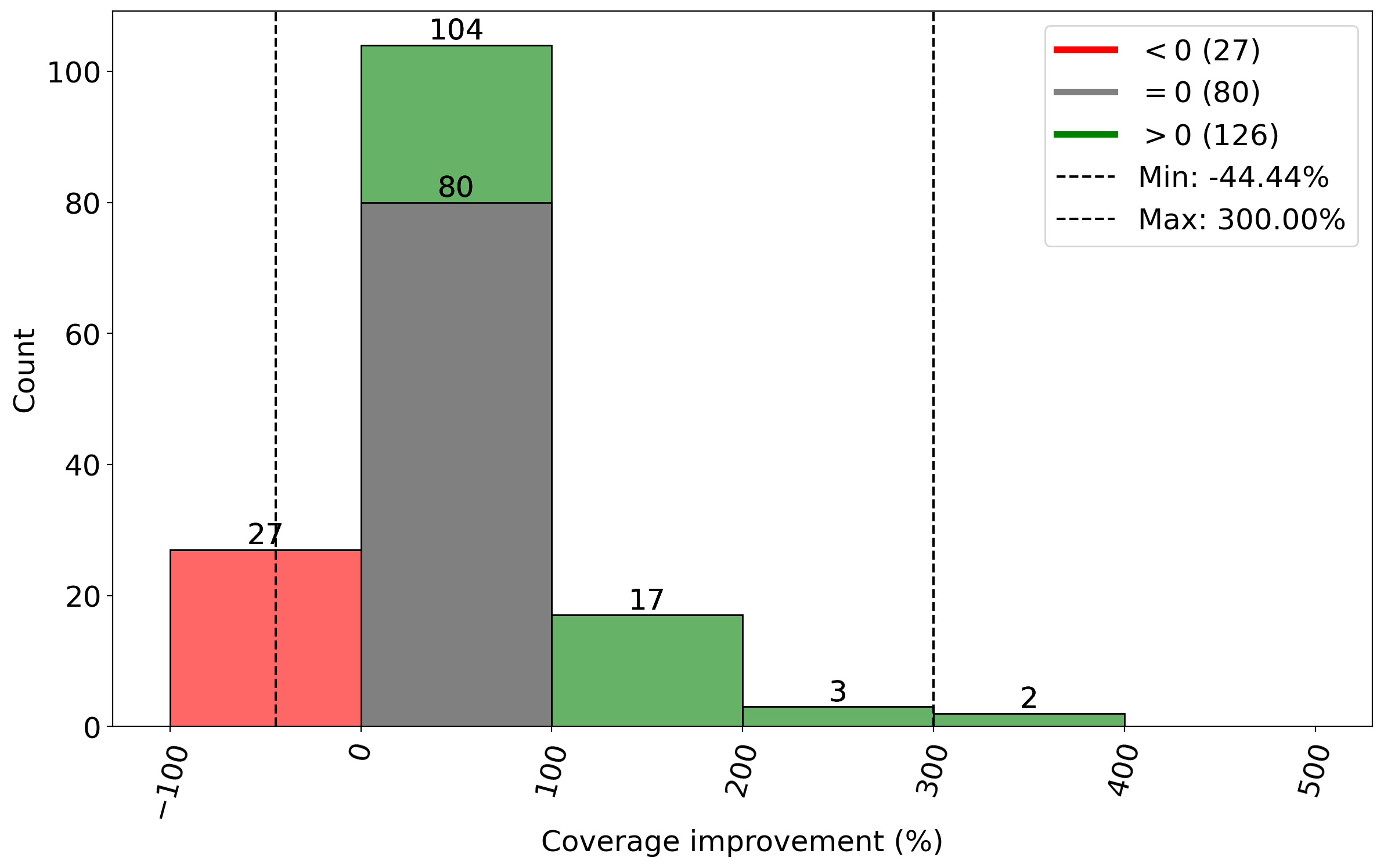}} 
    \subcaptionbox{PIMA dataset coverage}{\includegraphics[width=0.49\textwidth]{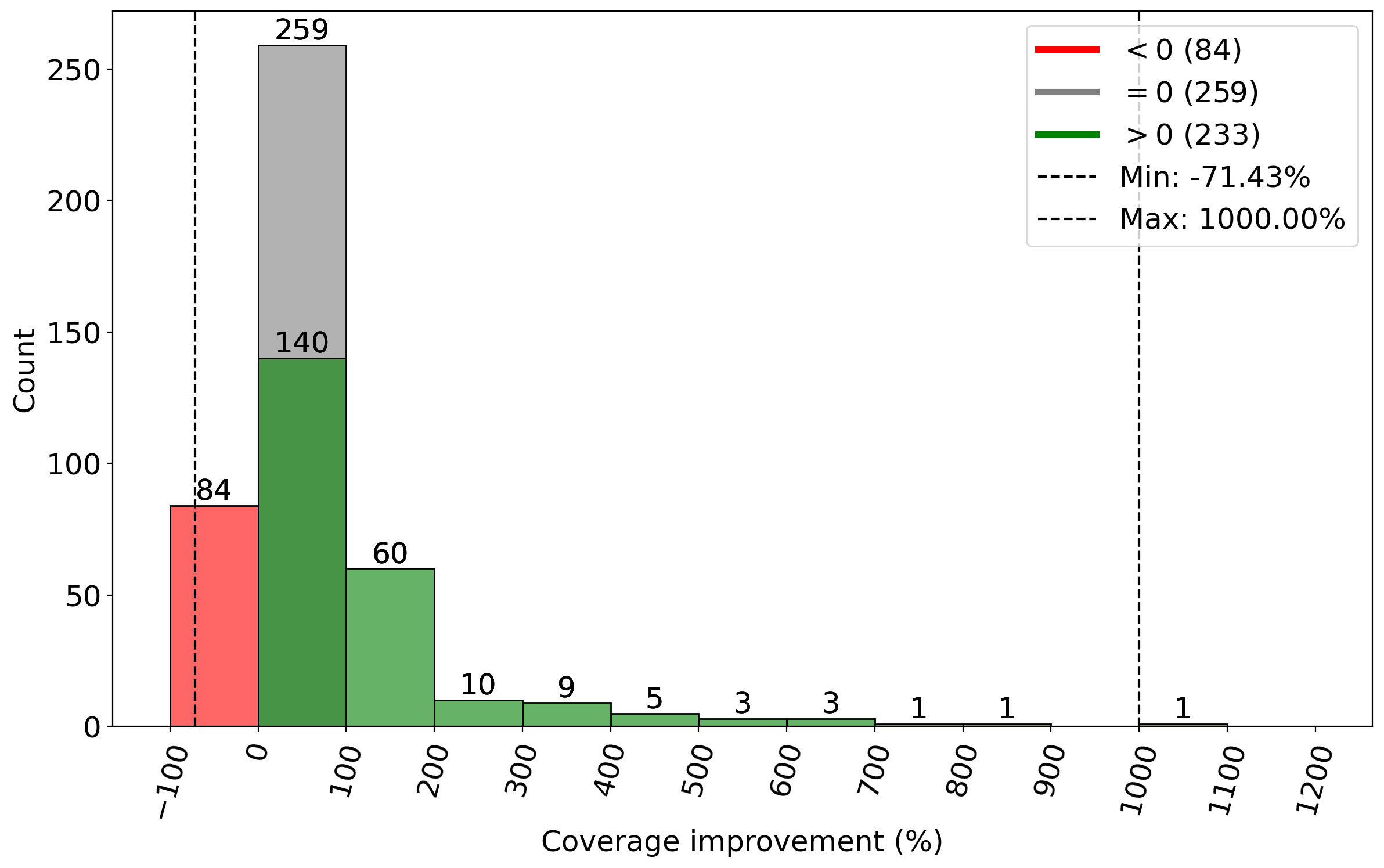}} \\
    \subcaptionbox{GLAS dataset coverage}{\includegraphics[width=0.49\textwidth]{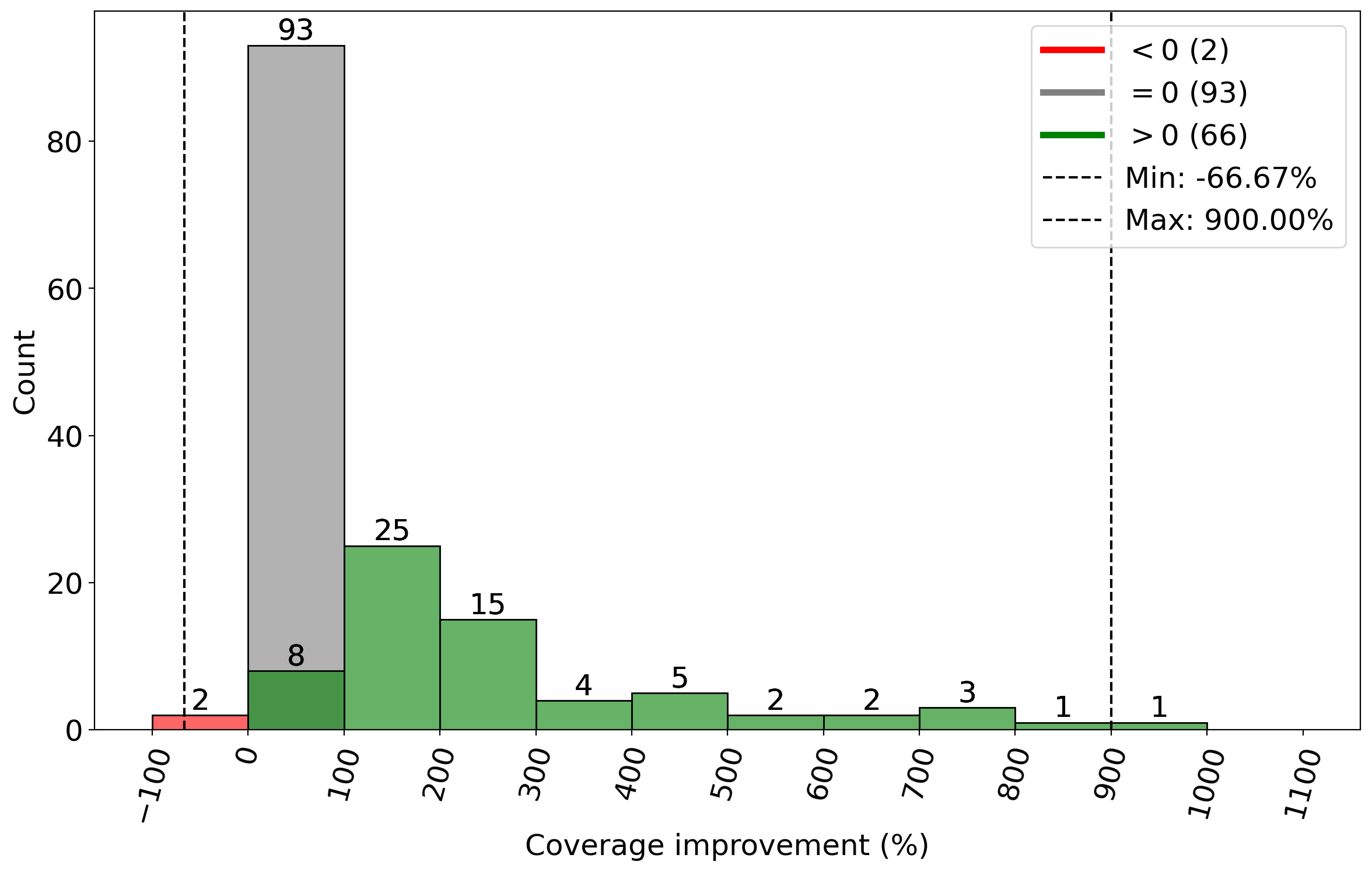}} 
    \subcaptionbox{WINE dataset coverage}{\includegraphics[width=0.49\textwidth]{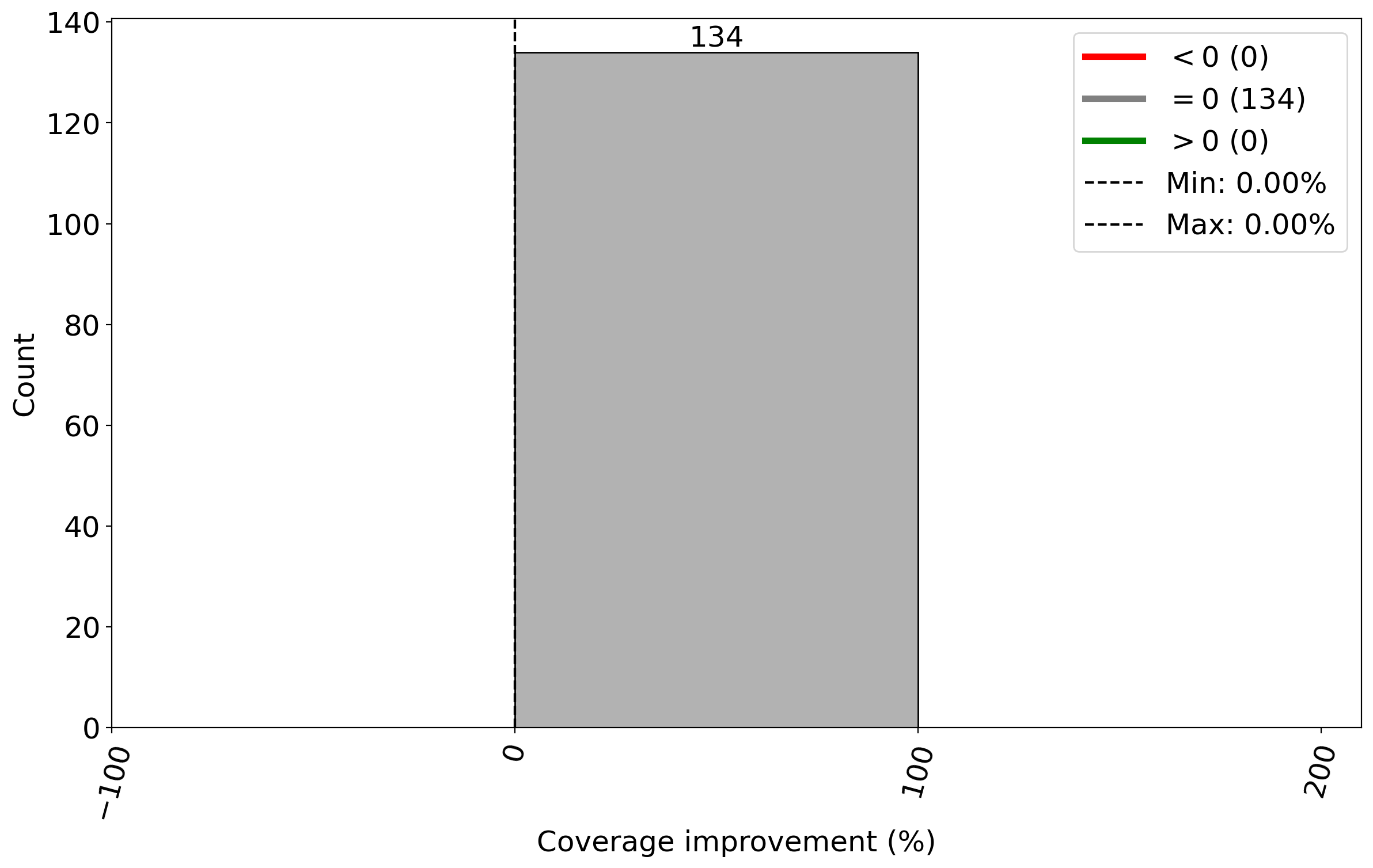}} \\
    \subcaptionbox{CLIM dataset coverage}{\includegraphics[width=0.49\textwidth]{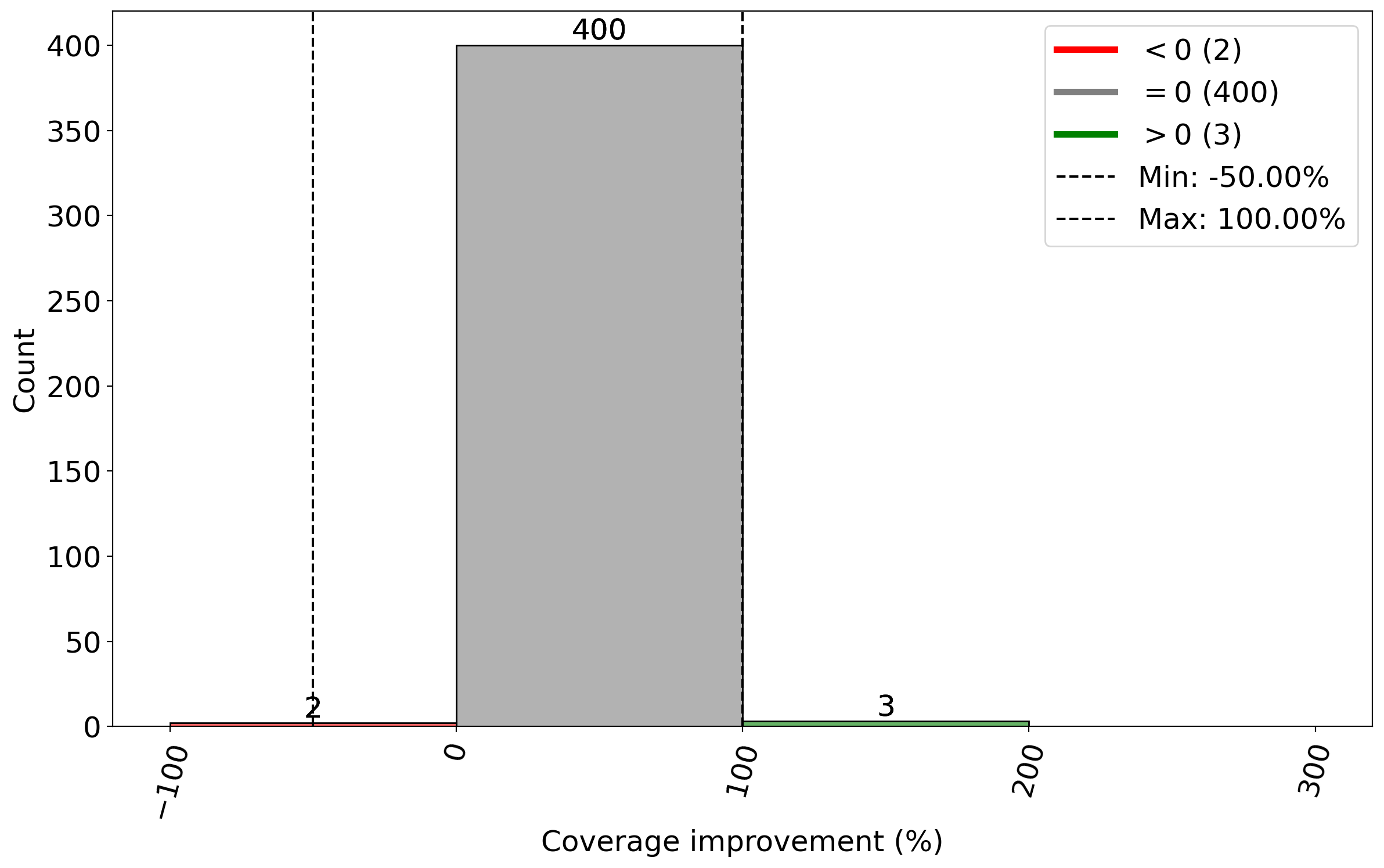}} 
    \subcaptionbox{PARK dataset coverage}{\includegraphics[width=0.49\textwidth]{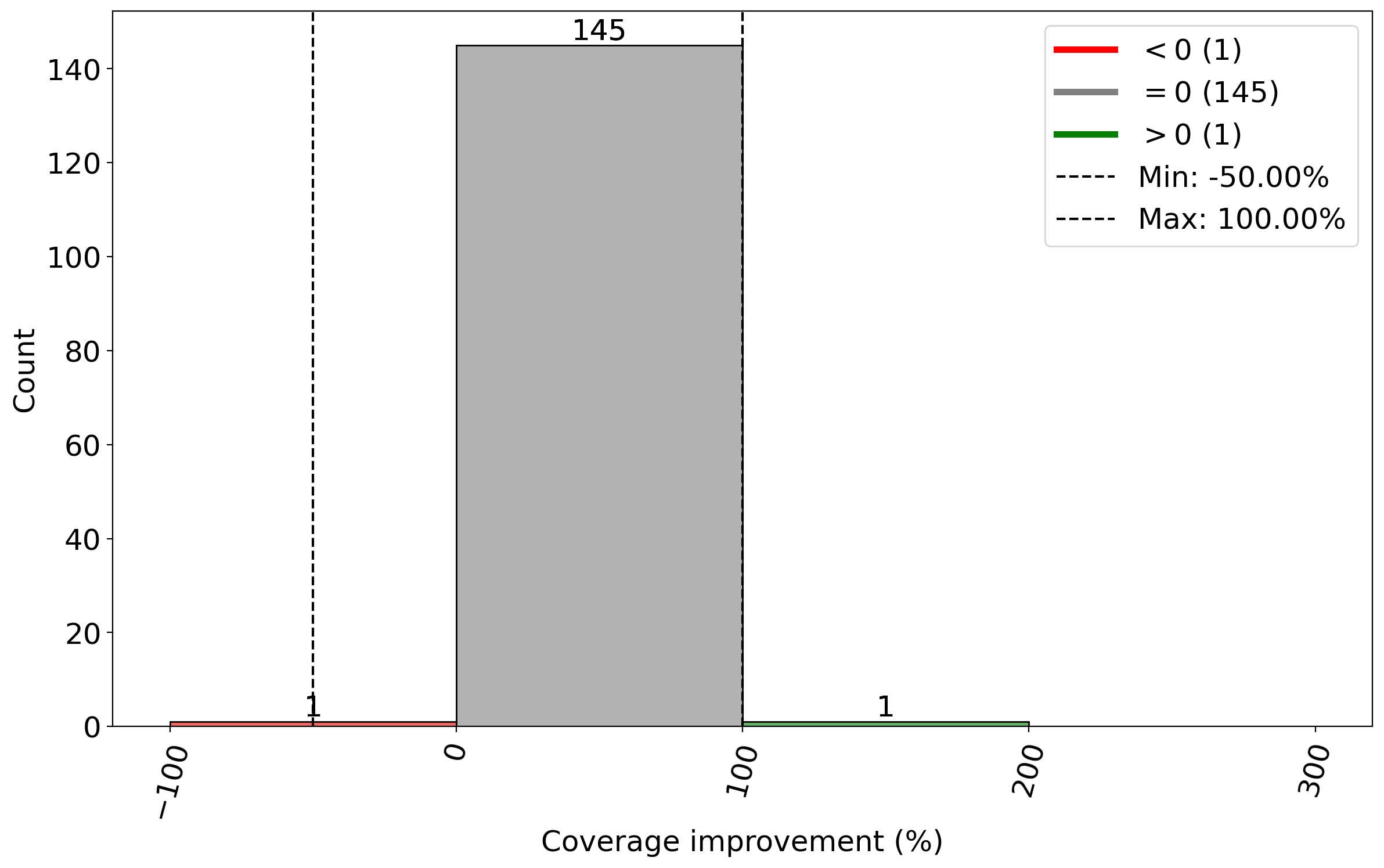}} \\
    \subcaptionbox{BRCW dataset coverage}{\includegraphics[width=0.49\textwidth]{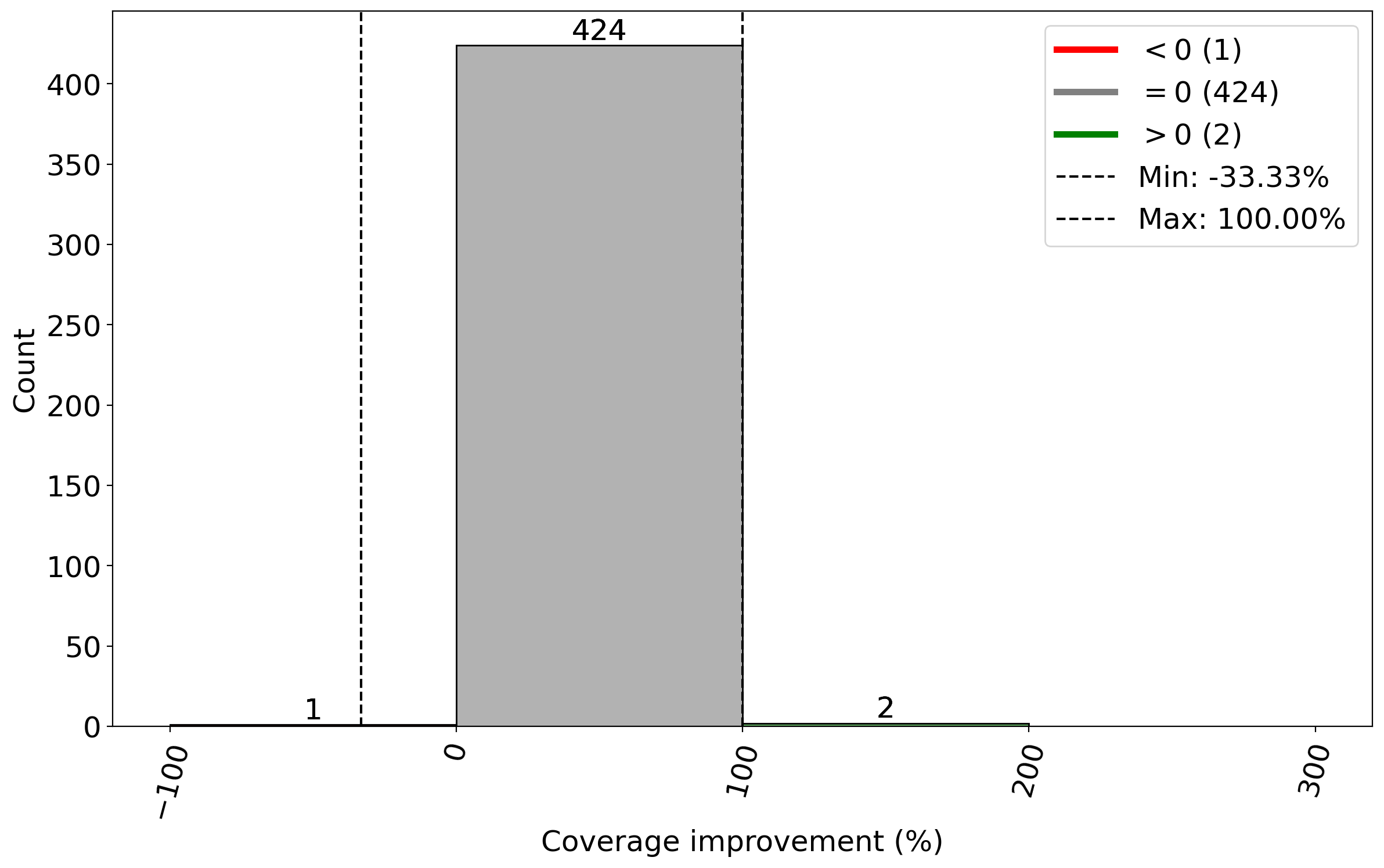}} 
    \subcaptionbox{IONS dataset coverage}{\includegraphics[width=0.49\textwidth]{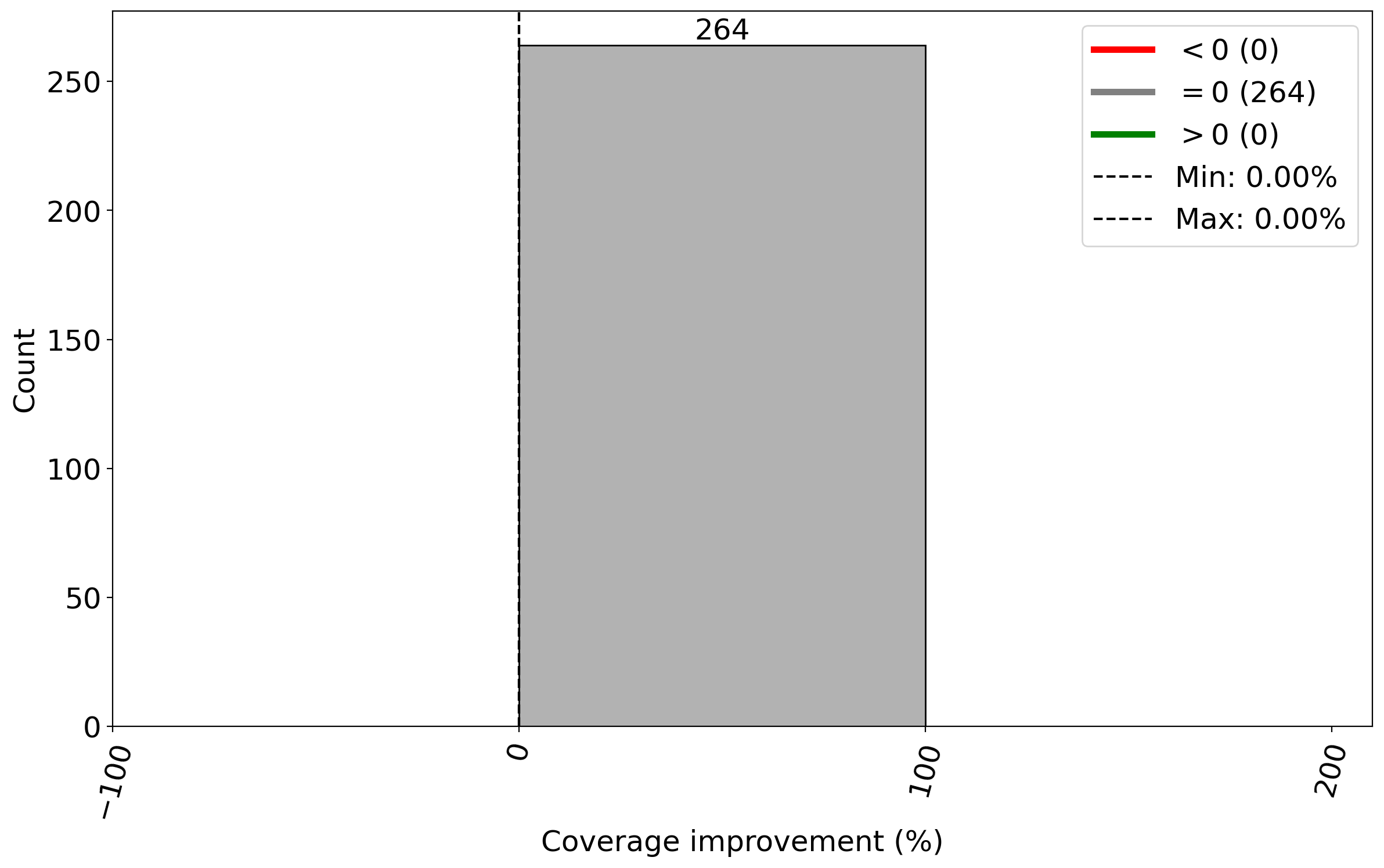}} \\
\end{figure}

The results for MLP in Figure \ref{fig:MLP_relative_coverage} reinforce the ability of Twostep to generate more general explanations. Twostep significantly outperforms Onestep in several datasets, with remarkable coverage improvements of up to 2000\% for the UKMO dataset and 2500\% for the BANK dataset, respectively. Similarly, in the GLAS and BLDT datasets, Twostep provides more general explanations in 66 against 2 and 246 against 99 cases, respectively. While Twostep achieves milder improvements for the CLIM, PARK, and BRCW datasets, it still performs better overall. Twostep does not provide additional coverage for the WINE and IONS datasets. 



\subsubsection{Results for Synthetic Data Coverage}


Assessing coverage on real datasets can be challenging when they contain only a few instances, potentially limiting a proper evaluation of generalization. To address this issue, we analyze coverage using artificially generated instances, allowing for a more comprehensive comparison between Twostep and Onestep.

Figures \ref{fig:SVM_artificial_coverage} and \ref{fig:MLP_artificial_coverage} present the synthetic data coverage results for SVM and MLP, respectively. In all datasets, for both SVM and MLP, Twostep shows a higher number of cases with improved coverage than with reduced coverage. Moreover, in half of the datasets, for both SVM and MLP, Twostep shows more cases with improved coverage than cases with decreased coverage, as well as more than cases where coverage remains the same.

For the datasets with a higher number of features, BRCW and IONS, Twostep achieved a significantly greater number of cases with improved coverage in the synthetic data evaluation compared to the dataset coverage results with MLP. Specifically, while no improvements were observed for IONS and only two cases showed improvement for BRCW in the dataset coverage analysis, the synthetic data coverage results show 22 cases of improvement for IONS and 136 for BRCW. Notably, in some instances, Twostep achieved up to a 10,000\% improvement in coverage. These findings highlight the importance of using synthetic data to assess the generalization properties of the explanation methods.

\begin{figure}[H]
    \centering
    \subcaptionbox{IRIS synthetic dataset coverage}{\includegraphics[width=0.49\textwidth]{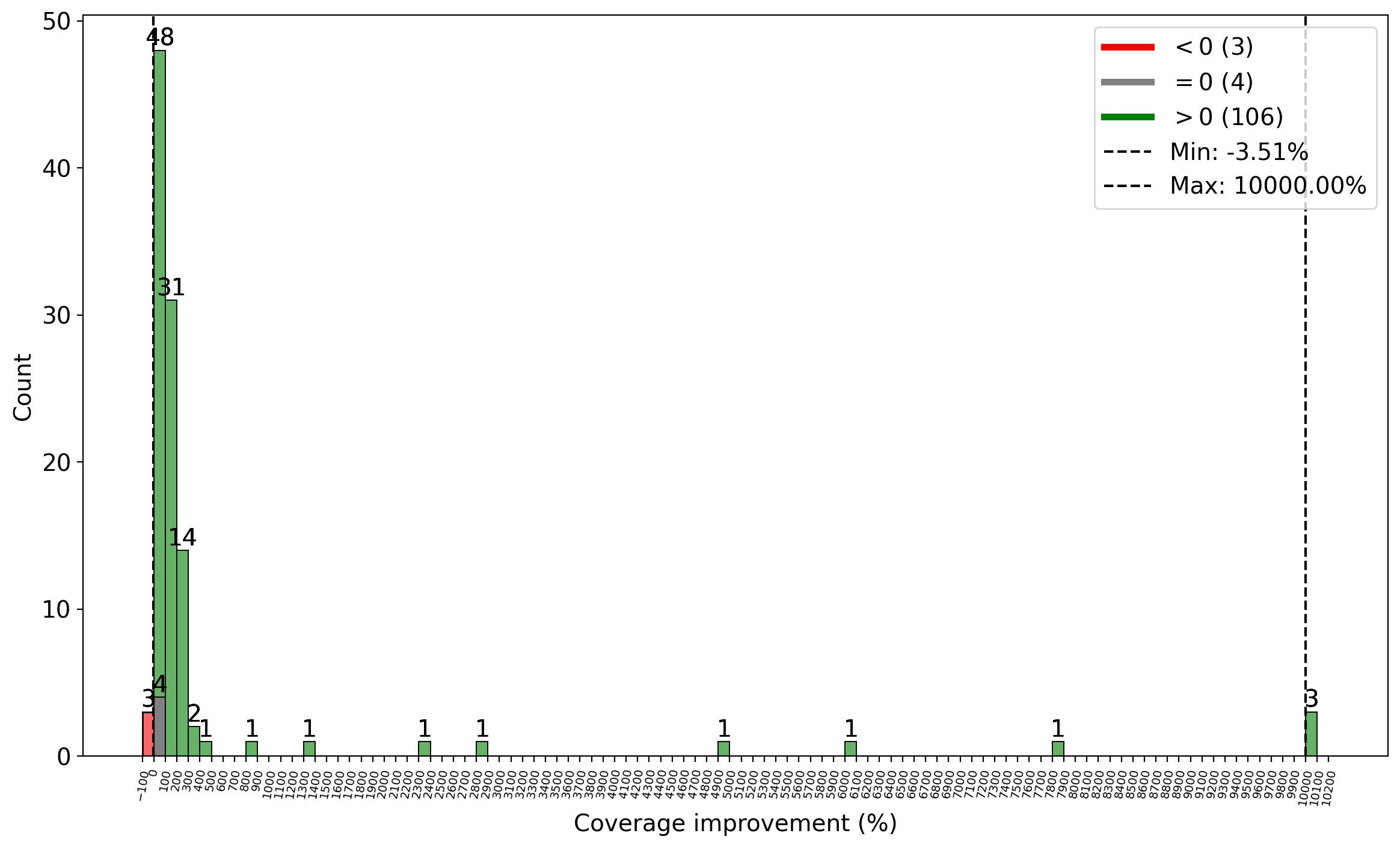}} 
    \subcaptionbox{BLDT synthetic dataset coverage}{\includegraphics[width=0.49\textwidth]{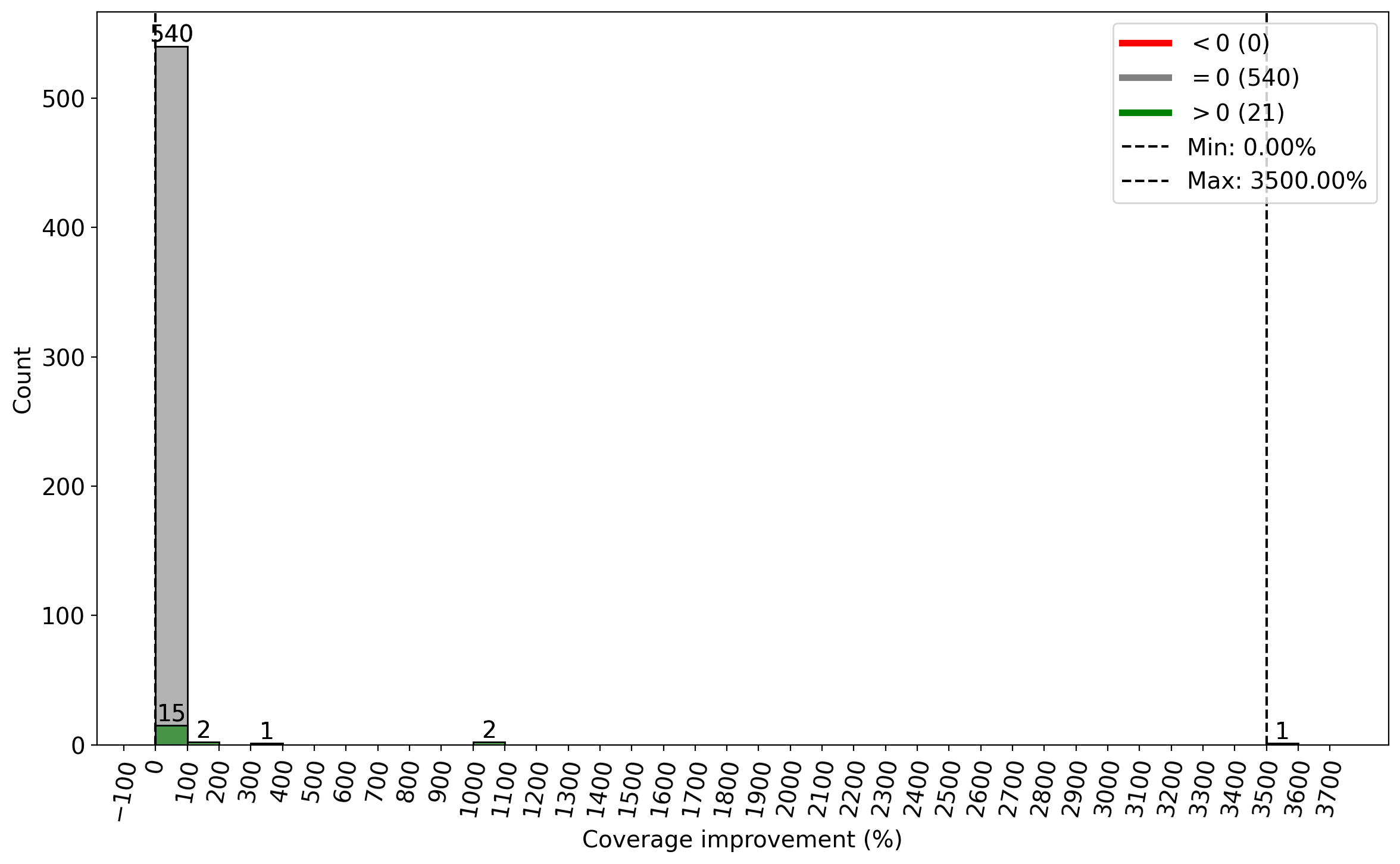}} \\
    \subcaptionbox{BANK synthetic dataset coverage}{\includegraphics[width=0.49\textwidth]{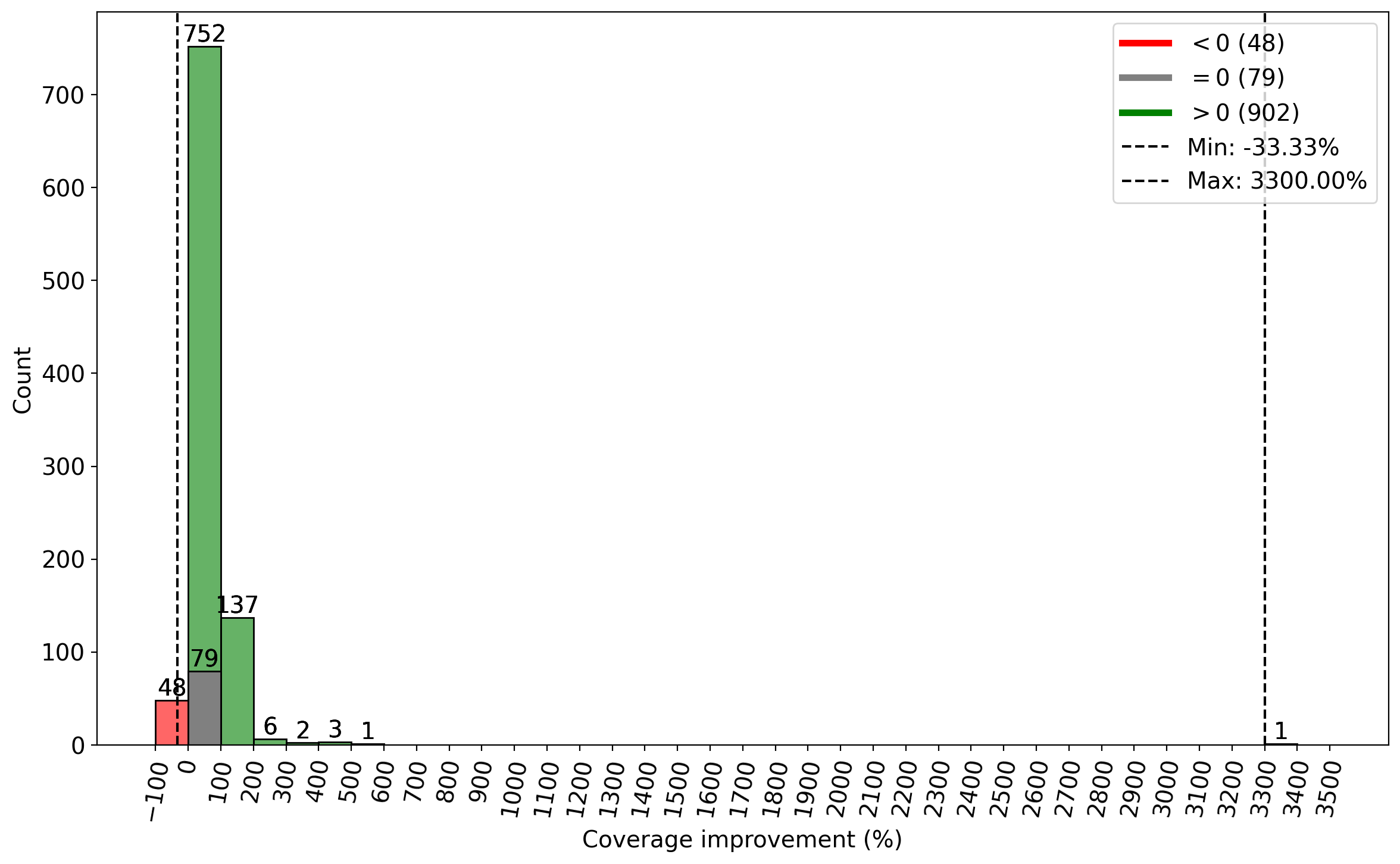}} 
    \subcaptionbox{UKMO synthetic dataset coverage}{\includegraphics[width=0.49\textwidth]{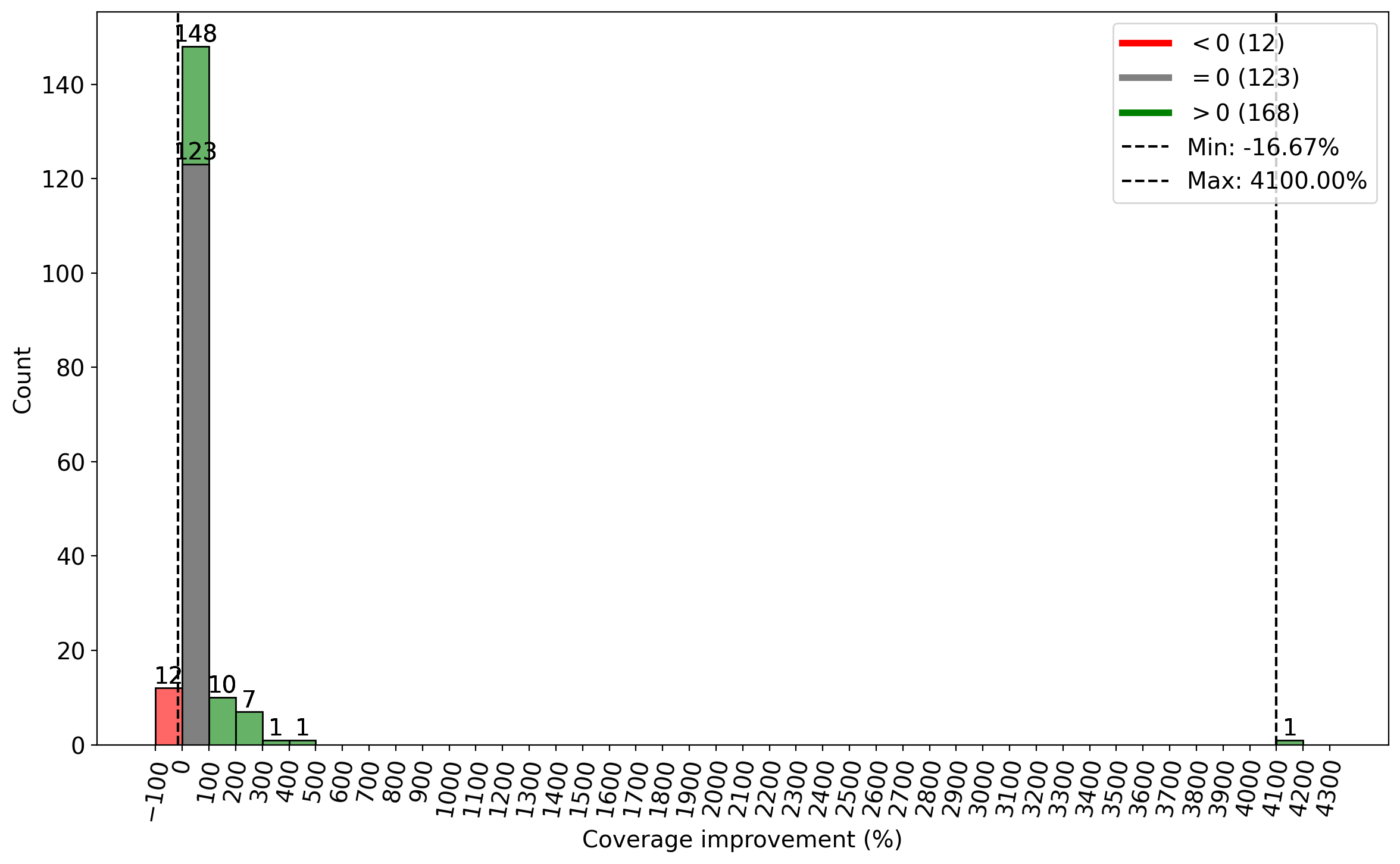}} \\
    \subcaptionbox{VRTC synthetic dataset coverage}{\includegraphics[width=0.49\textwidth]{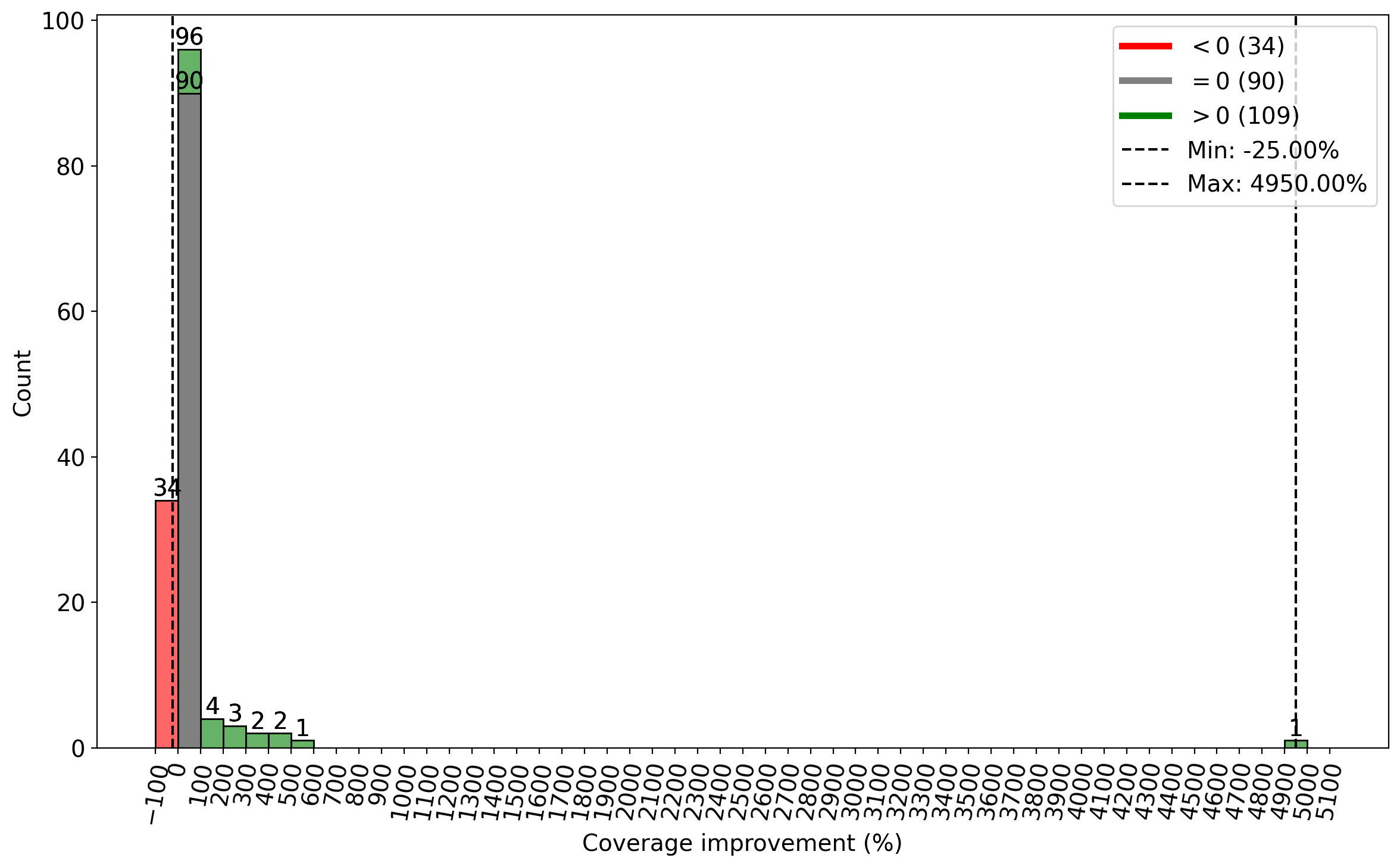}} 
    \subcaptionbox{PIMA synthetic dataset coverage}{\includegraphics[width=0.49\textwidth]{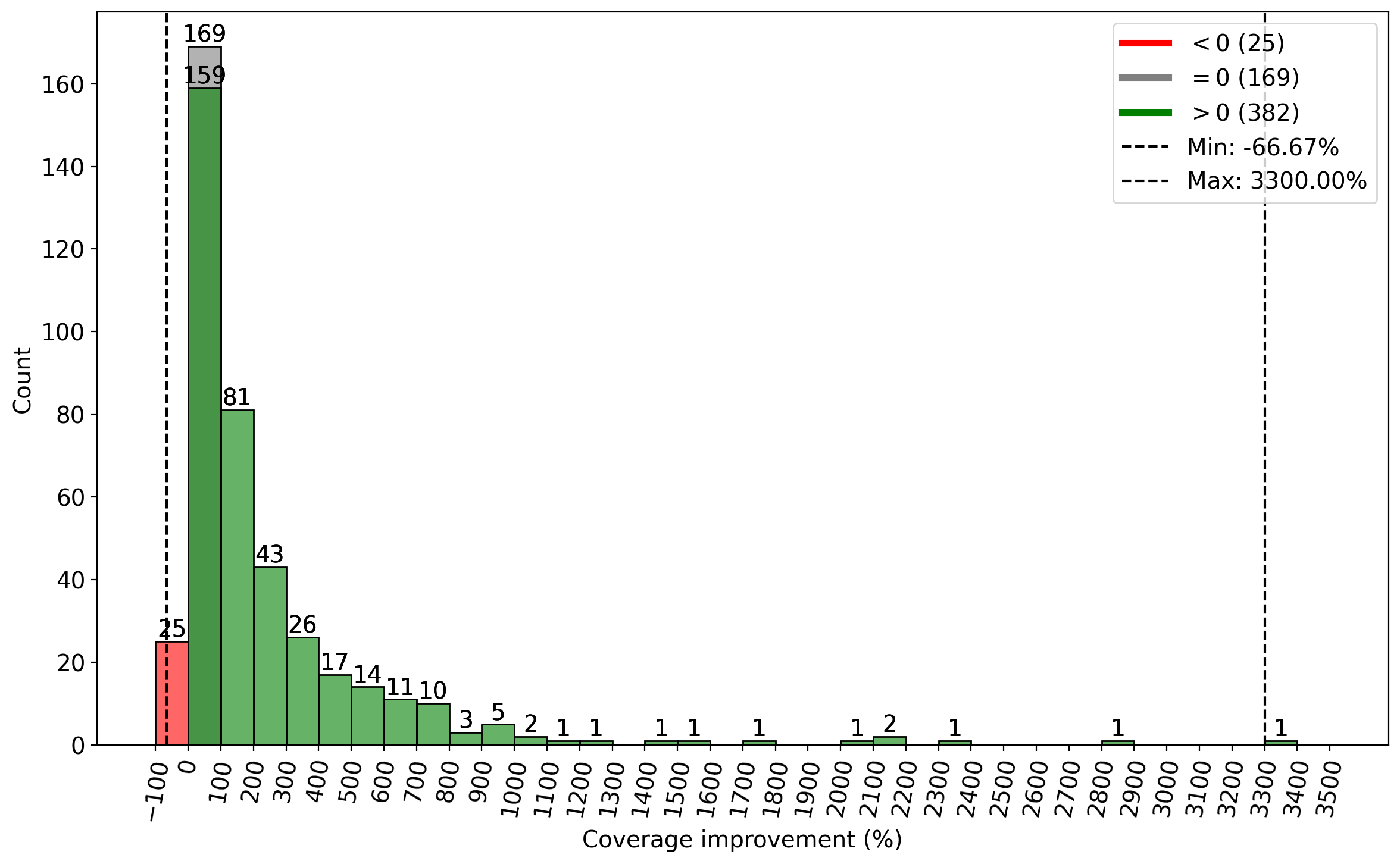}} \\
    \subcaptionbox{GLAS synthetic dataset coverage}{\includegraphics[width=0.49\textwidth]{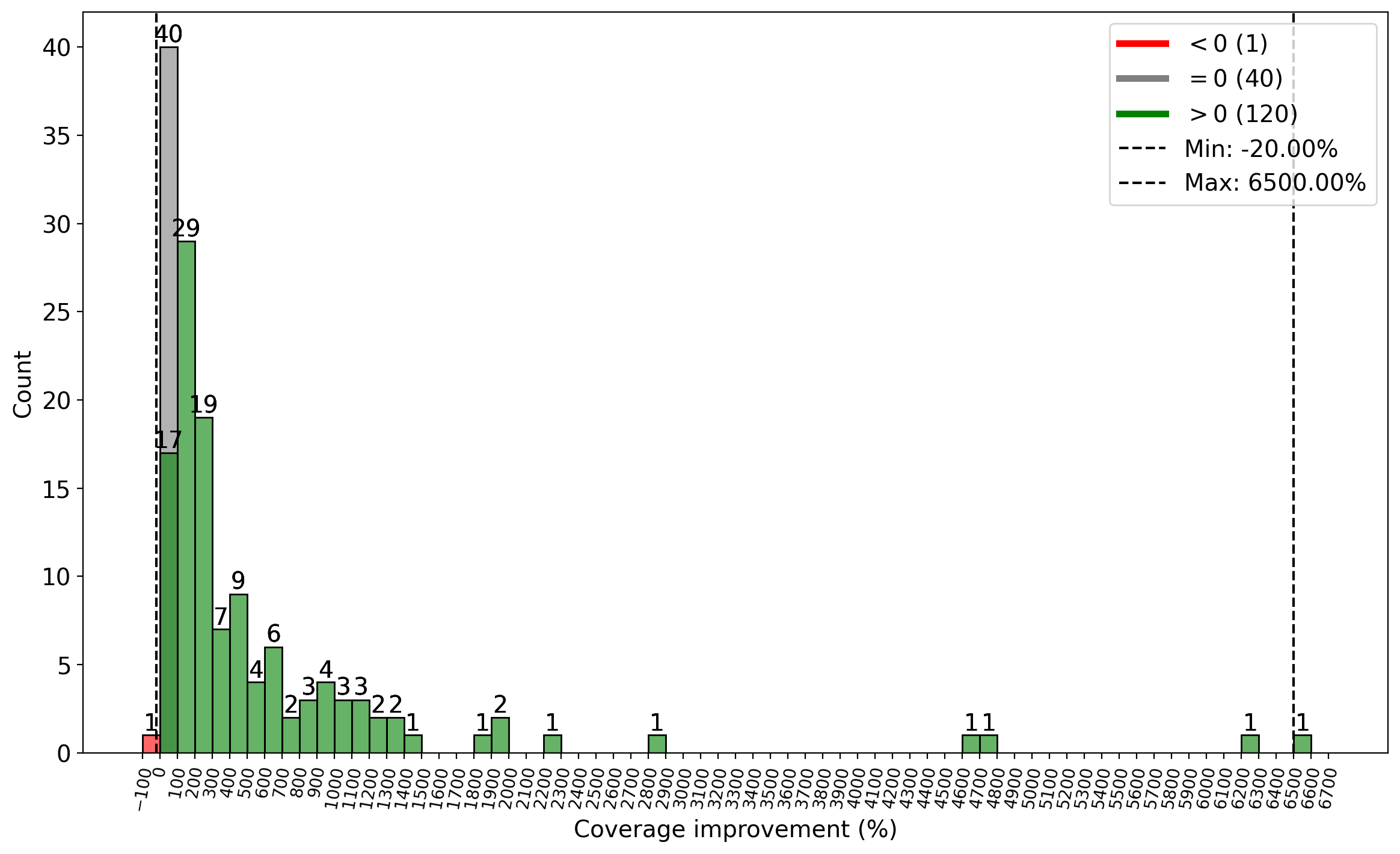}} 
    \subcaptionbox{WINE synthetic dataset coverage}{\includegraphics[width=0.49\textwidth]{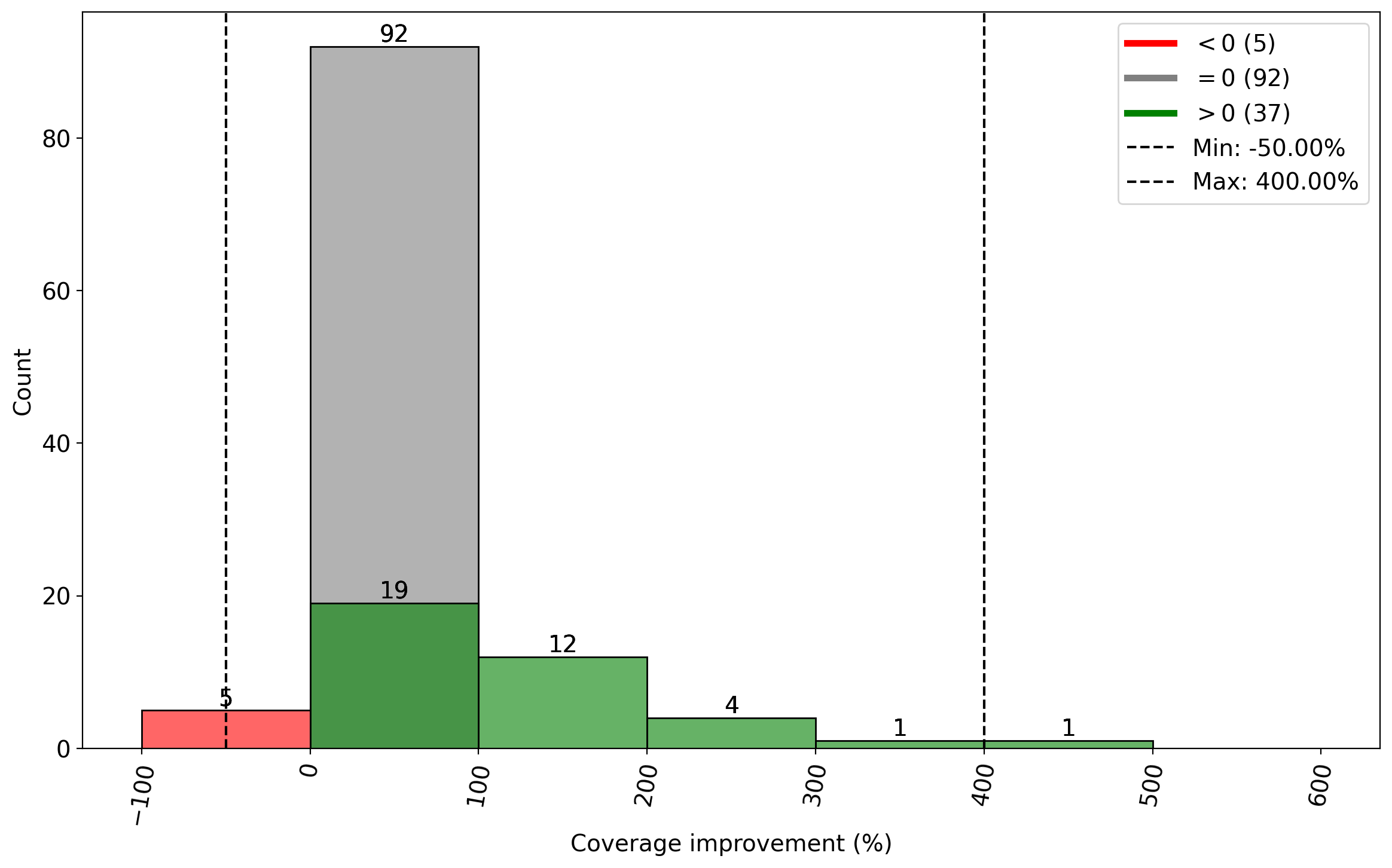}} \\
    \caption{Histogram of coverage improvement (\%) achieved by Twostep over Onestep for synthetic data, with the SVM model and Twostep parameter fixed at $p=0.25$. Red bars represent cases with worsened coverage, gray bars represent cases with same coverage, and green bars represent cases with improved coverage. The best improvement and the worst deterioration are highlighted.}
    \label{fig:SVM_artificial_coverage}
\end{figure}

\begin{figure}[H]
    \ContinuedFloat  
    \centering
    \subcaptionbox{CLIM synthetic dataset coverage}{\includegraphics[width=0.49\textwidth]{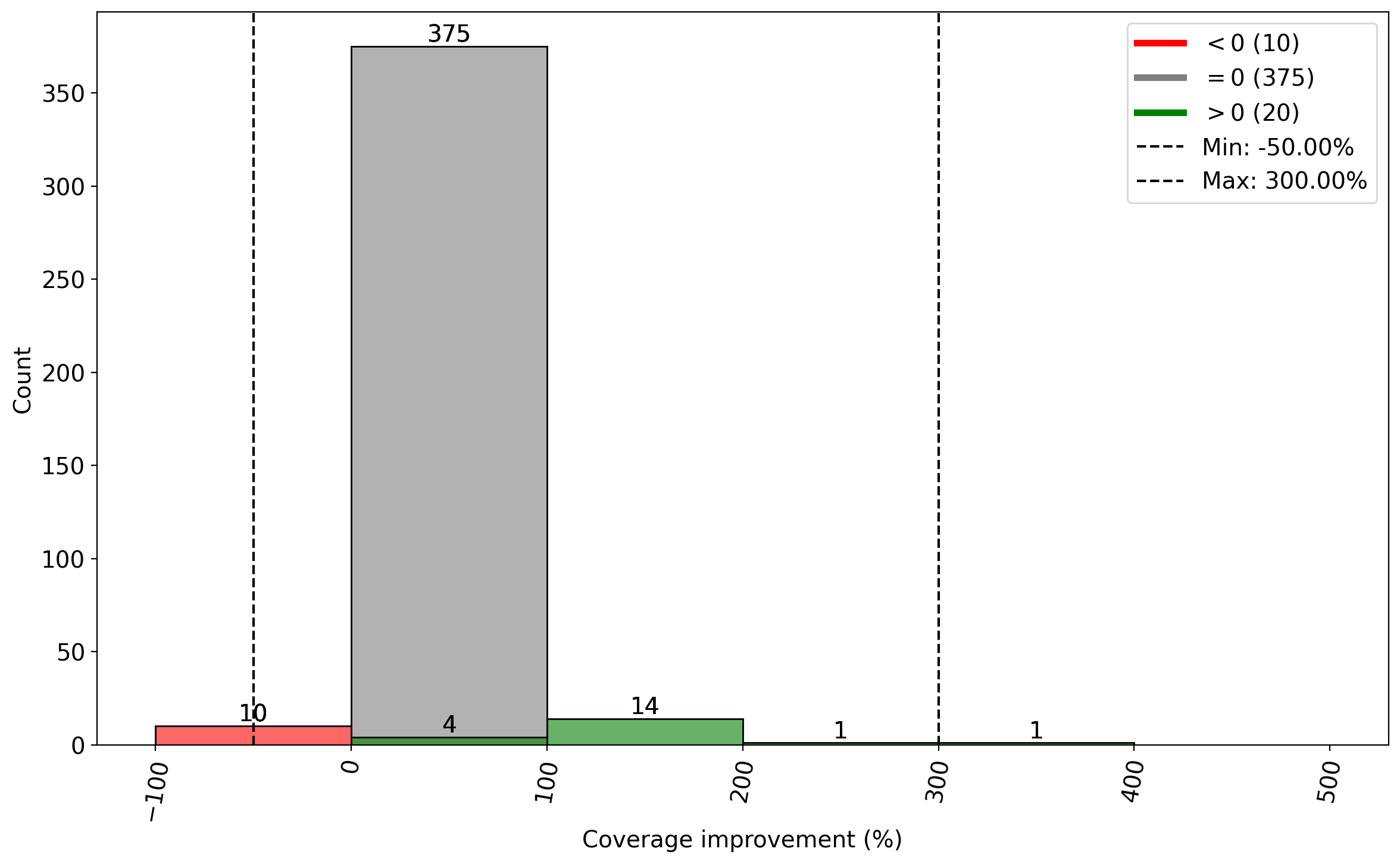}} 
    \subcaptionbox{PARK synthetic dataset coverage}{\includegraphics[width=0.49\textwidth]{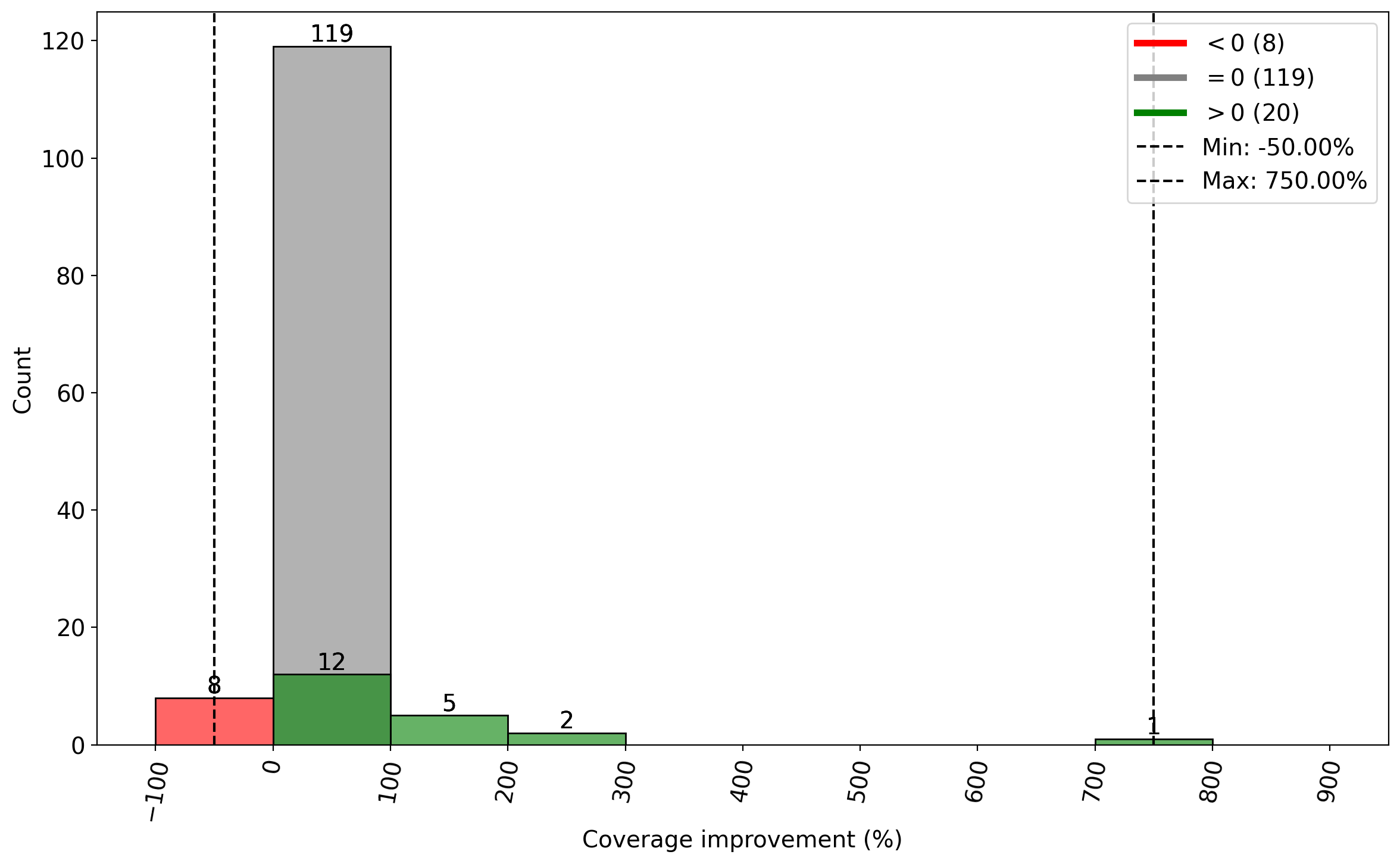}} \\
    \subcaptionbox{BRCW synthetic dataset coverage}{\includegraphics[width=0.49\textwidth]{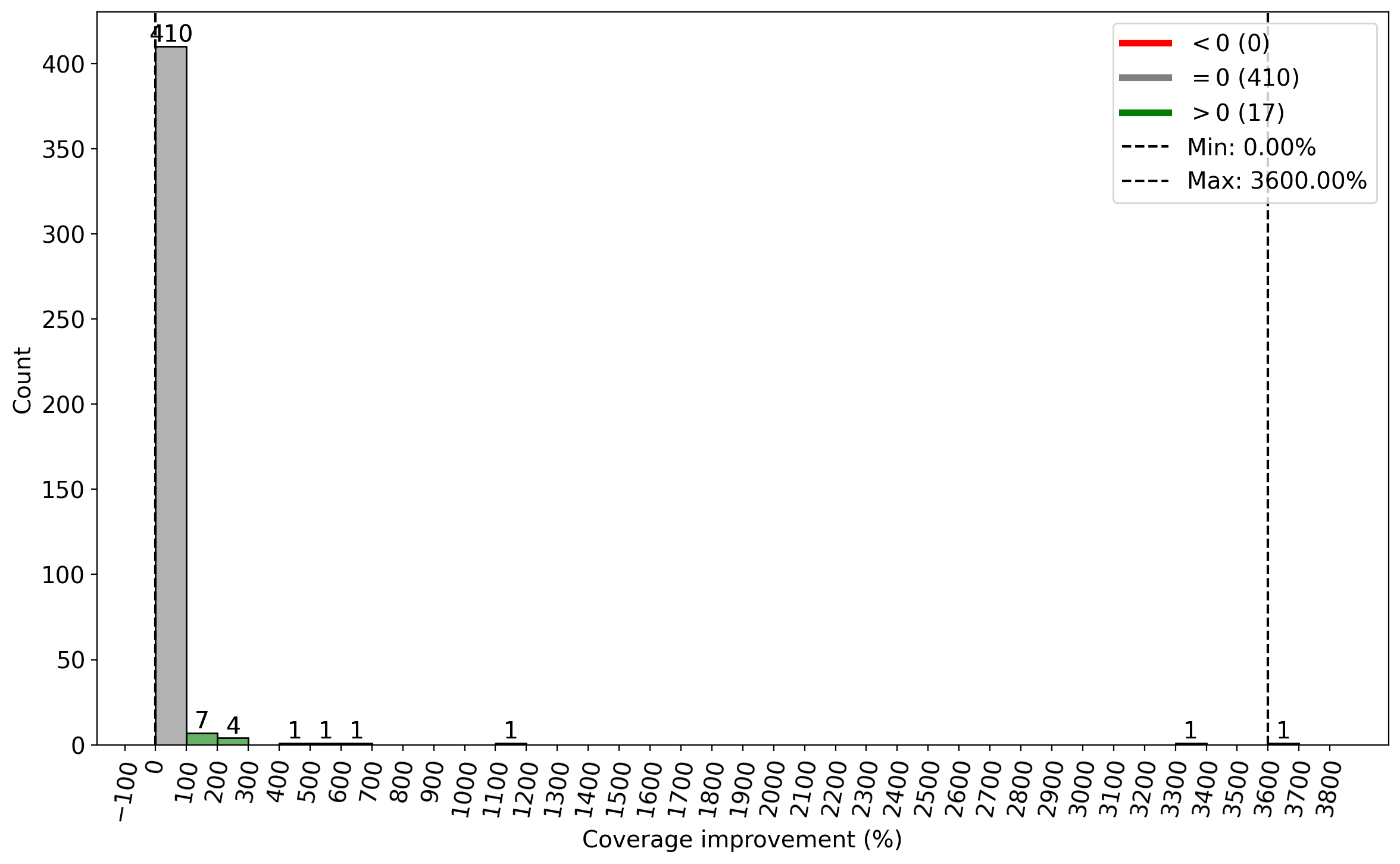}} 
    \subcaptionbox{IONS synthetic dataset coverage}{\includegraphics[width=0.49\textwidth]{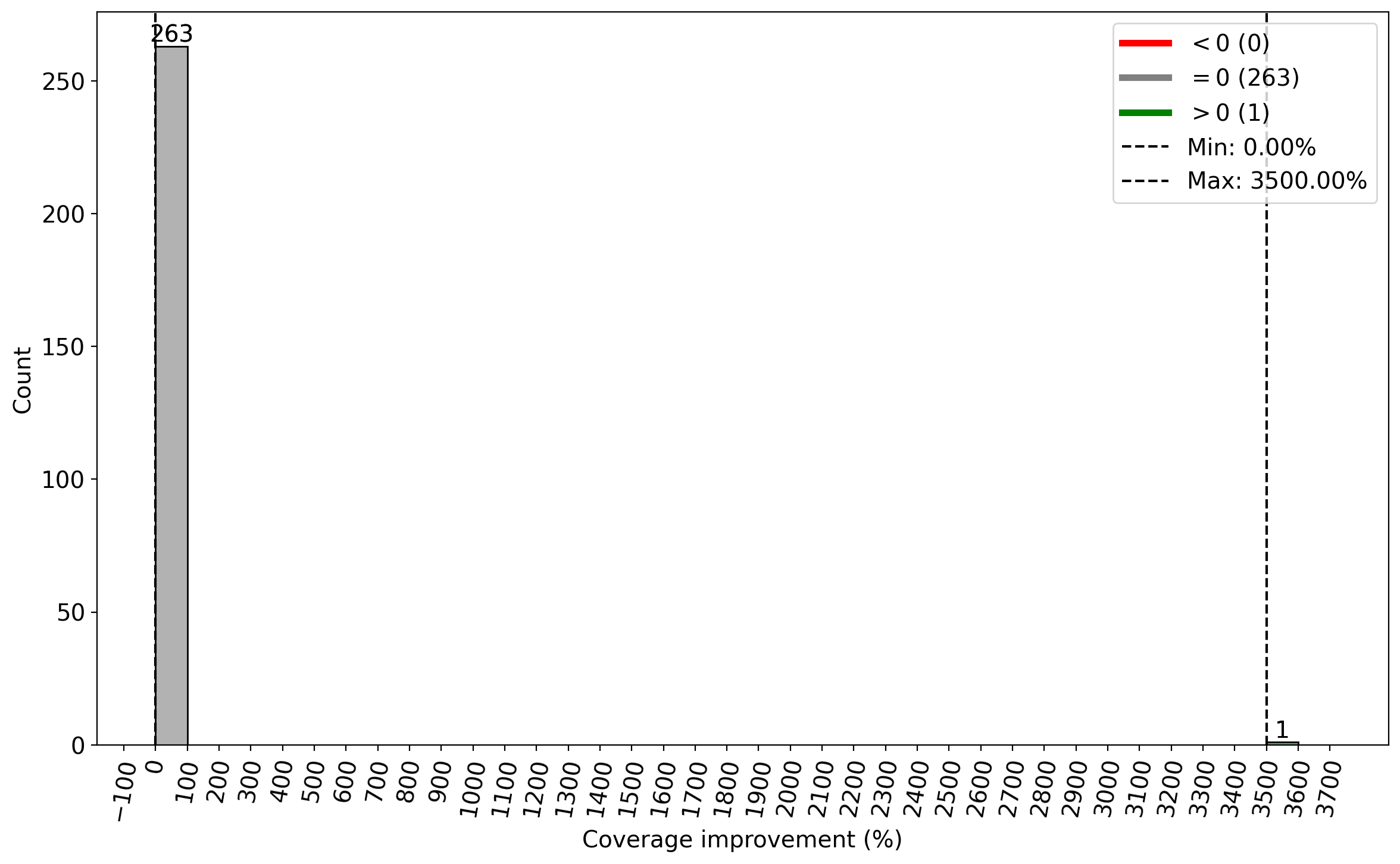}} \\
\end{figure}

\begin{figure}[H]
    \centering
    \subcaptionbox{IRIS synthetic dataset coverage}{\includegraphics[width=0.49\textwidth]{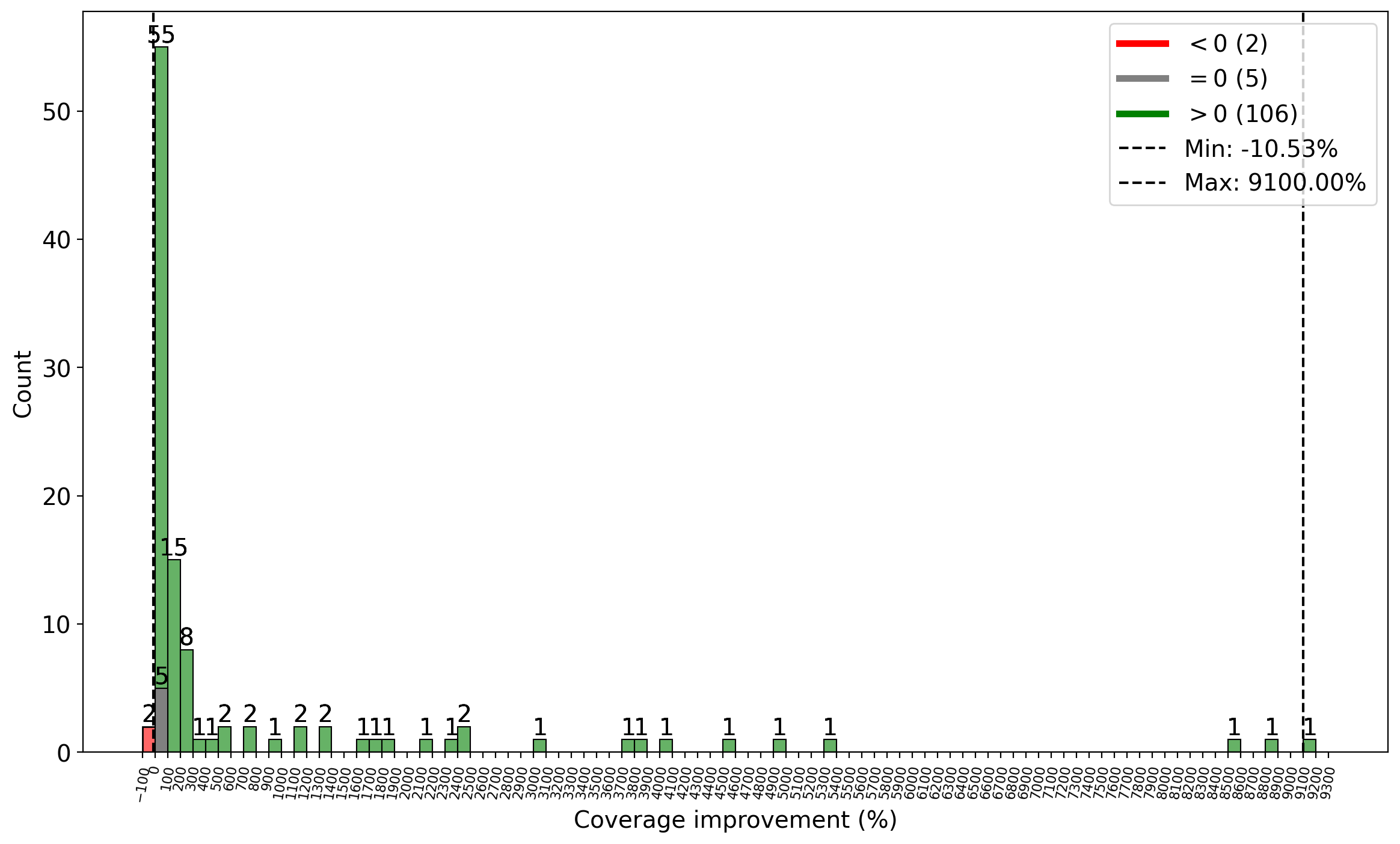}} 
    \subcaptionbox{BLDT synthetic dataset coverage}{\includegraphics[width=0.49\textwidth]{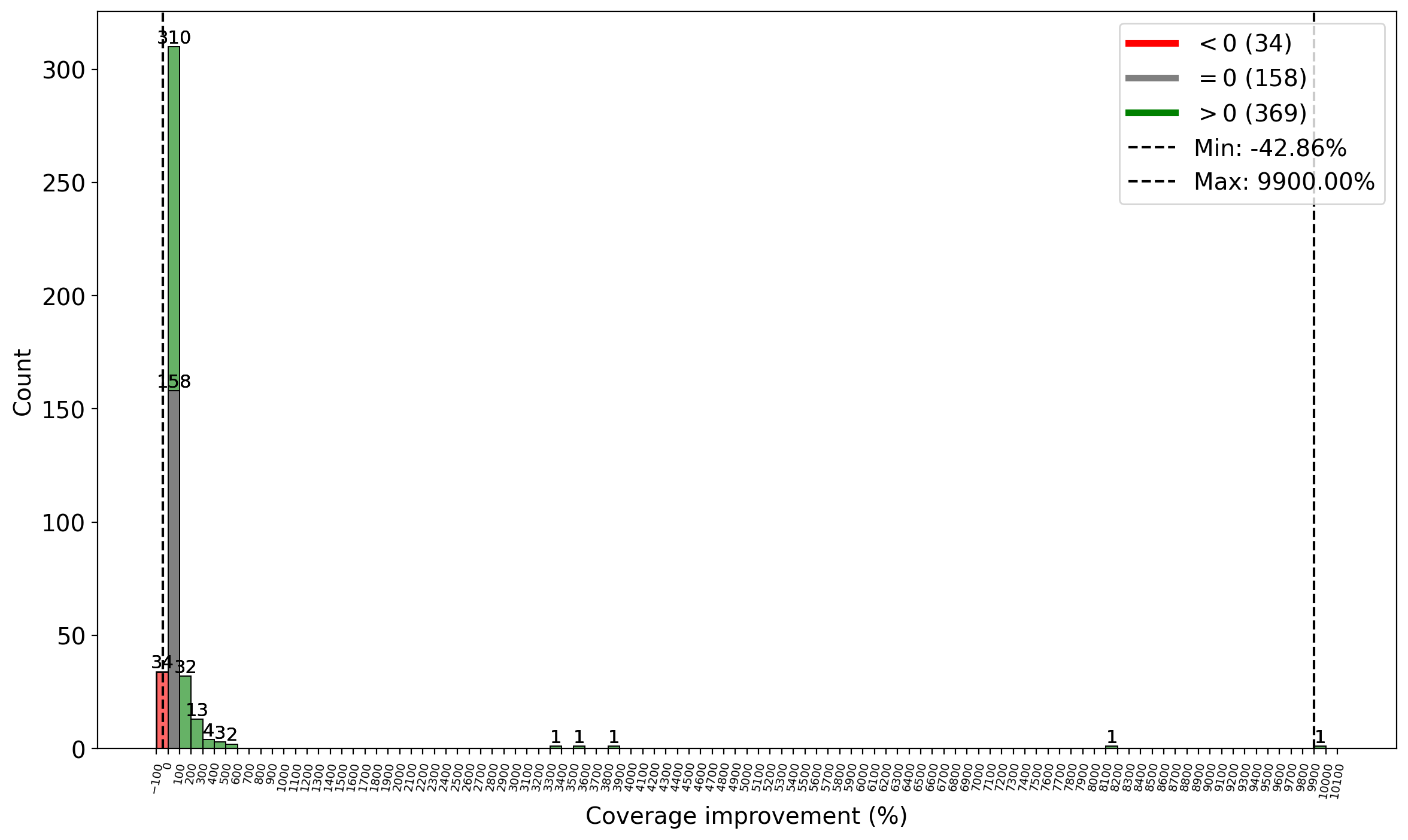}} \\
    \subcaptionbox{BANK synthetic dataset coverage}{\includegraphics[width=0.49\textwidth]{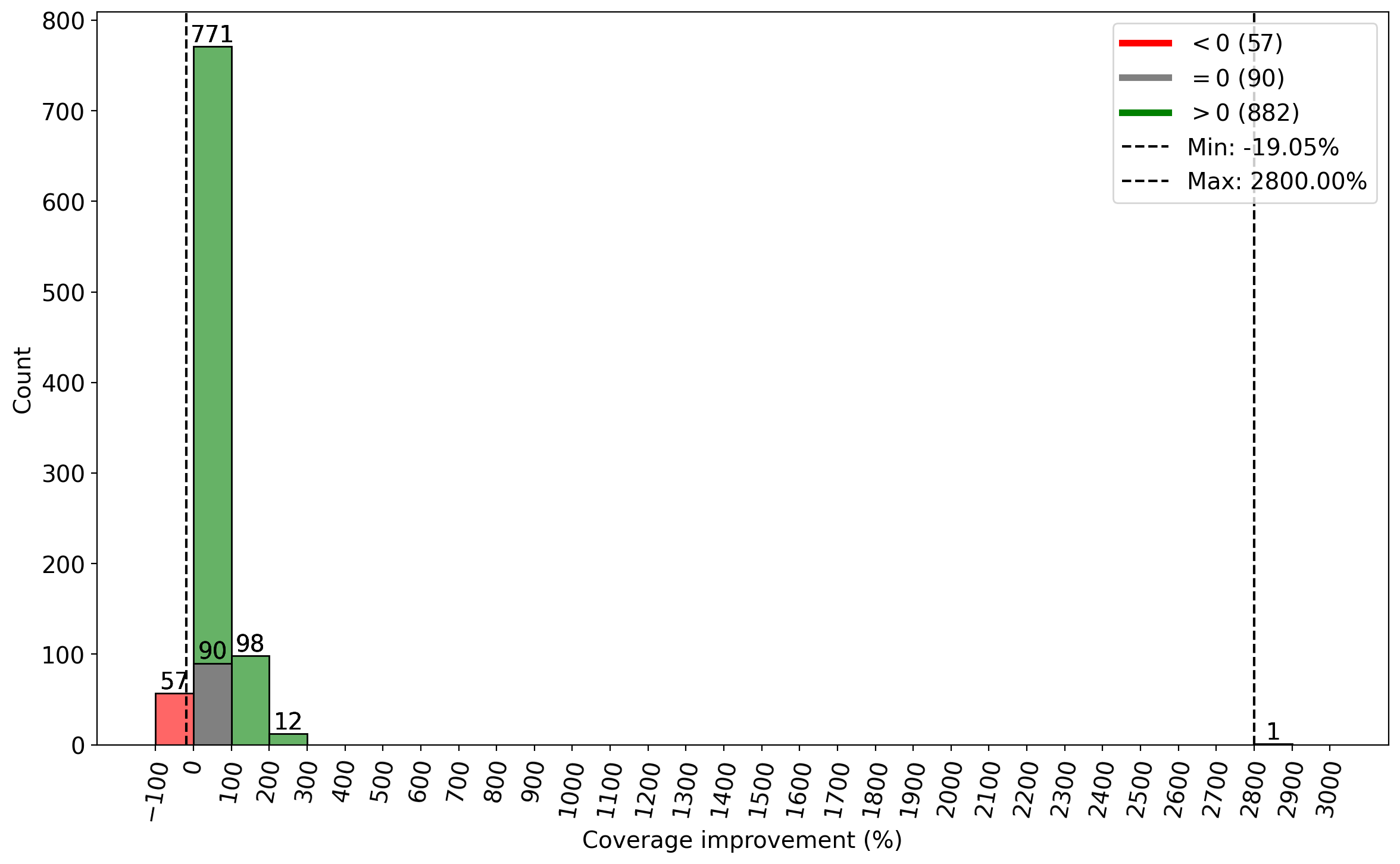}} 
    \subcaptionbox{UKMO synthetic dataset coverage}{\includegraphics[width=0.49\textwidth]{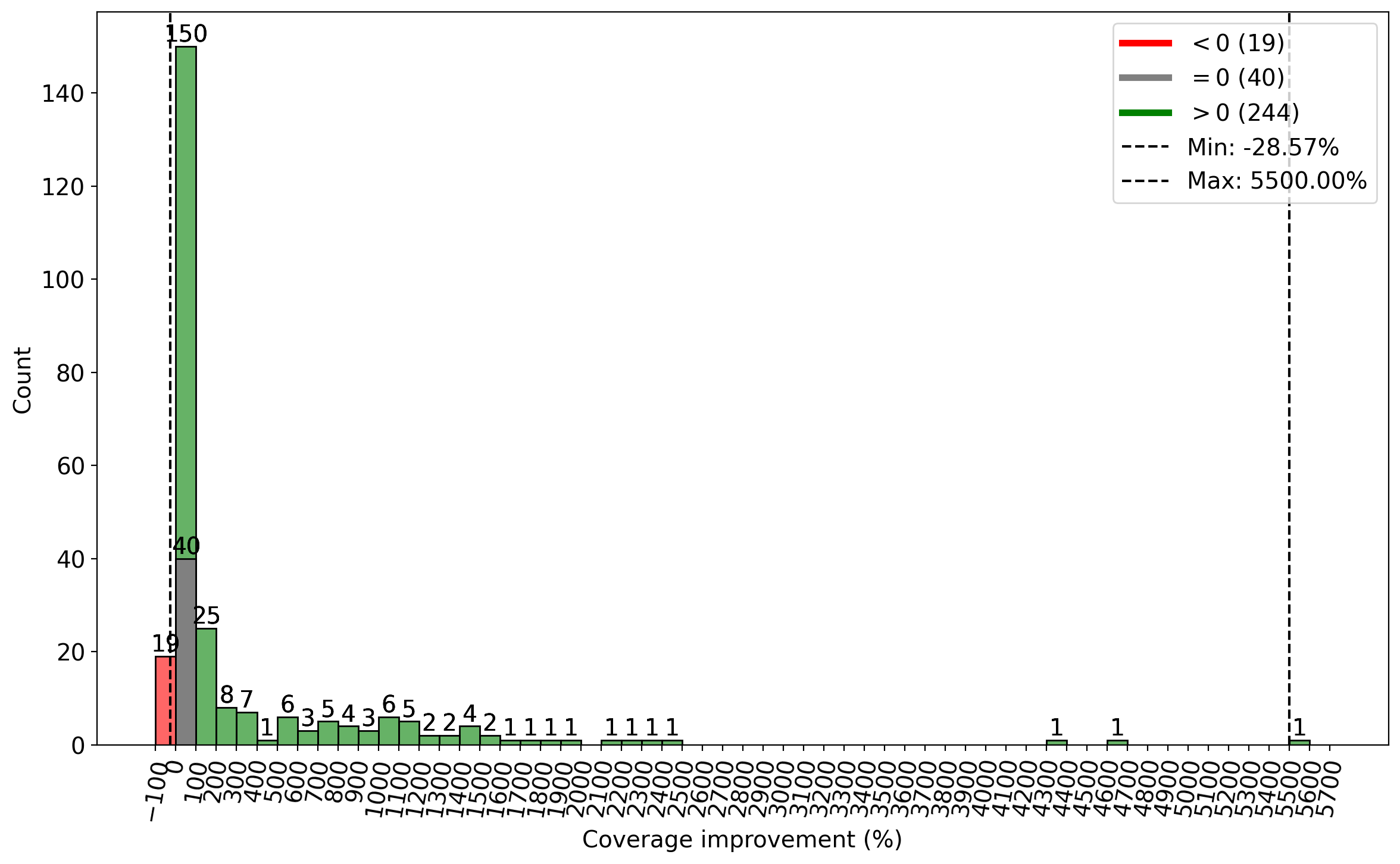}} \\
    
    \caption{Histogram of coverage improvement (\%) achieved by Twostep over Onestep for synthetic data, with the MLP model and Twostep parameter fixed at $p=0.25$. Red bars represent cases with worsened coverage, gray bars represent cases with same coverage, and green bars represent cases with improved coverage. The best improvement and the worst deterioration are highlighted.}
    \label{fig:MLP_artificial_coverage}
\end{figure}

\begin{figure}[H]
    \ContinuedFloat  
    \centering
    \subcaptionbox{VRTC synthetic dataset coverage}{\includegraphics[width=0.49\textwidth]{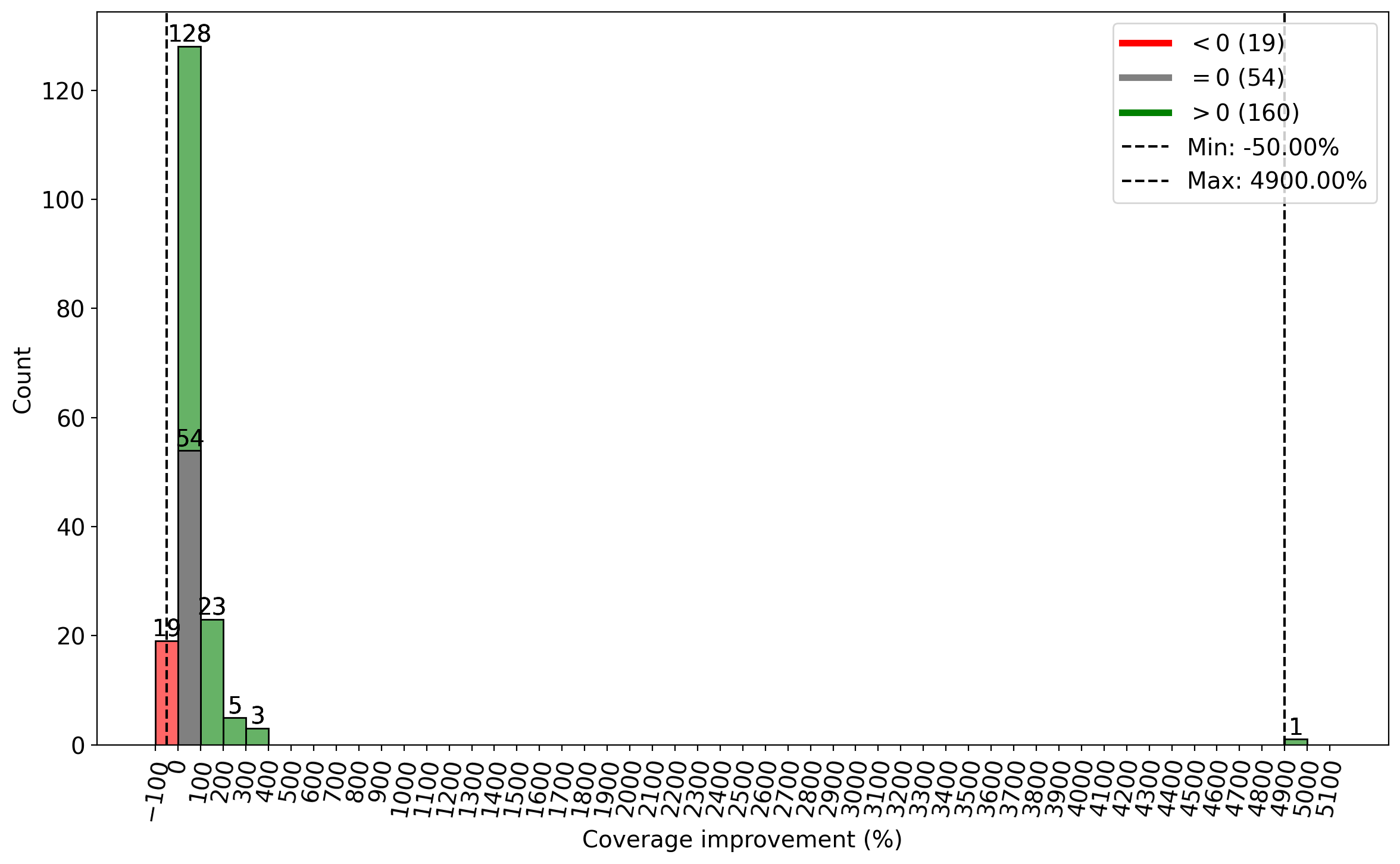}} 
    \subcaptionbox{PIMA synthetic dataset coverage}{\includegraphics[width=0.49\textwidth]{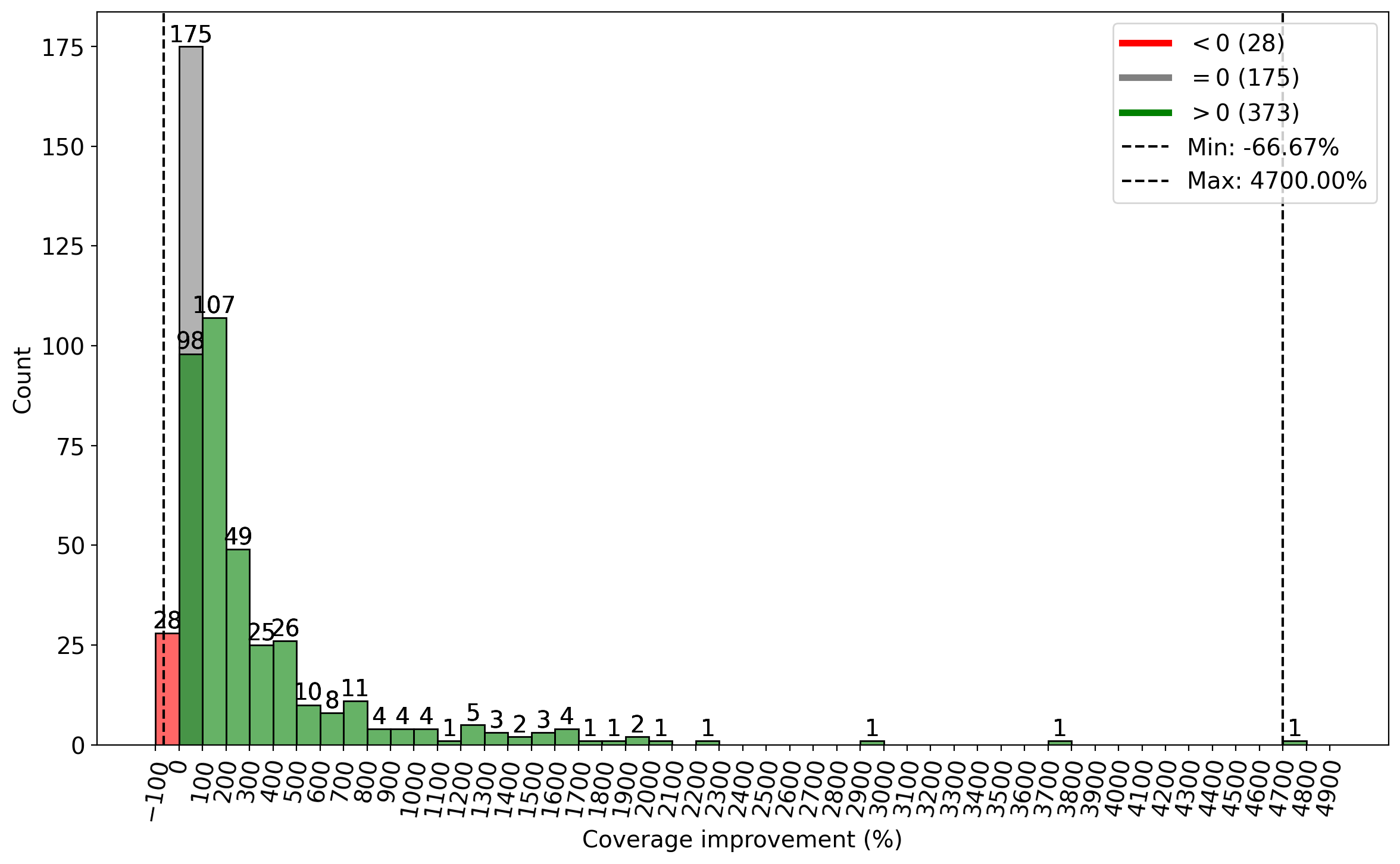}} \\
    \subcaptionbox{GLAS synthetic dataset coverage}{\includegraphics[width=0.49\textwidth]{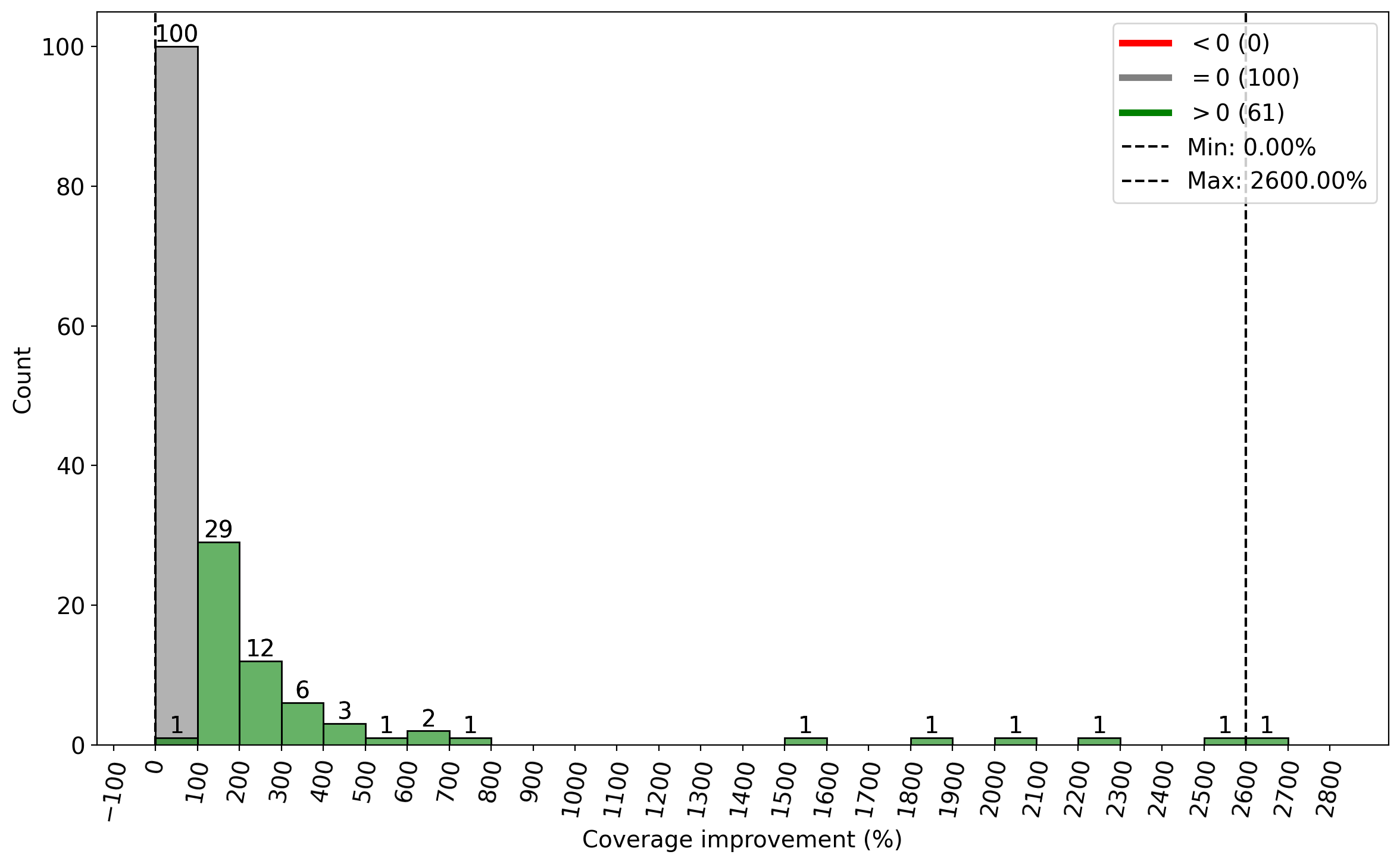}} 
    \subcaptionbox{WINE synthetic dataset coverage}{\includegraphics[width=0.49\textwidth]{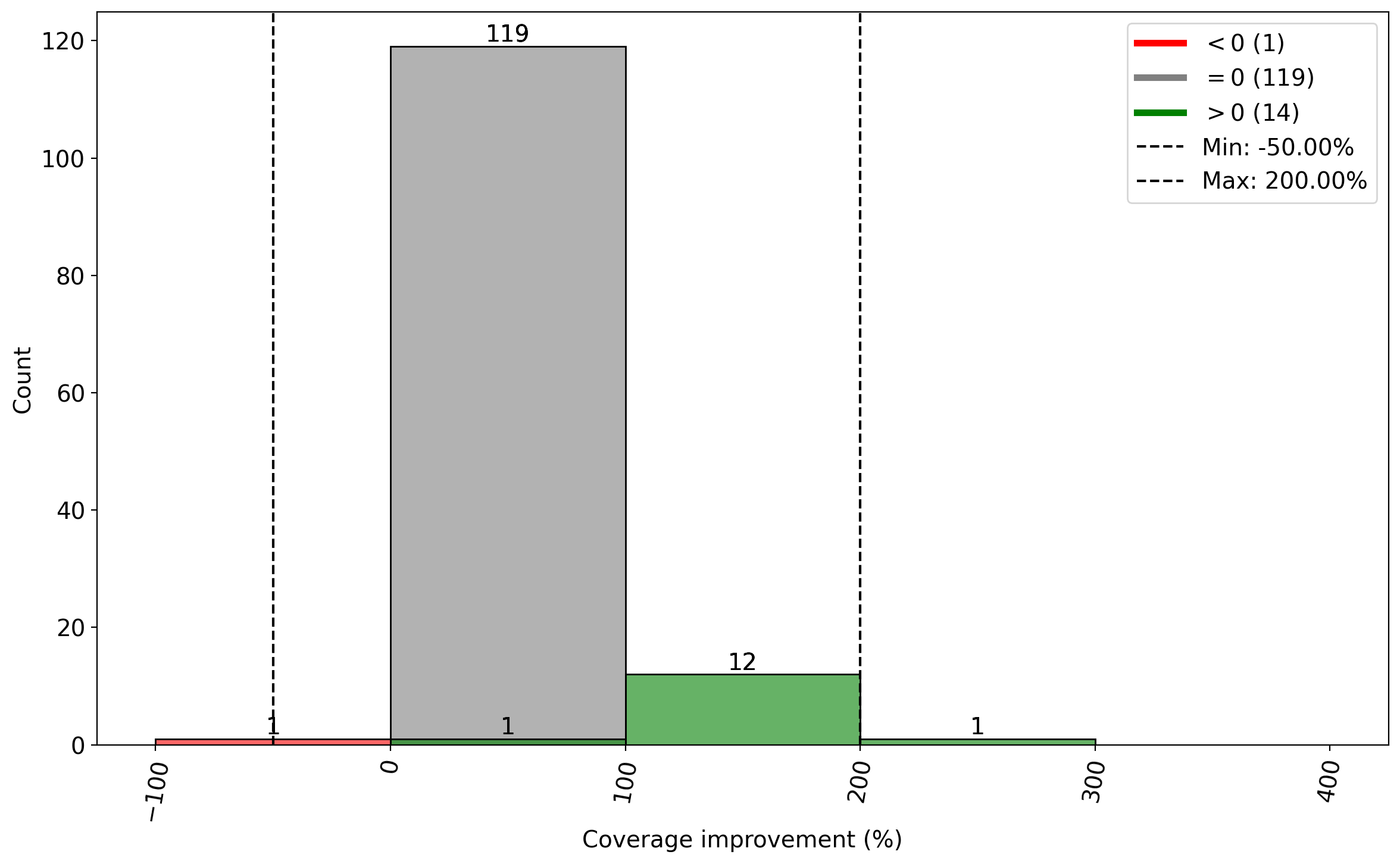}} \\
    \subcaptionbox{CLIM synthetic dataset coverage}{\includegraphics[width=0.49\textwidth]{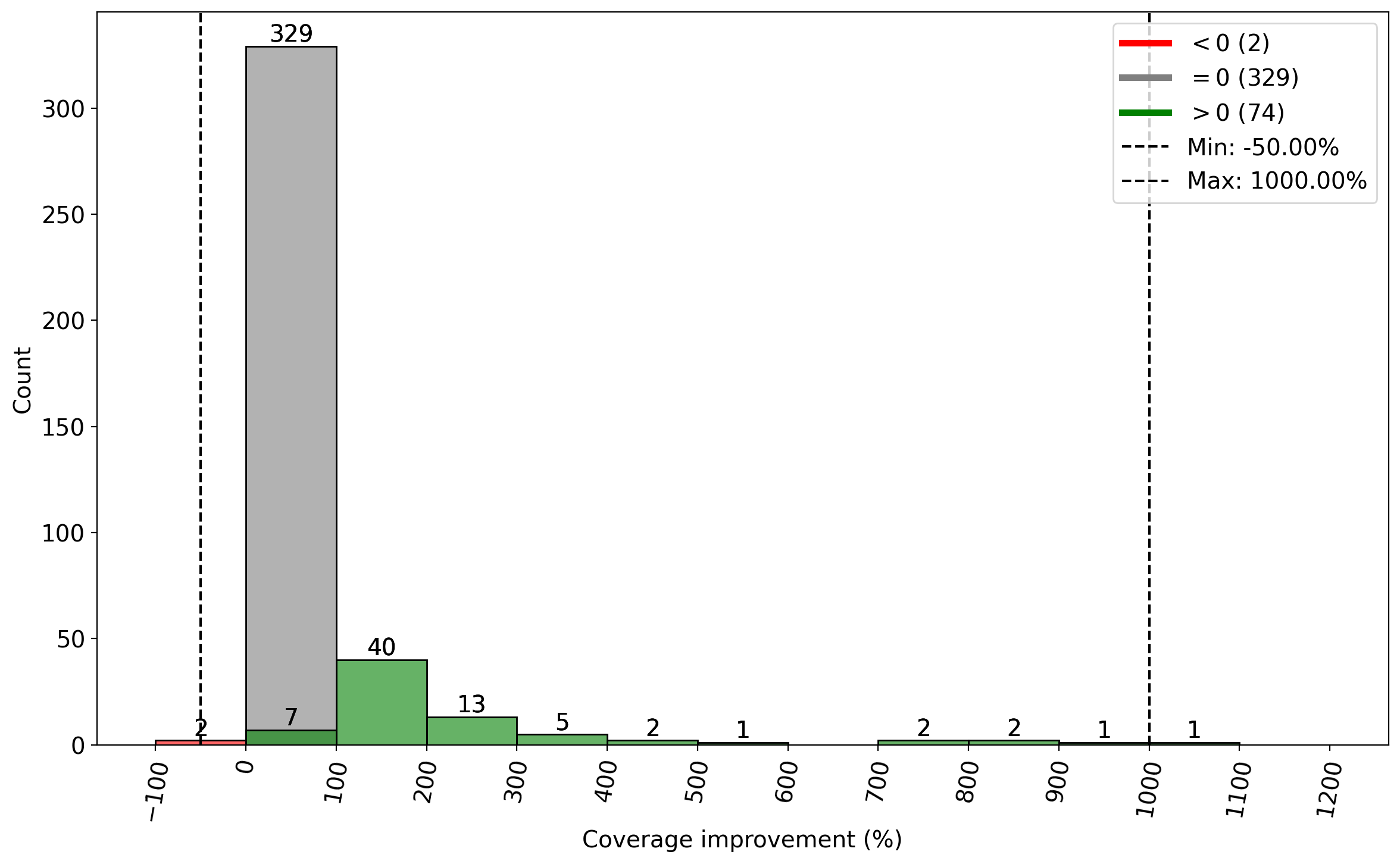}} 
    \subcaptionbox{PARK synthetic dataset coverage}{\includegraphics[width=0.49\textwidth]{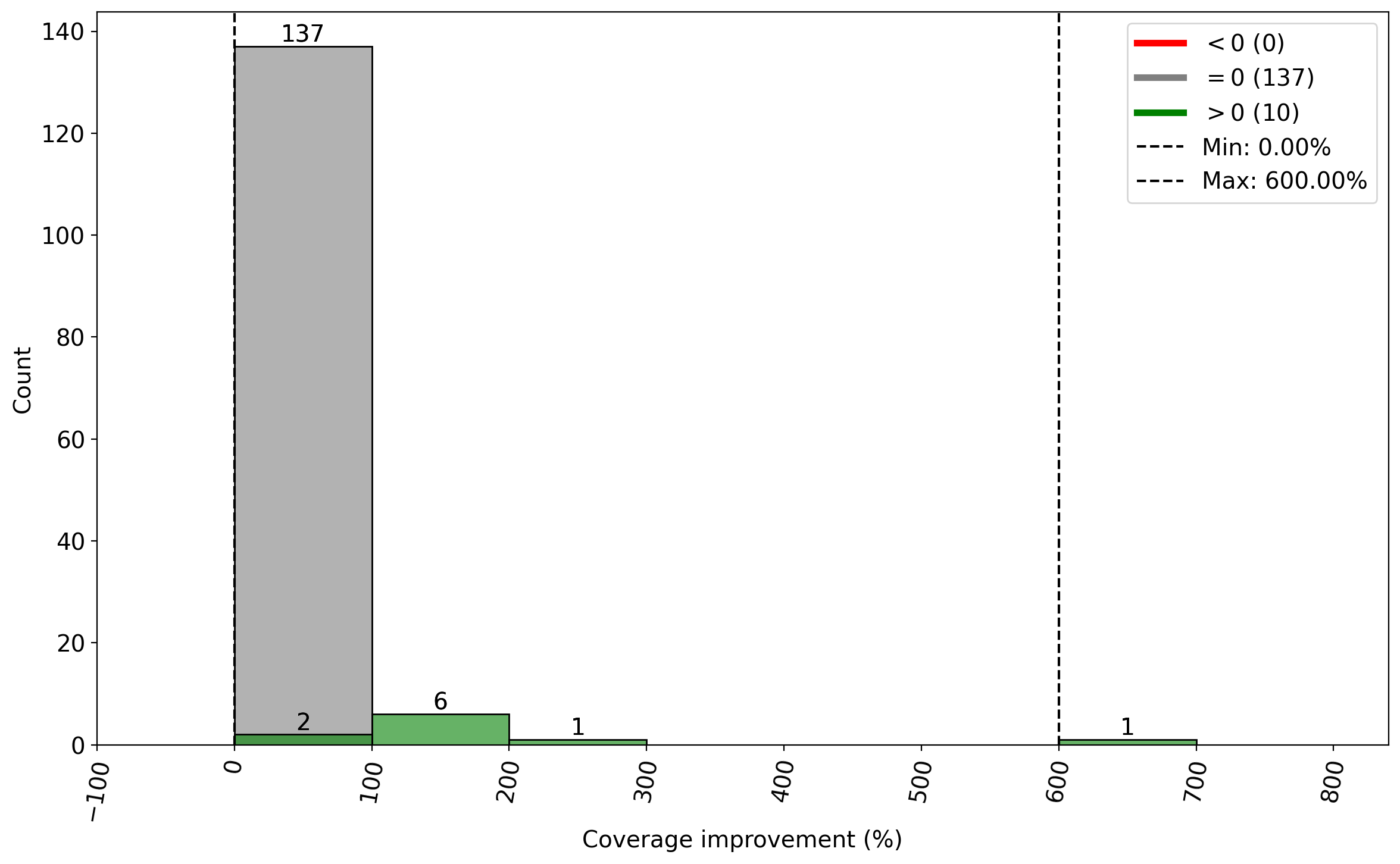}} \\
    \subcaptionbox{BRCW synthetic dataset coverage}{\includegraphics[width=0.49\textwidth]{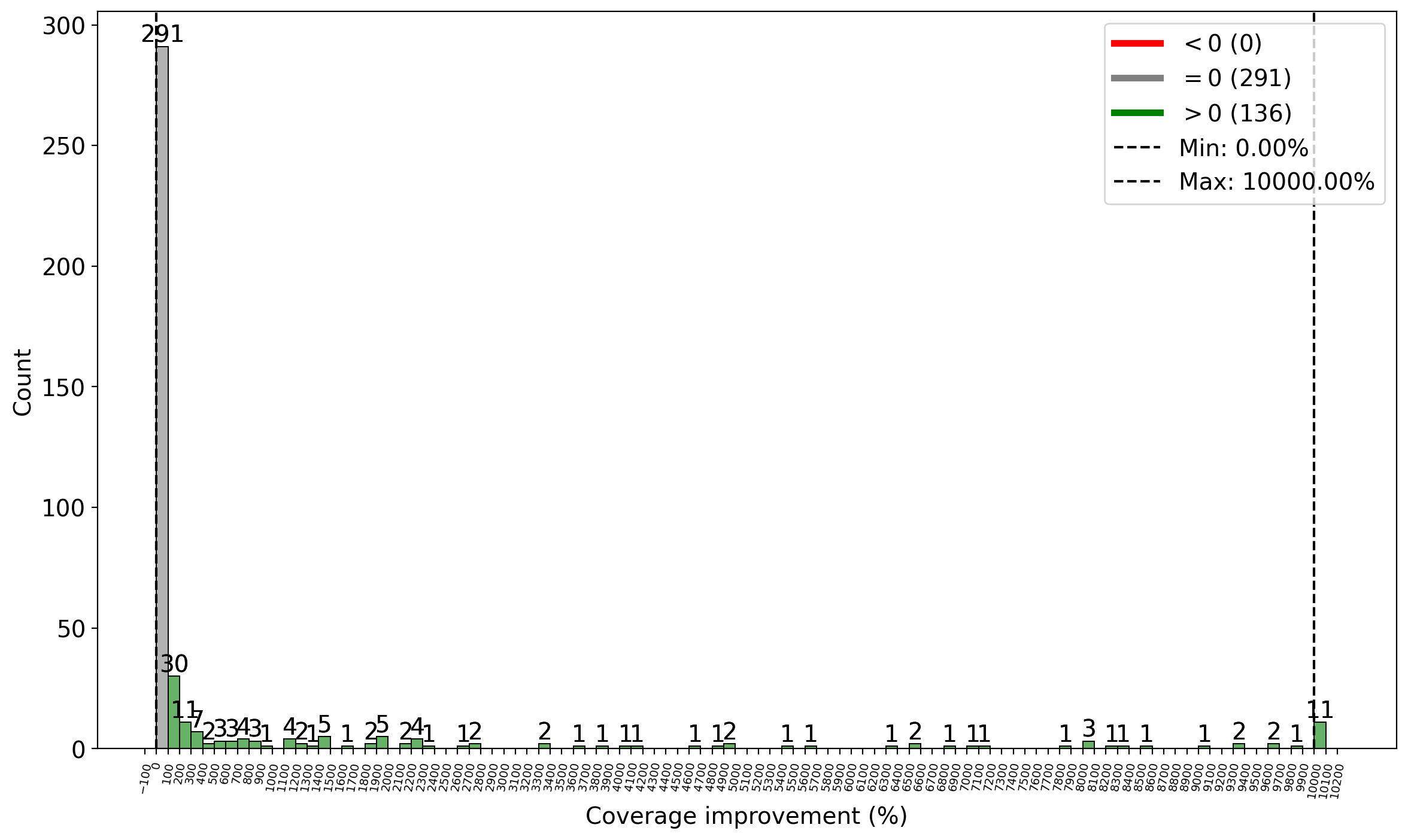}} 
    \subcaptionbox{IONS synthetic dataset coverage}{\includegraphics[width=0.49\textwidth]{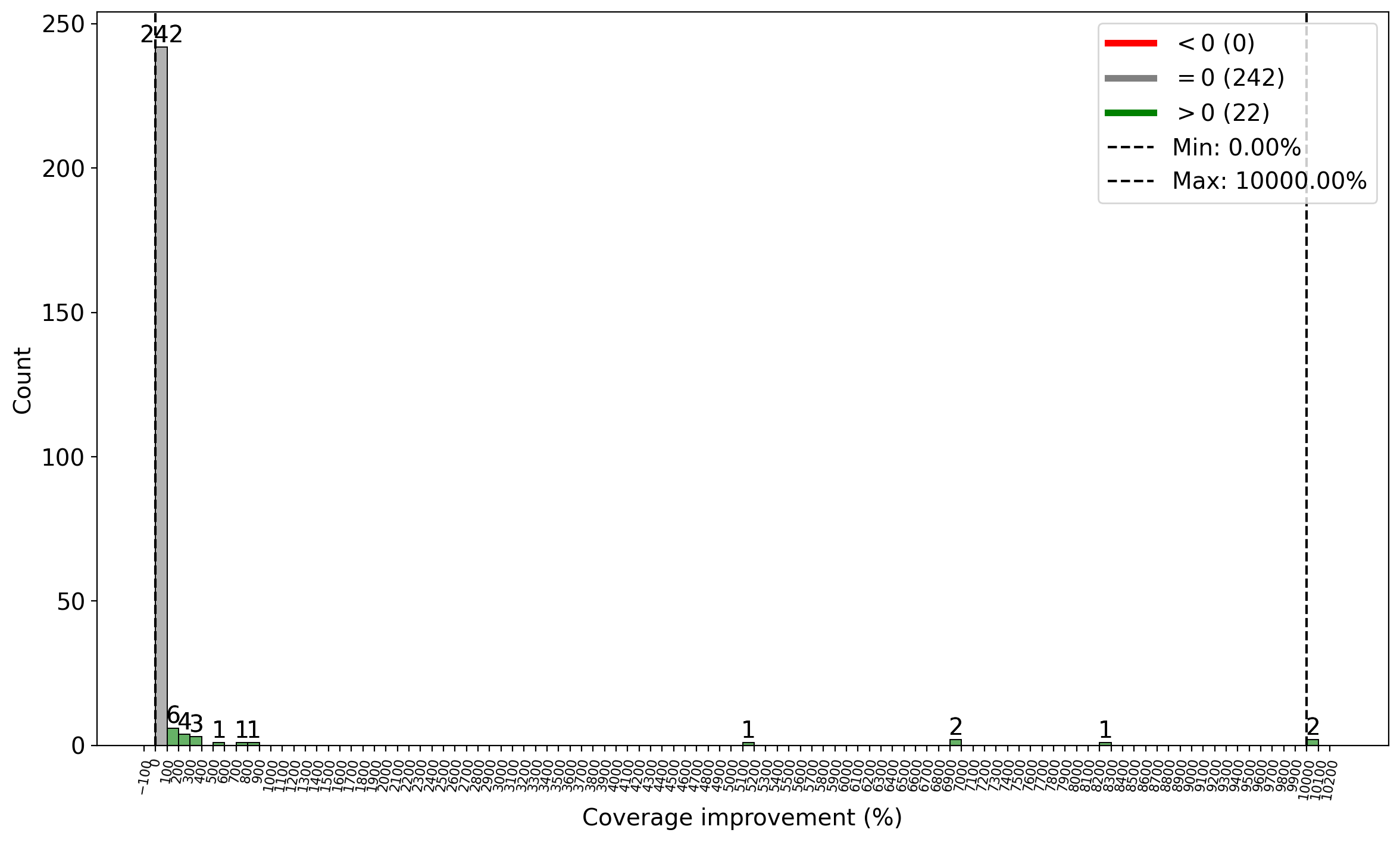}} \\
\end{figure}

\subsubsection{Results for Explanation Range Width}

Finally, the results related to the explanation range width are depicted in Table~\ref{tab:Sum_of_Ranges_SVM} and Table~\ref{tab:Sum_of_Ranges_MLP}. The results show that Twostep was able to increase the range width in most cases. MTwostep achieved results that were similar to or slightly better than Onestep for most datasets, including those with many features. However, Twostep with SVM performed worse for the IRIS and UKMO datasets and had similar results to Onestep for the VRTC and BLDT datasets.

\begin{table}[H]
\caption{Explanation range width results for SVM. $\#M$ represents the mean range width with standard deviation. $\%M$ is the percentage improvement in mean range width of Twostep over Onestep.}
\label{tab:Sum_of_Ranges_SVM}
\centering
\resizebox{1.0\textwidth}{!}{
    \begin{tabular}{ |c|c|c |c|c |c|c |c| }
    \hline
    \multicolumn{1}{|c|}{\multirow{3}{*}{Dataset}} & 
    \multicolumn{1}{c|}{\multirow{2}{*}{Onestep}} & 
    \multicolumn{6}{c|}{Twostep} \\  
    \cline{3-8}& & \multicolumn{2}{c|}{T-Step $0.25$} & \multicolumn{2}{c|}{T-Step $0.50$} & \multicolumn{2}{c|}{T-Step $0.75$} \\
    \cline{2-8}& $\#M$& $\#M$& $\%M$& $\#M$& $\%M$& $\#M$& $\%M$\\ \hline
                        
\rowcolor{lightgray} IRIS & 2.55 $\pm$ 0.43& 2.52 $\pm$ 0.44& -1.18& 2.52 $\pm$ 0.45& -1.18& 2.53 $\pm$ 0.44& -0.78\\\hline
\rowcolor{white} BLDT & 3.29 $\pm$ 0.11&  3.29 $\pm$ 0.11& 0.00& 3.29 $\pm$ 0.11& 0.00& 3.29 $\pm$ 0.11& 0.00\\\hline
\rowcolor{lightgray}BANK& 2.47 $\pm$ 0.13&  2.48 $\pm$ 0.13& 0.40& 2.48 $\pm$ 0.13& 0.40& 2.48 $\pm$ 0.13& 0.40\\\hline
\rowcolor{white} UKMO& 3.55 $\pm$ 0.63&  3.53 $\pm$ 0.62& -0.56& 3.53 $\pm$ 0.62& -0.56& 3.54 $\pm$ 0.62& -0.28\\\hline
\rowcolor{lightgray}VRTC & 4.45 $\pm$ 1.21&  4.45 $\pm$ 1.21& 0.00& 4.45 $\pm$ 1.21& 0.00& 4.45 $\pm$ 1.21& 0.00\\\hline
\rowcolor{white} PIMA & 4.75 $\pm$ 0.80&  4.79 $\pm$ 0.80& 0.84& 4.78 $\pm$ 0.80& 0.63& 4.76 $\pm$ 0.80& 0.21\\\hline
\rowcolor{lightgray}GLAS& 6.17 $\pm$ 0.60&  6.18 $\pm$ 0.59& 0.16& 6.18 $\pm$ 0.60& 0.16& 6.17 $\pm$ 0.60& 0.00\\\hline
\rowcolor{white} WINE & 8.33 $\pm$ 0.76&  8.35 $\pm$ 0.77& 0.24& 8.35 $\pm$ 0.76& 0.24& 8.34 $\pm$ 0.76& 0.12\\\hline
\rowcolor{lightgray}CLIM& 11.75 $\pm$ 1.16&  11.77 $\pm$ 1.16& 0.17& 11.76 $\pm$ 1.16& 0.09& 11.76 $\pm$ 1.16& 0.09\\\hline
\rowcolor{white} PARK & 16.07 $\pm$ 3.00&  16.09 $\pm$ 3.00& 0.12& 16.09 $\pm$ 3.00& 0.12& 16.08 $\pm$ 3.00& 0.06\\\hline
\rowcolor{lightgray}BRCW & 14.86 $\pm$ 4.07&  14.88 $\pm$ 4.07& 0.13& 14.87 $\pm$ 4.07& 0.07& 14.87 $\pm$ 4.07& 0.07\\\hline
\rowcolor{white} IONS & 17.52 $\pm$ 2.08&  17.54 $\pm$ 2.08& 0.11& 17.54 $\pm$ 2.08& 0.11& 17.53 $\pm$ 2.08& 0.06\\\hline
    \end{tabular} 
}
\\
      \makebox[\width]{}
\end{table}

The cases of IRIS, UKMO, and VRTC are particularly interesting because, in Table~\ref{tab:Coverage_SVM}, the IRIS dataset showed a considerable increase in average coverage with Twostep and SVM. Additionally, for these three datasets, Twostep with SVM demonstrated significant improvements in coverage on synthetic data, as shown in Figure~\ref{fig:SVM_artificial_coverage}. Similarly, the explanation range width for Twostep with MLP was worse or unchanged for the UKMO and VRTC datasets. Yet, Twostep achieved substantial improvements in coverage on synthetic data for these datasets, as seen in Figure~\ref{fig:MLP_artificial_coverage}, and also showed notable increases in average coverage, as reported in Table~\ref{tab:Coverage_MLP}. These results further supports the idea that a more strategic expansion of the ranges can lead to a better generalization of explanations, even if there is a reduction in the explanation range width.



\begin{table}[H]
\caption{Explanation range width results for MLP. $\#M$ represents the mean explanation range width, along with standard deviation. $\%M$ is the percentage improvement in mean range widht of Twostep over Onestep.}
\label{tab:Sum_of_Ranges_MLP}
\centering
\resizebox{1.0\textwidth}{!}{
    \begin{tabular}{ |c|c|c |c|c |c|c |c| }
    \hline
    \multicolumn{1}{|c|}{\multirow{3}{*}{Dataset}} & 
    \multicolumn{1}{c|}{\multirow{2}{*}{Onestep}} & 
    \multicolumn{6}{c|}{Twostep} \\  
    \cline{3-8}& & \multicolumn{2}{c|}{T-Step $0.25$} & \multicolumn{2}{c|}{T-Step $0.50$} & \multicolumn{2}{c|}{T-Step $0.75$} \\
    \cline{2-8}& $\#M$& $\#M$& $\%M$& $\#M$& $\%M$& $\#M$& $\%M$\\ \hline
                        
\rowcolor{lightgray} IRIS & 2.09 $\pm$ 0.69& 2.12 $\pm$ 0.65& 1.44& 2.14 $\pm$ 0.63& 2.39& 2.13 $\pm$ 0.65& 1.91\\\hline
\rowcolor{white} BLDT & 2.71 $\pm$ 0.27&  2.72 $\pm$ 0.27& 0.37& 2.73 $\pm$ 0.27& 0.74& 2.72 $\pm$ 0.27& 0.37\\\hline
\rowcolor{lightgray}BANK& 2.31 $\pm$ 0.29&  2.33 $\pm$ 0.29& 0.87& 2.33 $\pm$ 0.29& 0.87& 2.32 $\pm$ 0.29& 0.43\\\hline
\rowcolor{white} UKMO& 3.27 $\pm$ 0.65&  3.23 $\pm$ 0.64& -1.22& 3.23 $\pm$ 0.63& -1.22& 3.25 $\pm$ 0.62& -0.61\\\hline
\rowcolor{lightgray}VRTC & 3.87 $\pm$ 1.37&  3.87 $\pm$ 1.36& 0.00& 3.87 $\pm$ 1.36& 0.00& 3.87 $\pm$ 1.37& 0.00\\\hline
\rowcolor{white} PIMA & 4.40 $\pm$ 0.60&  4.43 $\pm$ 0.59& 0.68& 4.44 $\pm$ 0.59& 0.91& 4.43 $\pm$ 0.59& 0.68\\\hline
\rowcolor{lightgray}GLAS& 3.68 $\pm$ 0.78&  3.72 $\pm$ 0.78& 1.09& 3.70 $\pm$ 0.80& 0.54& 3.68 $\pm$ 0.80& 0.00\\\hline
\rowcolor{white} WINE & 5.87 $\pm$ 1.13&  5.98 $\pm$ 0.92& 1.87& 6.03 $\pm$ 0.79& 2.73& 6.06 $\pm$ 0.73& 3.24\\\hline
\rowcolor{lightgray}CLIM& 12.28 $\pm$ 1.17&  12.31 $\pm$ 1.11& 0.24& 12.32 $\pm$ 1.08& 0.33& 12.34 $\pm$ 1.02& 0.49\\\hline
\rowcolor{white} PARK & 13.70 $\pm$ 3.20&  13.78 $\pm$ 3.17& 0.58& 13.82 $\pm$ 3.10& 0.88& 13.87 $\pm$ 3.06& 1.24\\\hline
\rowcolor{lightgray}BRCW & 16.67 $\pm$ 2.76&  16.81 $\pm$ 2.44& 0.84& 16.96 $\pm$ 2.17& 1.74& 17.07 $\pm$ 2.00& 2.40\\\hline
\rowcolor{white} IONS & 14.36 $\pm$ 3.43&  14.93 $\pm$ 2.82& 3.97& 15.17 $\pm$ 2.51& 5.64& 15.43 $\pm$ 2.26& 7.45\\\hline
    \end{tabular} 
}
\\
      \makebox[\width]{}
\end{table}

\subsection{Brief Case Study}

We will present a brief case study based on the explanations generated by Onestep and Twostep. Let an instance from the IRIS dataset be $\mathbf{x} = \{$\textit{sepal\_length} $ = 4.6$, \textit{sepal\_width} $ = 3.4$, \textit{petal\_length} $ = 1.4$, \textit{petal\_width} $ = 0.3\}$ predicted as class \textit{setosa} by a SVC. The explanation obtained by Onestep that guarantees the predictions is:
\begin{align*}
   \textbf{IF} \quad & 2.51 \leq  \textit{ sepal\_width } \leq 4.40 \\
    \textbf{AND} \quad& 1.00 \leq  \textit{ petal\_length } \leq 1.40 \\
    \textbf{AND} \quad& 0.10 \leq  \textit{ petal\_width }  \leq 0.30 \\
    \textbf{THEN } &\textit{setosa}.
\end{align*}

Comparatively, the explanation obtained by Twostep is:
\begin{align*}
   \textbf{IF} \quad  & 2.84 \leq  \textit{ sepal\_width } \leq 4.40 \\
   \textbf{AND} \quad & 1.00 \leq  \textit{ petal\_length } \leq 1.73 \\
   \textbf{AND} \quad & 0.10 \leq  \textit{ petal\_width }  \leq 0.38 \\
   \textbf{THEN} \quad & \textit{setosa}.
\end{align*}

In this case, the explanations obtained by Onestep and Twostep have the feature \textit{sepal\_length} omitted since it can assume any value within the respective domain without changing the classification result. In comparison with the explanation obtained by Onestep, the explanation obtained by Twostep has a more narrow value interval for \textit{sepal\_width}, which allows a broader interval for both \textit{petal\_length} and \textit{petal\_width}. As a result, the explanation obtained by Twostep achieved coverage of 13 IRIS dataset instances against the coverage of 8 instances for the explanation obtained by Onestep. Therefore, Twostep was able to cover $62.5\%$ more instances than Onestep.

Moreover, let an instance from the VRTC dataset be $\mathbf{x} = \{$\textit{pelvic\_incidence} $= 63.03$, \textit{ pelvic\_tilt} $ = 22.55$, \textit{ lumbar\_lordosis\_angle} $ = 39.61$, \textit{ sacral\_slope} $ = 40.48$, \textit{ pelvic\_radius} $ = 98.67$,\textit{ degree\_spondylolisthesis} $ = -0.25\}$ predicted as class \textit{abnormal} by a SVC. The explanation obtained by Onestep for such a prediction is
\begin{align*}
   \textbf{IF} \quad  & 27.43 \leq \textit{lumbar\_lordosis\_angle} \leq 125.74 \\
   \textbf{AND} \quad & 13.37 \leq \textit{sacral\_slope} \leq 42.38 \\
   \textbf{AND} \quad & 70.08 \leq \textit{pelvic\_radius} \leq 98.67 \\
   \textbf{THEN} \quad & \textit{abnormal}.
\end{align*}

In comparison, the explanation provided by Twostep is 
\begin{align*}
   \textbf{IF} \quad  & 31.24 \leq \textit{lumbar\_lordosis\_angle} \leq 125.74 \\
   \textbf{AND} \quad & 13.37 \leq \textit{sacral\_slope} \leq 51.47 \\
   \textbf{AND} \quad & 70.08 \leq \textit{pelvic\_radius} \leq 98.79 \\
   \textbf{THEN} \quad & \textit{abnormal}.
\end{align*}

For this case, the explanations obtained by Onestep and Twostep have the features \textit{pelvic\_incidence}, \textit{ pelvic\_tilt}, and \textit{ degree\_spondylolisthesis} omitted since they can assume any value within their respective domain without changing the predictions. The method Twostep shortened the value interval of \textit{lumbar\_lordosis\_angle} in favor of broadening the intervals of \textit{sacral\_slope} and \textit{pelvic\_radius}. This change enabled the explanation obtained by Twostep to achieve coverage of 6 VRTC dataset instances against the coverage of 4 instances for the explanation obtained by Onestep, covering $50\%$ more instances.

These cases make more evident how Twostep differs from Onestep, often shortening the ranges of some features and widening the ranges of others whenever possible. It is also significant for Twostep, as it highlights the potential it has to achieve greater instance coverage compared to Onestep.

\subsection{More Examples of Explanations}

In the previous section, we presented explanations for instances from the IRIS and VRTC datasets. In this subsection, we now show examples of explanations generated by Twostep and Onestep for the PIMA and WINE datasets. These datasets are used to visualize explanations for datasets with more features.

\noindent For the PIMA dataset, the original instance has the following values:

\vspace{-1cm}
\begin{center}
\resizebox{1\textwidth}{!}{
\begin{minipage}{0.99\textwidth}
\begin{align*}
   \textbf{Original Instance:} \\
    \text{Pregnancies} = 7.00, \quad &\text{Glucose} = 194.00, \\
    \text{Blood Pressure} = 68.00, \quad &\text{Skin Thickness} = 28.00, \\
    \text{Insulin} = 0.00, \quad &\text{BMI} = 35.90, \\
    \text{Diabetes Pedigree Function} = 0.745, \quad &\text{Age} = 41.00
\end{align*}
\end{minipage}
}
\end{center}

\vspace{1cm}
\noindent The explanations generated by both method are as follows:
\vspace{-1cm}
\begin{center}
\resizebox{1.0\textwidth}{!}{
\begin{minipage}{0.48\textwidth}
    \begin{align*}
    &\textbf{Onestep Explanation:}  \\
     \textbf{IF} \quad &188.96 \leq \text{Glucose} \leq 199.00 \\
     \textbf{AND} \quad &35.66 \leq \text{BMI} \leq 67.10 \\
     \textbf{THEN} \quad &\textit{Positive.}
    \end{align*}
\end{minipage}
\hfill
\begin{minipage}{0.48\textwidth}
    \begin{align*}
    &\textbf{Twostep Explanation:} \\
     \textbf{IF} \quad &190.56 \leq \text{Glucose} \leq 199.00 \\
     \textbf{AND} \quad &34.45 \leq \text{BMI} \leq 67.10 \\
     \textbf{THEN} \quad &\textit{Positive.}
    \end{align*}
\end{minipage}
}
\end{center}

\vspace{1cm}

\noindent For the WINE dataset, the original instance has the following values:
\vspace{-1cm}
\begin{center}
\resizebox{1.0\textwidth}{!}{
\begin{minipage}{0.99\textwidth}
\begin{align*}
   \textbf{Original Instance:} \\
    \text{alcohol} = 13.88, \quad &\text{malic acid} = 5.04, \\
    \text{ash} = 2.23, \quad &\text{alcalinity of ash} = 20.00, \\
    \text{magnesium} = 80.00, \quad &\text{total phenols} = 0.98, \\
    \text{flavanoids} = 0.34, \quad &\text{nonflavanoid phenols} = 0.40, \\
    \text{proanthocyanins} = 0.68, \quad &\text{color intensity} = 4.90, \\
    \text{hue} = 0.58, \quad &\text{od280/od315} = 1.33, \\
   & \text{proline} = 415.00
\end{align*}
\end{minipage}
}
\end{center}
\pagebreak
\noindent The explanations generated by both methods are:
\vspace{-1cm}
\begin{center}
\resizebox{1.0\textwidth}{!}{
\begin{minipage}{0.48\textwidth}
    \begin{align*}
    &\textbf{Onestep Explanation:} \\  
     \textbf{IF} \quad &0.98 \leq \text{total phenols} \leq 1.95 \\
     \textbf{AND} \quad &0.34 \leq \text{flavanoids} \leq 0.34 \\
     \textbf{AND} \quad &1.27 \leq \text{od280/od315} \leq 1.33 \\
     \textbf{AND} \quad &278.00 \leq \text{proline} \leq 415.00 \\
     \textbf{THEN} \quad &\textit{Positive.} 
    \end{align*}
\end{minipage}
\hfill
\begin{minipage}{0.48\textwidth}
    \begin{align*}
    &\textbf{Twostep Explanation:} \\  
     \textbf{IF} \quad &0.98 \leq \text{total phenols} \leq 1.57 \\
     \textbf{AND} \quad &0.34 \leq \text{flavanoids} \leq 0.48 \\
     \textbf{AND} \quad &1.27 \leq \text{od280/od315} \leq 1.37 \\
     \textbf{AND} \quad &278.00 \leq \text{proline} \leq 423.01 \\
     \textbf{THEN} \quad &\textit{Positive.} 
    \end{align*}
\end{minipage}
}
\end{center}


\section{Conclusions}\label{Conclusions}

In this work, we proposed two approaches, Onestep and Twostep, aimed at improving the generalization of logic-based explanations through optimization. Both methods offer guarantees of correctness and minimality, ensuring that the explanations remain reliable and succinct. Onestep serves as a reference point, improving upon the method proposed in earlier work \citep{izza2023delivering} by eliminating the incremental process, which makes it more efficient and results in more comprehensive explanations. Twostep builds on this by introducing a more flexible expansion process. It aims to further improve generalization by addressing cases where feature ranges remain too narrow.




We compared Onestep and Twostep across 12 different datasets. Onestep, improving upon previous work \citep{izza2023delivering}, serves as a reference for assessing the improvements brought by Twostep. By comparing Twostep with Onestep, we indirectly evaluate its performance relative to prior methods. Both methods were applied to the following classifiers: MLP and linear SVC. The results suggest that the more strategic expansion by Twostep, though slightly slower than Onestep, achieved higher coverage while maintaining similar or even better explanation range width.


Both Onestep and Twostep have clear limitations. One notable issue is the loss of coverage as data dimensionality increases. While our methods achieve a good degree of generalization, they may still be insufficient for more complex datasets. Additionally, since both methods build on a minimal explanation, they may miss explanations with fewer features, which we cannot guarantee to find. These smaller explanations could potentially offer higher coverage, especially when considering that we are only evaluating the explanations through instances present in the datasets.

Future work can explore several directions to enhance both methods. One possible improvement is to analyze the instance distribution beforehand to gain a deeper understanding of how features are structured within the instance space. This analysis could help enhance generalization by prioritizing feature ranges that provide broader coverage. Another promising direction is formulating the range expansion task as a cost optimization problem, where a cost function evaluates the quality of value ranges for each feature. This would enable a solver to systematically refine the lower and upper bounds, potentially leading to more efficient range distributions. Additionally, our techniques could be adapted to generate contrastive explanations, which could offer clearer insights by highlighting why a particular decision was made instead of an alternative. Ultimately, these enhancements could contribute to more insightful explanations, improving the interpretability and understanding of the models.

\section{Acknowledgments}
The authors thank FUNCAP, CNPq, and CAPES for partially supporting our research work.

\bibliography{refs}

\end{document}